\documentclass[aps,prc,preprint,groupedaddress,amsmath,floatfix]{revtex4-1}

\usepackage{graphicx}% Include figure files
\usepackage{braket}
\usepackage{bm}% bold math
\usepackage{longtable}

\usepackage{amsmath}	% required for `\align' (yatex added)
\begin{document}

\title{Antisymmetrized molecular dynamics studies \\for exotic clustering phenomena in neutron-rich
nuclei} 
%\subtitle{}
\author{M. Kimura}
\affiliation{Department of Physics, Hokkaido University, Sapporo 060-0807, Japan} 
\affiliation{Nuclear Reaction Data Centre, Faculty of Science, Hokkaido University, Sapporo
060-0810, Japan} 
\author{T. Suhara}                     % Do not remove
\affiliation{Matsue College of Technology, Matsue 690-8518, Japan}
\author{Y. Kanada-En'yo}
\affiliation{Department of Physics, Kyoto University, Kyoto 606-8502, Japan}
%
%\titlerunning{AMD studies for exotic clustering phenomena in neutron-rich nuclei}
\date{Received: date / Revised version: date}
% The correct dates will be entered by Springer
%
\begin{abstract}
We present a review of recent works on clustering phenomena in unstable nuclei studied by 
antisymmetrized molecular dynamics (AMD). The AMD studies in these decades have uncovered novel
types of clustering phenomena brought about by the excess neutrons. Among them, this review
focuses on the molecule-like structure of unstable nuclei. \\    
One of the earliest discussions on the clustering in unstable nuclei was made for neutron-rich Be
and B isotopes. AMD calculations predicted that the ground state clustering is enhanced or
reduced depending on the number of excess neutrons. Today, the experiments are confirming this
prediction as the change of the proton radii.
Behind this enhancement and reduction of the clustering, there are underlying
shell effects called molecular- and atomic-orbits. These orbits form covalent and ionic bonding of
the clusters analogous to the atomic molecules. It was found that this ``molecular-orbit picture''
reasonably explains the low-lying spectra of Be isotopes.  The molecular-orbit picture is extended
to other systems having parity asymmetric cluster cores and to the 
three cluster systems. O and Ne isotopes are the candidates of the former, while the $3\alpha$
linear chains in C isotopes are the latter. For both subjects, many intensive
studies are now  in progress.\\ 
We also pay a special attention to the observables which are the fingerprint of the clustering. In
particular, we focus on the monopole and dipole transitions which are recently regarded as good
probe for the clustering. We discuss how they have and will reveal the exotic clustering.  
\end{abstract}

\pacs{Valid PACS appear here}% PACS, the Physics and Astronomy
\maketitle

\section{Introduction} \label{sec:intro}
\begin{figure}[h] 
 \begin{center}
  \resizebox{0.7\textwidth}{!}{
  \includegraphics{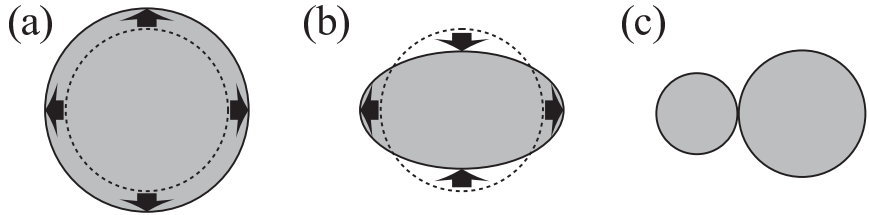}}
  \caption{(a) Nuclear excitation mode that changes the matter density. (b) Nuclear shape
  fluctuation  and (c) Clustering are the excitation modes that conserve matter density.} 
  \label{fig:int1}       % Give a unique label
 \end{center}
\end{figure}

\begin{figure}[h]
 \begin{center}
  \resizebox{0.8\textwidth}{!}{
  \includegraphics{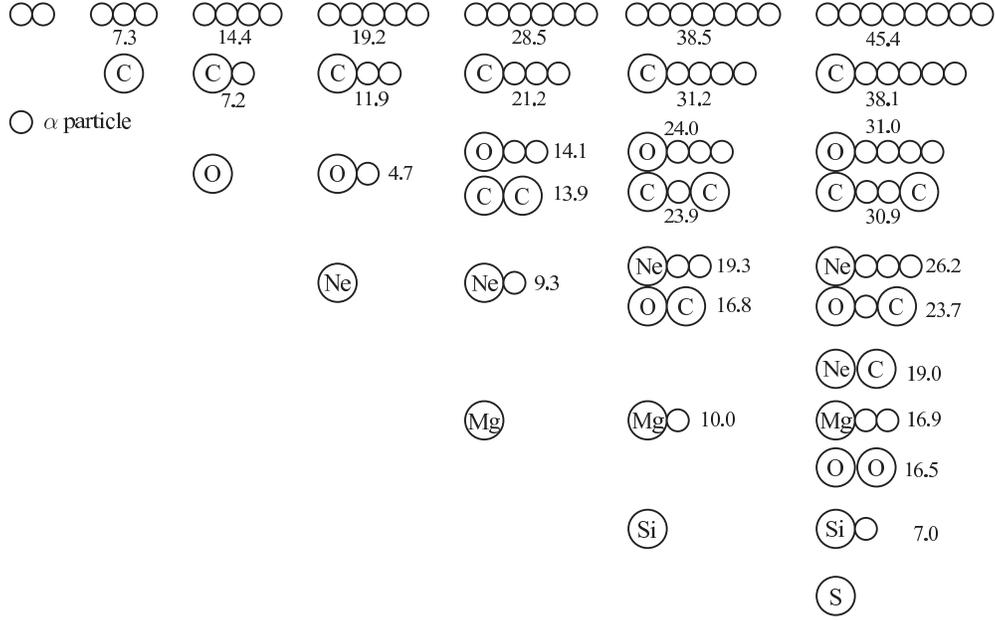}}
  \caption{Ikeda threshold rule \cite{Ikeda:68a} suggests that cluster states of the stable nuclei
  should appear approximately at the excitation energies that are required to decompose the
  nucleus into clusters.}  
  \label{fig:int2}       % Give a unique label
 \end{center}
\end{figure}

Atomic nuclei and symmetric nuclear matter have large incompressibility, and hence, both of the
 matter density and the energy density are kept almost constant in stable nuclei
(saturation of energy and density). As a result, in addition to the single-particle excitations,
the excitation modes that conserve the matter density are dominant in the low-lying states of
stable nuclei, because the excitation modes which change the matter density (Fig. \ref{fig:int1}
(a)) require much larger energy. There may be two possible ways to  excite nucleus without
changing the matter density. The first is the fluctuation of nuclear shape (Fig. \ref{fig:int1}
(b)) known as collective vibrations and the other is the decomposition of nucleus into subunits
(clusters) called clustering (Fig. \ref{fig:int1} (c)). Based on the saturation of energy and
matter densities, Ikeda threshold rule \cite{Ikeda:68a} (Fig. \ref{fig:int2}) tells that the
cluster 
states should appear in the vicinity of the  excitation energy required to decompose a nucleus
into clusters. A variety of cluster states in stable nuclei has been studied in detail and
experimentally identified up to $A\simeq 20$ stable nuclei
\cite{Wildermuth:1977a,Fujiwara-supp,Furutani:1980a}. Today, the cluster models and experimental 
studies are verifying this picture in heavier nuclei up to the beginning of the $pf$-shell region
\cite{Michel98a,Yamaya98a,Horiuchi:2010aa}.  

\begin{figure}[h]
 \begin{center}
  \resizebox{0.7\textwidth}{!}{
  \includegraphics{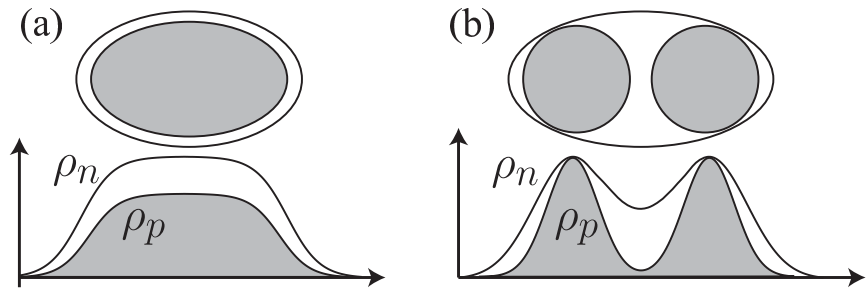}}
  \caption{(a) Density profile of neutron-rich nuclei where the symmetry energy is {\it
  globally} minimized. (b) Density profile of clustered neutron-rich nuclei where the symmetry
  energy is  {\it locally} minimized. } 
  \label{fig:int3}       % Give a unique label
 \end{center}
\end{figure}

In the last decades, the study of unstable nuclei has altered our basic knowledge on atomic
nuclei: Both of the saturation of density and energy are broken near the neutron drip line
\cite{Tanihata:1986kh,Tanihata:2013jwa}. Shell structure is changed from the ordinary one and
magic numbers disappear or migrate \cite{Sorlin:2008jg}. 
As a natural consequence, our interests to nuclear clustering are greatly 
renewed. Since both of the energy and matter density saturation are broken, Ikeda threshold rule
cannot be straightforwardly applied to unstable nuclei and must be reconsidered. For example, an
extension of Ikeda diagram to unstable nuclei called ``extended Ikeda diagram'' was suggested by
von Oertzen \cite{oertzen-ne}.  Furthermore, because of the imbalance of the proton and neutron
numbers, the effect of the symmetry energy  (difference between the proton and neutron densities)
must be taken into account for the discussion of the clustering in unstable nuclei. For
example, it is interesting to consider if the excess neutrons (protons) will diminish or enhance
the clustering. One may consider that the clustering will be diminished to minimize the symmetry 
energy as illustrated in  Fig. \ref{fig:int3} (a). However, this is not the unique solution. One
can also find another possibility as illustrated in Fig. \ref{fig:int3} (b) where the nucleus is
clustered to minimize  the symmetry energy {\it locally} \cite{Horiuchi:2000aa}. 
Thus, the evolution of the nuclear clustering toward neutron drip line and the effect of excess
neutrons to the clustering are non-trivial fascinating problems and must be examined carefully.   

In the above consideration, an important ingredient, the quantum shell effect, is neglected.
Since the shell structure and magic numbers are changed, the shell correction energy
is different from stable nuclei and must affect the clustering of unstable nuclei. In
some cases, the clustering can be  enhanced with the assist of shell effect, but in some other
cases it will be diminished. Furthermore, we expect that a special class of the shell structure is
formed  around the clustered core, which is different from the ordinary shell formed around
spherical core.  
\begin{figure}[h]
 \begin{center}
  \resizebox{0.7\textwidth}{!}{
  \includegraphics{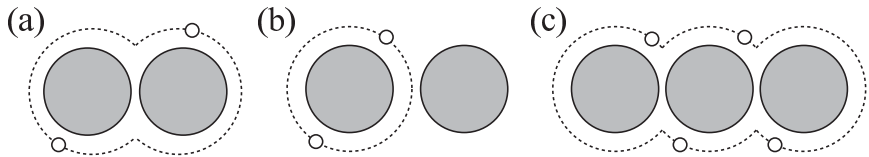}}
  \caption{(a) Molecular orbit in which excess neutron orbit around the clustered core leading to
  the covalent bonding of clusters. (b) Atomic orbit in which the motion of excess neutron is
  localized to one of clusters leading to the ionic bonding of clusters. (c) Trimer of
  clusters. Such exotic cluster state might be realized in neutron-rich nuclei with the assist of
  covalent neutrons.} 
  \label{fig:int4}       % Give a unique label
 \end{center}
\end{figure}
The possible formation of such novel type of shell structure in neutron-rich nuclei has long been
discussed by many authors \cite{SEYA,OERTZENa,OERTZENb,oertzen03-rev,Oertzen-rev,Freer:2007rev},
and the orbits are, roughly 
speaking, classified into two 
types; so-called ``molecular orbit'' and ``atomic orbit''. The former extends to the entire system
leading to the covalent bonding of clusters analogous to the atomic molecules (Fig. \ref{fig:int4}
(a)) , while the latter is localized to one of clusters leading to the ionic bonding of
clusters (Fig. \ref{fig:int4} (b)). With the formation of those orbits, one can imagine that 
very exotic cluster states which cannot be realized in stable nuclei are stabilized with the assist
of excess neutrons. One such example is the linear-chain of clusters in which three or more
clusters are linearly aligned (Fig. \ref{fig:int4} (c)) 
\cite{Morinaga:1956zza,Morinaga:1966,Oertzen:2004,Itagaki:2001mb,Suhara:2010ww,Maruhn:2010dtc,Baba:2014lsa,Freer:2014gza,Fritsch:2016vcq}. 
In addition to this, formation of molecular or atomic orbits will bring about a novel type of
correlations between valence neutrons. They will also yield new excitation modes where the
interplay between the inter-cluster motion and excess  neutrons plays the central role. 

In the last decades, numerous studies have been devoted to answer those questions
\cite{Oertzen-rev,Freer:2007rev,KanadaEnyo:2010aa,Ikeda:2010aa,Oertzen:2014aa}. 
Today, we know that the above-mentioned phenomena do realize in many isotopes. The purpose of
this article is to review those clustering phenomena in unstable nuclei from a theoretical
point-of-view provided by the studies based on antisymmetrized molecular dynamics (AMD)
\cite{KanadaEnyo:2010aa,KanadaEn'yo:2001qw,KanadaEnyo:2003aa,KanadaEn'yo:2012bj}. The AMD studies
predicted that the excess neutrons enhances the clustering in Be and B isotopes toward neutron
drip line as illustrated in Fig. \ref{fig:int3} (b), which was recently confirmed from the
measurement of the proton and matter radii. A similar cluster enhancement in Ne isotopes was also
predicted. Behind this enhancement of the clustering, the shell effect plays
an important role. It was shown that the molecular orbits are formed in Be and Ne
 isotopes and it is the driving force to enhance (and reduce in some cases)
the clustering. This finding is motivating the search for very exotic cluster structure of 
linear-chain states in neutron-rich C isotopes which is composed of  linearly
aligned three $\alpha$ particles. AMD calculations predicted that the assist of the
valence neutrons stabilizes the linear-chain configuration and builds a couple of rotational bands
. Recently, a couple of experiments found rather convincing data for the predicted linear-chain
bands in $^{14}{\rm C}$ and the discussion is going to be extended to heavier systems such as
$^{16}{\rm C}$ .

This article is organized as follows. The next section gives a brief explanation of the
theoretical framework of AMD on which the discussions in this review rely. In the section
\ref{sec:3}, we discuss the evolution of the clustering in the ground states of light elements
from Be to Ne as function of neutron number. In the section \ref{sec:4}, the formation of
molecular and atomic orbits are discussed. We first review the molecular and atomic orbits in Be
and B isotopes. Then, analogous orbits in O, F and Ne isotopes are discussed. By extending the
concept of molecular orbit to three cluster systems, the formation of the linear-chain
configurations in C isotopes is also discussed. In the section  \ref{sec:5}, we discuss the
excitation modes peculiar to the clustering. We focus on the  monopole and dipole excitations
which are known to be sensitive to the clustering. We will see how the excess neutrons affect
those excitation modes. In the last section, we summarize present status. 

\section{Theoretical framework of antisymmetrized molecular dynamics}\label{sec:2}
Since AMD was firstly applied to the nuclear structure problems in 1990's
\cite{KanadaEnyo:1995amd}, its 
framework has been continuously extended and developed. As a result, there are many different
versions of AMD which are tailored for various nuclear structure and reaction
problems. We here explain the most basic framework which is commonly used. The variation from
this basic framework will be explained in  each section.

\subsection{Hamiltonian and variational wave function}
In the AMD framework, we employ the microscopic $A$-body Hamiltonian which reads,
\begin{align}
 H = \sum_{i=1}^A t(i) - t_{cm} + \sum_{i<j}^A v_n(ij) + \sum_{i<j}^Z v_C(ij).
\end{align}
Here $t(i)$ and $t_{cm}$ respectively denote the kinetic energies of the $i$th nucleon and the
center-of-mass. The center-of-mass wave function is analitycally separable from the variational
wave function and $t_{cm}$ is exactly removed. Hence, the AMD framework is free from the spurious
center-of-mass motion. For the nucleon-nucleon interaction $v_n(ij)$, various effective
interactions have been utilized. In this article, the Volkov interaction \cite{Volkov:1965zz}
combined with the spin-orbit part of the G3RS interaction \cite{Tamagaki:1968zz} was used in the
sections \ref{sec:3.1},  \ref{sec:4.3} and  \ref{sec:5.2}. The modified Volkov interaction
\cite{Ando:1980hp} combined with the G3RS spin-orbit interaction was 
used in the sections \ref{sec:3.1}, \ref{sec:4.1},  \ref{sec:5.1}, and \ref{sec:5.2}. 
The Gogny D1S interaction
\cite{Berger:91} was used in the sections \ref{sec:3.2}, \ref{sec:4.2} and \ref{sec:5.1}.
In calculations using the Volkov and modified Volkov interactions, it is difficult to 
globally fit binding energies in a wide mass-number region with a fixed parametrization.
Therefore, the parameters $w$, $m$, $b$, and $h$ for the Wigner, Majorana, Bartlett, 
and Heisenberg terms are often modified and treated as adjustable parameters. In this article, $w$
is set as $w=1-m$.  The strength parameters $u_1$ and $u_2$  
of the G3RS spin-orbit interaction are modified from the original values and adjusted 
for each central interaction to reproduce low-energy spectra.
The detail of the adopted parametrization is explained in
each section. The Coulomb interaction $v_C(ij)$ is approximated by the sum of Gaussians.  

The intrinsic wave function $\Phi_{int}$  is represented by a Slater determinant of single 
particle wave packets,
\begin{align}
 \Phi_{int}&=\frac{1}{\sqrt{A!}}{\mathcal A} \{\varphi_1,\varphi_2,...,\varphi_A \},
\end{align}
where $\varphi_i$ denotes the single nucleon wave packet expressed by a Gaussian oriented at the
complex valued three dimensional vector $\bm Z_i$,
\begin{align}
 \varphi_i({\bm r}) &=  \left(\frac{2\nu}{\pi}\right)^{3/4}
 \exp\Set{-\nu\left(\bm r - \frac{\bm Z_{i}}{\sqrt{\nu}}\right)^2+\frac{1}{2}Z_i^2}\chi_i\xi_i.
 \label{eq:singlewf} 
\end{align}
Here $\chi_i$ is the spinor and $\xi_i$ is the isospin fixed to proton or neutron. The use of the
Gaussian wave packets make the AMD wave function very flexible. Without any assumption for the
formation of constituent clusters, multi-cluster structures can be described by the spatially
localized groups of Gaussian wave packets. On the other hand, if all of the Gaussian centroids
gather at the same position, the AMD wave function becomes equivalent to a harmonic oscillator
shell-model wave function owing to the antisymmetrization effect. Thus, the AMD wave function is
able to describe both of the shell and cluster structures in an unified way. The energy
minimization explained below determines which structure, cluster or shell, is energetically
favored. This advantage and feature of AMD is common to  fermionic molecular dynamics
(FMD) \cite {Feldmeier:2000cn,Neff:2008aa} which also employs the Gaussian wave packets. 

Instead of the spherical Gaussian, triaxially deformed Gaussian wave packet is also used as the
single-particle wave packet \cite{Kimura:2003uf},
\begin{align}
 \varphi_i({\bm r}) &=  \prod_{\sigma=x,y,z}\left(\frac{2\nu_\sigma}{\pi}\right)^{1/4}\nonumber\\
 &\times
 \exp\Set{-\nu_\sigma\left(r_\sigma - \frac{ Z_{i\sigma}}{\sqrt{\nu_\sigma}}\right)^2+\frac{1}{2}Z_{i\sigma}^2}\chi_i\xi_i.
 \label{eq:singlewf} 
\end{align}
The use of the deformed Gaussian is effective to describe the formation of the deformed mean-field
in the ground states of $sd$-$pf$-shell nuclei.

The intrinsic
wave function is projected to the eigenstate of parity and angular momentum,
\begin{align}
 \Phi^\pi&=P^\pi\Phi_{int}=\frac{1+\pi P_x}{2}\Phi_{int},\\
 \Phi^{J^\pi}_{MK} &= P^{J}_{MK}\Phi^\pi=
 \frac{2J+1}{8\pi^2}\int d\Omega D^{J*}_{MK}(\Omega)R(\Omega)\Phi^{\pi}.
\end{align}
Here, $P^\pi$ and $P^{J}_{MK}$ are the parity and angular momentum
projectors. $D^{J}_{MK}(\Omega)$, $P_x$ and $R(\Omega)$ are the Wigner $D$ function, parity and
rotation operators, respectively.  

\subsection{Energy minimization and superposition of the wave functions}
The centroids of the Gaussian wave packets $\bm Z_i$, Gaussian width $\nu$ and spinors
$\chi_i$ are the parameters of the variational wave function, which are determined by the energy 
minimization using the frictional cooloing equation, 
\begin{align}
 i\hbar\frac{d}{dt} X &= (\lambda+i\mu)\frac{\partial \mathcal{H}}{\partial X^*},
 \label{eq:form:fric}
 \end{align}
where $X$ denotes $\bm Z_i$, $\nu$ or $\chi_i$.  $\lambda$ and $\mu$ are arbitrary numbers, but
$\mu$ must have negative sign. The energy of the system $\mathcal H$ is defined as   
 \begin{align}
 {\mathcal H} &\equiv \left\{
 \begin{array}{l}
  \displaystyle{\frac{\braket{\Phi^\pi|H|\Phi^\pi}}{\braket{\Phi^\pi|\Phi^\pi}}}
   \quad\text{for VBP},\\
  \displaystyle{\frac{\braket{\Phi^{J^\pi}_{MK}|H|\Phi^{J^\pi}_{MK}}}
   {\braket{\Phi^{J^\pi}_{MK}|\Phi^{J^\pi}_{MK}}}}
   \quad\text{for VAP},
 \end{array}
 \right.
\end{align}
As time $t$ being evolved by the Eq. (\ref{eq:form:fric}), the energy of the system is decreased 
and we obtain the set of parameters which minimizes the energy of the system.
The calculation which employs the eigenstate of the parity $\Phi^\pi$ to evaluate the energy
is called variation before angular momentum projection (VBP), while the calculation
employing the eigenstate of parity and angular momentum $\Phi^{J\pi}_{MK}$ is called variation
after angular momentum projection (VAP) \cite{KanadaEn'yo:1998rf}. In the case of VBP, the angular
momentum projection is performed after the energy minimization. 

To obtain the wave functions which have the configurations different from the energy minimum, the 
constrained energy minimization is often performed in VBP calculations. For example, the
constraint on the quadrupole deformation parameter $\beta$ is imposed by adding the constraint
potential to the energy as
\begin{align}
 \mathcal H&=\frac{\langle \Phi^\pi| H|\Phi^\pi\rangle}{\langle
  \Phi^\pi|\Phi^\pi\rangle} +  v_\beta(\langle\beta\rangle-\beta_0)^2.
\end{align}
Here, the magnitude of constraint potential $v_\beta$ is chosen large enough so that
the deformation parameter $\braket{\beta}$ of the wave function approximately equals to $\beta_0$
after the energy minimization. Another way to obtain the excited configurations is
orthogonalization method \cite{KanadaEn'yo:1998rf} that is used in VAP calculatons in which the
variational wave function is orthogonalized to the energy minimum state $\Phi_{min}$,
\begin{align}
 \widetilde{\Phi}^{J^\pi}_{MK} = \Phi^{J^\pi}_{MK} - \braket{\Phi_{min}|\Phi^{J^\pi}_{MK}}\Phi_{min}.
\end{align}
By applying this procedure successively, the second, third and more excited configurations are
obtained. Hereafter, the wave functions obtained by the constrained energy minimization or
the orthogonalization method are denoted by $\Phi^{J^\pi}_{MK;i},\ i=1,2,...,M$, where $M$ denotes
the number of wave functions.

Finally, thus-obtained wave functions are superposed to take the configuration mixing into
account (generator coordinate method; GCM \cite{Hill:1952jb}). The superposed wave function reads 
\begin{align}
 \Psi_{nM}^{J^\pi} = \sum_{Ki} c_{Kin}\Phi^{J^\pi}_{MK;i},\label{eq:form:gcmwf}
\end{align}
where the coefficient of superposition $c_{Kin}$ and eigenenergy $E^{J^\pi}_n$ are determined
by the diangonalization of the Hamiltonian, which reduces to the generalized eigenvalue problem;
\begin{align}
 &\sum_{Lj}{H^{J^\pi}_{KiLj}c_{Ljn}} = E^{J^\pi}_n \sum_{Lj}{N^{J^\pi}_{KiLj}c_{Ljn}},\\  
 &H^{J^\pi}_{KiLj} 
 = \braket{\Phi^{J^\pi}_{MK;i}|H|\Phi^{J^\pi}_{ML;j}}, \\
 &N^{J^\pi}_{KiLj} 
 = \braket{\Phi^{J^\pi}_{MK;i}|\Phi^{J^\pi}_{ML;j}}.
\end{align}
Physical observables are calculated by using thus-obtained wave functions ginven in
Eq. (\ref{eq:form:gcmwf}).  

\subsection{Single particle levels}
The single-particle configurations of a variational wave function can be investigated by the
nucleon-single particle energy and orbits. In particular, it is very suggestive and useful for 
understanding the motion of valence neutrons in neutron-rich nuclei. 

For this purpose, we construct the single-particle Hamiltonian from the intrinsic wave function 
obtained by the energy minimization \cite{Dote:1997zz}. We first transform the single particle wave
packets of the intrinsic wave function $\Phi_{int}$ to the orthonormalized basis,  
\begin{align}
 \widetilde{\varphi}_p(\bm r) =
 \frac{1}{\sqrt{\lambda_p}}\sum_{i=1}^{A}d_{ip}\varphi_i(\bm r).   
\end{align}
Here, $\lambda_p$ and $d_{ip}$ are the eigenvalues and eigenvectors of the
overlap matrix $B_{ij}=\langle\varphi_i|\varphi_j\rangle$. Using this basis, the single-particle
Hamiltonian $h_{pq}$ is constracted as,
\begin{align}
 h_{pq} &= 
  \langle\widetilde{\varphi}_p|t|\widetilde{\varphi}_q\rangle + 
  \sum_{r=1}^{A}\langle
  \widetilde{\varphi}_p\widetilde{\varphi}_r|
  {v_n+v_C}|
  \widetilde{\varphi}_q\widetilde{\varphi}_r -
 \widetilde{\varphi}_r\widetilde{\varphi}_q\rangle \nonumber\\ 
 &+\frac{1}{2}\sum_{r,s=1}^{A}
 \langle\widetilde{\varphi}_r\widetilde{\varphi}_s
|\widetilde{\varphi}_p^*\widetilde{\varphi}_q
\frac{\delta v_n}{\delta \rho}|\widetilde{\varphi}_r
\widetilde{\varphi}_s - \widetilde{\varphi}_s  \widetilde{\varphi}_r
\rangle.\label{eq:form:sph}
\end{align}
The last term of Eq. (\ref{eq:form:sph}) arizes from the density dependent terms contained in
Gogny interaction. The eigenvectors $f_{q\alpha}$ of the single particle Hamiltonian
defines the occupied sigle particle orbits,
\begin{align}
 \phi_i(\bm r) &= \sum_{p=1}^{A}f_{p\alpha}\widetilde{\varphi}_p(\bm r)\nonumber\\
 &=\sum_{i=1}^A\left(\sum_{p=1}^A d_{ip}\frac{1}{\sqrt{\lambda_p}}f_{p\alpha}\right)
 \varphi_i(\bm r),
\end{align}
and the eigenvalue $\varepsilon_p$ is the corresponding single-particle energy. In this review, we
define the weakly bound neutrons as valence neutrons and remaining nucleons as the core.

\section{Growth of the ground state clustering toward neutron drip line}\label{sec:3}
Ordinary clustering phenomena in stable nuclei are governed by the Ikeda threshold rule which is
based on the saturation of the energy and density. Since these saturation properties are broken,
the clustering in neutron-rich nuclei should have different aspects and obey different rule.

One of the interesting problem is the variation of the ground state clustering in an isotope
chain. For example, it is well known that the ground states of  $^{9}{\rm Be}$ ($2\alpha+n$) and
$^{20}{\rm Ne}$  ($\alpha+{}^{16}{\rm O}$) are clustered. What will happen when we add excess
neutrons to these stable nuclei? AMD calculations predicted that the addition of neutrons first
reduces the clustering of nuclei close to $N=Z$ line, but then, further addition of neutrons
enhances the clustering toward the neutron-drip line. Behind this enhancement of clustering, the
breakdown of neutron magic numbers $N=8$ and 20 plays an important role. It also predicted that
the clustering of neutron-rich B isotopes is also enhanced, although the stable nuclei
$^{10,11}{\rm B}$ are not clustered. These predictions are experimentally supported by the recent
systematic measurements of the proton radii in the isotope chains. 

In this section, we summarize these AMD studies and recent data. In the section \ref{sec:3.1}, we
discuss how the ground state clustering in Be, B and C isotopes are varied as function of neutron
number, and how it is correlated to the ground state deformation and proton, neutron and matter
radii. The survey is extended to O, Ne and Mg isotopes which are discussed in the section
\ref{sec:3.2}.

\subsection{ground state clustering in Be and B isotopes}\label{sec:3.1}

In Be isotopes, cluster structures develop in the ground and excited states 
\cite{SEYA,OERTZENa,OERTZENb,oertzen03-rev,Oertzen-rev,KanadaEn'yo:2001qw,KanadaEn'yo:2012bj,Dote:1997zz,KanadaEnyo:1995tb,Arai:1996dq,Ogawa:1998et,KanadaEn'yo:1999ub,Itagaki:1999vm,Itagaki:2000nn,Itagaki:2001az,Descouvemont:2001wek,Descouvemont:2002mnw,KanadaEn'yo:2002ay,KanadaEn'yo:2002rh,KanadaEn'yo:2003ue,Ito:2003px,Neff:2003ib,Arai:2004yf,Ito:2005yy,Ito:2008zza,Suhara:2009jb,Dufour:2010dmf,Ito:2012zza,KanadaEn'yo:2012rm,Ito2014-rev,Hamada:1994zz,Soic:1995av,Curtis:2001sd,Fletcher:2003wk,Curtis:2004wr,Ahmed:2004th,Millin05,Freer:2006zz,Bohlen:2007qx,Curtis:2009zz,Suzuki:2013mga,bohlen98,bohlen02,korsheninnikov95,Freer:1999zz,Freer:2001ef,SAITO04,Yang:2014kxa}.  
Along the series of isotopes, the cluster structure
in the ground state changes rapidly depending on the neutron number $N$. 
The remarkably developed cluster structure
plays an important role in the breaking of the neutron magic number $N=8$
in neutron-rich Be. Experimentally, the breaking of the $N=8$ magic number 
has been observed by the abnormal spin-parity $1/2^+$ of the $^{11}$Be ground state and supported
for $^{12}$Be by indirect and direct evidences  
\cite{Barker76,Suzuki:1997zza,Alburger:1978zza,Fortune:1994zz,Iwasaki:2000gh,Iwasaki:2000gp,Navin:2000zz,shimoura03,Pain:2005xw,Imai:2009zza,Meharchand:2012zz}.
%Because of the remarkable cluster structure with intruder neutron configurations, 
%these nuclei have more largely deformed ground states than the neighboring isotope, $^{10}$Be. 
The enhanced clustering has been theoretically 
predicted also in neutron-rich B isotopes such as $^{15}$B and $^{17}$B \cite{KanadaEnyo:1995ir}, 
whereas no development of clustering is predicted for neutron-rich C isotopes
at least in the ground states \cite{KanadaEn'yo:2001qw,thiamova2004,KanadaEn'yo:2004bi}.
It means that the cluster structure in the ground states sensitively depends 
on the neutron and proton numbers.

%%%%%%%%%%%%%%%%%%%%%%%%%%%%%%
\begin{figure}
 \includegraphics[width=\hsize]{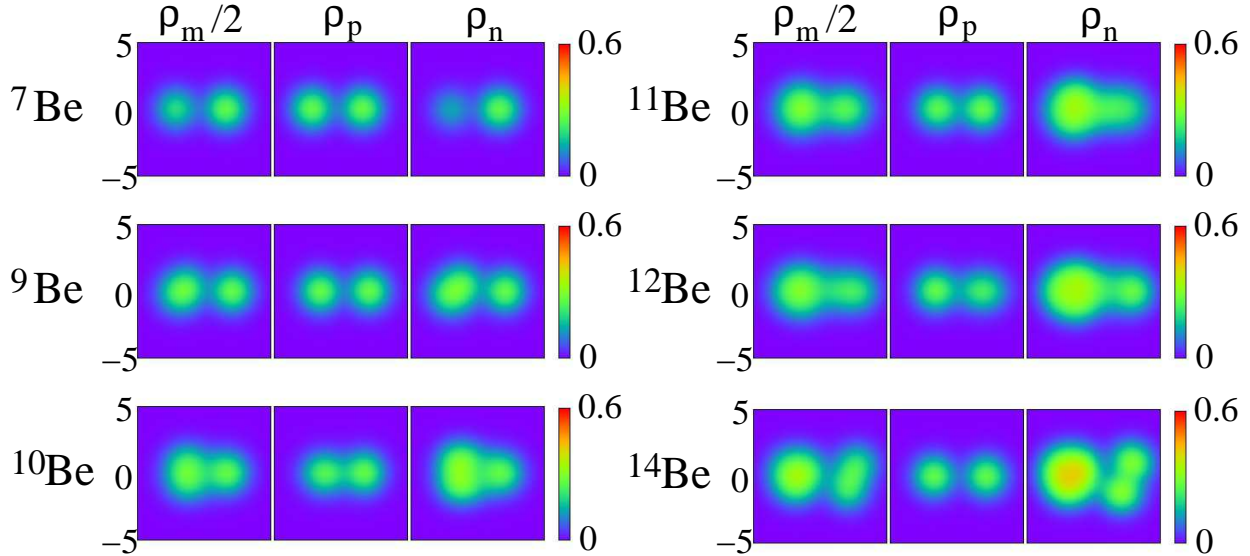}
%\vspace{0.5cm}
  \caption{(Color online) 
Distributions of matter (left), proton (middle), and neutron (right) 
densities of
Be isotopes obtained by the AMD+VAP calculation
using the MV1 ($m=0.65$, $b=h=0$) central and G3RS 
($u_1=-u_2=-3700$ MeV) spin-orbit interactions.
The densities  of intrinsic states are integrated with respect to the $z$ axis and
plotted on the $x$-$y$ plane in the unit of  fm$^{-2}$. 
The axes of the intrinsic frame are chosen so as to be
$\langle x^2\rangle \ge \langle y^2\rangle \ge \langle z^2\rangle$.  
The figure is taken from Ref.~\cite{Kanada-En'yo:2014toa}.
\label{fig:be-dense}}
\end{figure}
%%%%%%%%%%%%%%%%%%%%%%%%%%%%%

Figure \ref{fig:be-dense} shows distributions of matter, proton, and neutron
densities of the intrinsic wave functions for Be isotopes
obtained by the variation after the angular-momentum and parity 
projections in the framework of AMD (AMD+VAP) 
using the MV1 central and G3RS spin-orbit interactions \cite{Kanada-En'yo:2014toa}.
The adopted interaction parameters are $m=0.65$, $b=h=0$, and $u_1=-u_2=-3700$ MeV,
which reproduce the $1/2^+_1$ and $1/2^-_1$ states of $^{11}$Be.
As seen in a dumbbell shape of  proton density, 
the $2\alpha$ cluster core is formed in the ground states of Be isotopes. 
With increase of $N$, the neutron structure changes rapidly.
Following the rapid change of the deformation of the neutron density, 
the proton distribution changes showing
enhancement and reduction of the $2\alpha$ 
cluster structure.
Namely, the $2\alpha$ clustering is remarkably enhanced in $^9$Be at $N=5$, 
but it reduces in $^{10}$Be because of the less deformed 
neutron density at $N=6$. In the $N>6$ region, from $^{11}{\rm Be}$ to the
drip-line nucleus $^{14}$Be, the enhanced clustering with large prolate deformations of neutron
density can be seen. The largely deformed ground states of $^{11}$Be and $^{12}$Be 
at $N\sim 8$ indicate the disappearance of the $N=8$ magic number in Be isotopes.

%%%%%%%%%%%%%%%%%%%%%%%%%%%%%%
\begin{figure}[htb]
\begin{center}
\resizebox{0.8\textwidth}{!}{%
\includegraphics{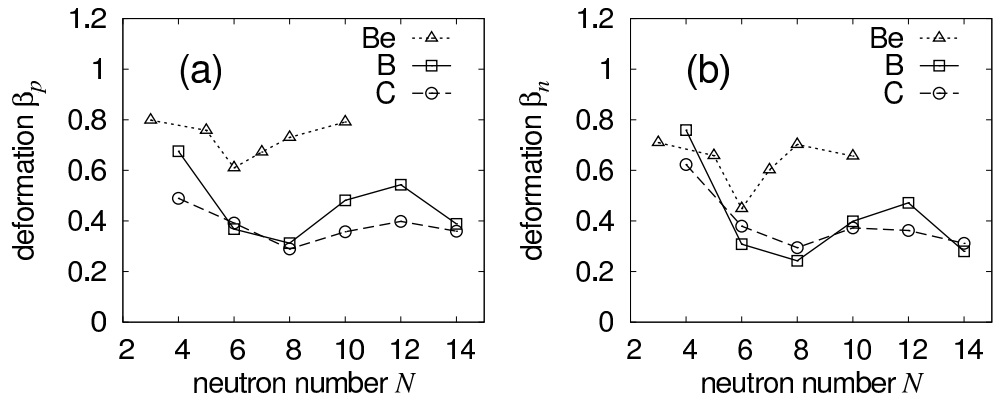} }	
%\includegraphics[width=12.5cm]{bec-defo-fig.eps} 	
%\vspace{0.5cm}
  \caption{Deformation parameters of Be, B, and  C isotopes obtained by 
the AMD+VAP using the MV1 ($m=0.65$, $b=h=0$) central and G3RS 
($u_1=-u_2=-3700$ MeV) spin-orbit interactions for Be and B isotopes and
the MV1 ($m=0.62$, $b=h=0$) central and G3RS 
($u_1=-u_2=-2600$ MeV) spin-orbit interactions for C isotopes.
$\beta_p$($\beta_n$) for the proton(neutron)
density is shown in the left(right). 
The figure is modified from the original figure in Ref.~\cite{Kanada-En'yo:2014toa}.
\label{fig:bebc-defo}}
\end{center}
\end{figure}
%%%%%%%%%%%%%%%%%%%%%%%%%%%%%

For more quantitative discussion of intrinsic deformations,
we show deformation parameters $\beta_p$ 
and $\beta_n$ for the proton and neutron densities in Be, B, and C isotopes
calculated using the AMD+VAP in Fig.~\ref{fig:bebc-defo}. 
The MV1  ($m=0.65$, $b=h=0$) central and G3RS ($u_1=-u_2=-3700$ MeV) spin-orbit interactions are used
for Be and B isotopes. For C isotopes, we use 
the MV1  ($m=0.62$, $b=h=0$) central and G3RS ($u_1=-u_2=-2600$ MeV) spin-orbit interactions
which can describe systematics of the $2^+_1$ excitation energies of C isotopes. 

In Be isotopes,
the neutron deformation is smallest at $N=6$ for $^{10}$Be, whereas it increases in 
$^{11}$Be and $^{12}$Be. Reflecting the change of the neutron deformation, 
the proton deformation $\beta_p$ shows the $N$ dependence similar to that of 
$\beta_n$. It means that, the proton
deformation is smallest not at $N=8$ but
at $N=6$ for $^{10}$Be. The coherent change of $\beta_p$ and $\beta_n$
indicates the enhancement and reduction of the clustering.

Also in B isotopes, $\beta_p$ changes rapidly and shows the $N$ dependence similar to that of
$\beta_n$  because of the enhancement and reduction of the clustering. $\beta_p$ and $\beta_n$ are
smallest at the neutron magic number $N=8$ meaning the reduction of the clustering in $^{13}$B.
This is consistent with the observed magicity of the neutron number $N=8$ in B isotopes
differently 
from Be isotopes. $^9$B is the mirror nucleus of $^9$Be and has a remarkably large deformation 
because of the developed cluster structure. In the neutron-rich region, 
the deformation increases in $^{15}$B and $^{17}$B at $N=10$ and $12$, and it decreases
again in $^{19}$B at $N=14$ because of the $d_{5/2}$ sub-shell closure.

Compared with Be and B isotopes, the proton deformation in C isotopes is not so sensitive to 
the neutron number and no prominent clustering is seen in neutron-rich isotopes. The stability of
the proton structure in the ground states of C isotopes is a feature peculiar to the proton number 
$Z=6$, and has been discussed in relation with different deformations between proton and neutron 
densities in $Z\ne N$ C isotopes \cite{KanadaEn'yo:2012bj,KanadaEn'yo:2004bi,KanadaEn'yo:1996hi}.

%%%%%%%%%%%%%%%%%%%%%%%%%%%%%%
\begin{figure}[htb]
\begin{center}
\resizebox{0.8\textwidth}{!}{%
\includegraphics{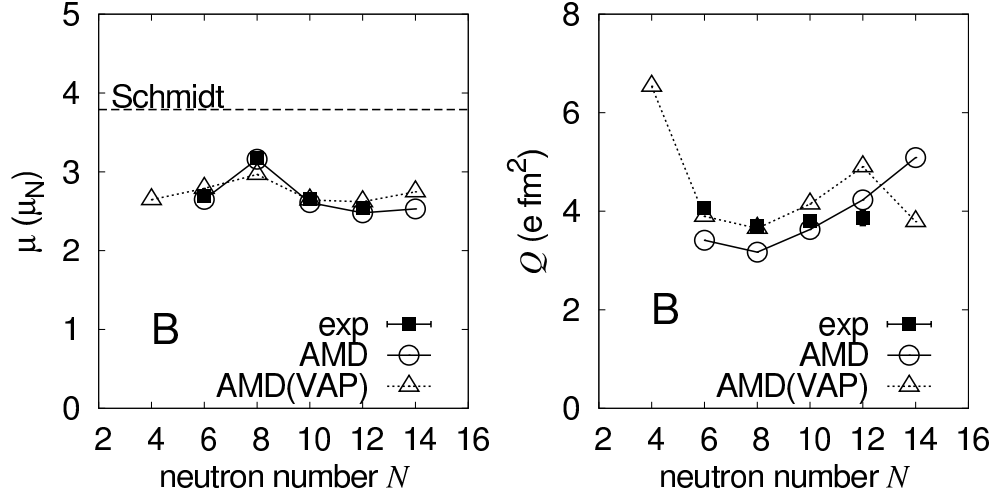} }	
%\includegraphics[width=10cm]{b-qmu-fig.eps} 	
%\vspace{0.5cm}
  \caption{The magnetic moments $\mu$ and the electric quadrupole moments 
$Q$ of B isotopes. 
The theoretical values calculated with the AMD in Ref.~\cite{KanadaEnyo:1995ir}
and those with the AMD+VAP in Ref.~\cite{Kanada-En'yo:2014toa} are shown 
by open circles and triangles, respectively.
The dashed line in the left figure shows the Schmidt value for the $p_{3/2}$ proton orbit. 
The MV1  ($m=0.65$, $b=h=0$) central and G3RS ($u_1=-u_2=-3700$ MeV) spin-orbit interactions are used
in the AMD+VAP calculation. For the AMD calculation,  
the MV1  ($m=0.576$, $b=h=0$) central interaction with mass number 
dependent parameters and the G3RS ($u_1=-u_2=-900$ MeV) spin-orbit interaction
(see Ref.~\cite{KanadaEnyo:1995ir}) are used.
The experimental data shown by filled squares are taken from
 Refs.~\cite{Okuno1995,Ueno:1995hp,Izumi:1995mm,Ogawa:2003gp,Tilley:2004zz,Ajzenberg90}.
\label{fig:b-qmu}}
\end{center}
\end{figure}
%%%%%%%%%%%%%%%%%%%%%%%%%%%%%

How one can observe the structure change along the isotopes, in particular, the enhancement and
reduction of the cluster structure with increase  of $N$?
For neutron-rich B isotopes, the  measurements of  magnetic ($\mu$) and electric quadrupole ($Q$)
moments  have been intensively performed in 1990s
\cite{Okuno1995,Ueno:1995hp,Izumi:1995mm,Ogawa:2003gp}. 
Figure \ref{fig:b-qmu} shows $\mu$ and $Q$ moments 
of B isotopes.
The theoretical  values are those 
obtained by the AMD \cite{KanadaEnyo:1995ir} and AMD+VAP calculations \cite{Kanada-En'yo:2014toa}.
The $N$ dependence of two calculations are in principle consistent with each other
except for $^{19}$B. The calculated $\mu$ moments reproduce well the experimental data 
of $^{11-17}$B. The deviation of $\mu$ moments from the 
Schmidt value for the $p_{3/2}$ proton  is smallest at $N=8$ for $^{13}$B because of the
$N=8$ shell closure, whereas it is relatively large for other B isotopes, $^{11}$B, $^{15}$B, and
$^{17}$B, because of significant contribution from the non-zero angular momentum of the core 
brought about by the developed clustering \cite{KanadaEnyo:1995ir}.
For the $Q$ moments, two calculations are in reasonable agreement 
with the experimental values of $^{11-17}$B.
The calculated $Q$ moment is smallest at $N=8$ for $^{13}$B because of the neutron shell closure,
whereas it  increases in $^{15}$B and $^{17}$B because of the enhanced cluster structure. 
Note that, if there is no spatial development of the cluster structure, the $Q$ moment should 
be largest in $^{13}$B and relatively small in $^{15}$B and $^{17}$B consistently 
with the $N$ dependence of $\mu$ moments as demonstrated in Ref.~\cite{KanadaEnyo:1995ir}.
The experimental $Q$ moments do not show the strong $N$ dependence, 
however, slight increase of the $Q$ moments from $^{13}$B to $^{15}$B and $^{17}$B
might be a signature of the spatially developed clustering in $^{15}$B and $^{17}$B. 
For $^{19}$B, the simple AMD calculation in Ref.~\cite{KanadaEnyo:1995ir} suggested
a prolate deformation with a remarkable clustering, whereas the latest 
calculation of the AMD+VAP in Ref.~\cite{Kanada-En'yo:2014toa} predicted a less deformation 
with weak clustering. The  measurement of the $Q$ moment of $^{19}$B
is required to determine the ground state deformation at the neutron drip line and 
to understand evolution of deformation (clustering) in B isotopes. 

Thus, magnetic and electric moments are useful information to discuss ground state properties of
odd-mass nuclei. However, they are not available for spin-zero ground states of even-even nuclei, 
in which moments are trivially zero. Furthermore, moments are sensitive not only to spatial
development of clustering but also to  the angular momentum coupling. 

In contrast to the electromagnetic moments, charge radius is the observable which is directly
related to the proton density distribution, and is available for both even- and odd-mass nuclei.
As shown in Fig.~\ref{fig:be-dense}, the enhancement and reduction of the cluster structure  
can be clearly seen in the proton density distribution. The $N$ dependence of charge radii 
should reflect this change of clustering via the proton density distribution, and therefore it 
can be a good probe to pin down the clustering.
%Indeed, the $N$ dependence of the proton radius reflects the cluster structure 
%in the ground state. 
Recently, root mean square (rms) charge radii of neutron-rich Be isotopes 
up to $^{12}$Be
have been precisely measured by 
means of isotope shift \cite{Nortershauser:2008vp,Krieger:2012jx}.
For B and C isotopes, isotope shift measurements are 
still limited to nuclei near the stability line. However, a new experimental approach has been 
recently applied to determine 
rms radii of point-proton density (proton radii) from charge changing interaction cross 
sections \cite{Yamaguchi:2011zz,Estrade:2014aba,Terashima:2014wca}, and the  data 
of proton radii of Be and B isotopes are now available.

%%%%%%%%%%%%%%%%%%%%%%%%%%%%%%
\begin{figure}[htb]
\begin{center}
\resizebox{0.8\textwidth}{!}{%
\includegraphics{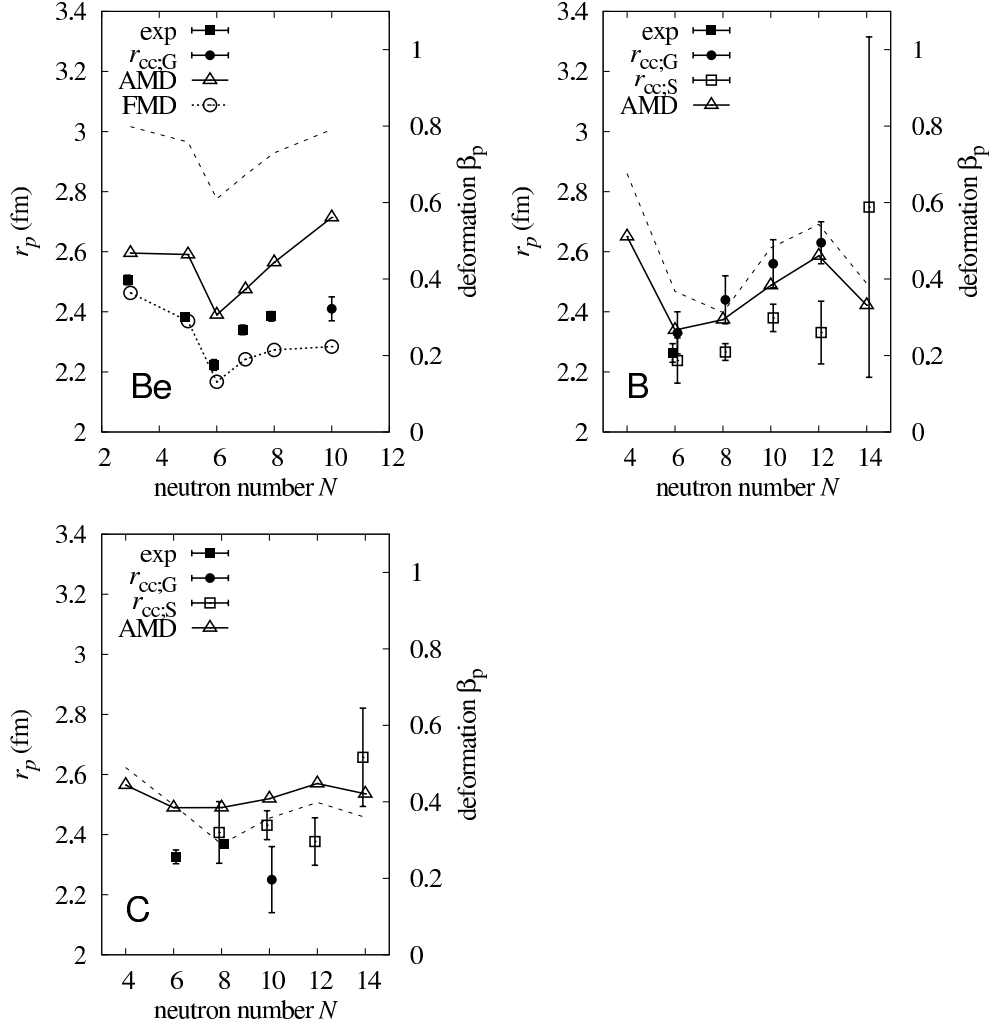} }	
%\includegraphics[width=15cm]{bec-rp-fig.eps} 	
%\vspace{0.5cm}
  \caption{Proton radii of Be, B, and C isotopes.
Open triangles
connected by solid lines indicate 
the theoretical values obtained by the AMD+VAP 
using the MV1 ($m=0.65$, $b=h=0$) central and G3RS 
($u_1=-u_2=-3700$ MeV) spin-orbit interactions for Be and B isotopes and
the MV1 ($m=0.62$, $b=h=0$) central and G3RS 
($u_1=-u_2=-2600$ MeV) spin-orbit interactions for C isotopes.
The deformation parameter ($\beta_p$) of the proton density is shown by dashed lines.
Experimental proton radii of $^{9,10,11,12}$Be, $^{11}$B, and $^{12,14}$C 
reduced from experimental charge radii are shown by filled squares
\cite{Nortershauser:2008vp,Krieger:2012jx,angeli04}. 
Experimental proton radii $r_{\rm cc;G}$ deduced from charge changing
 interaction cross sections 
$\sigma_{\rm cc}$ by the Glauber analysis in Refs.~\cite{Yamaguchi:2011zz,Estrade:2014aba,Terashima:2014wca}
are shown by filled circles. 
Proton radii $r_{\rm cc:S}$ evaluated from $\sigma_{\rm cc}$ in Ref.~\cite{chulkov00} 
using a simple ansatz
are shown by open squares (see Ref.~\cite{Kanada-En'yo:2014toa}).
Values of FMD calculations for 
Be isotopes are also shown \cite{Krieger:2012jx}. 
The figure is modified from the original figure in Ref.~\cite{Kanada-En'yo:2014toa}
\label{fig:bec-rp}}
\end{center}
\end{figure}
%%%%%%%%%%%%%%%%%%%%%%%%%%%%%

Figure \ref{fig:bec-rp} shows the $N$ dependence of proton radii in Be, B, and C isotopes 
combined with those of deformation parameters $\beta_p$ and $\beta_n$.
In Be isotopes, the proton radius is relatively large in $^{9}$Be
because of the remarkable cluster structure, whereas it decreases 
at $N=6$ for $^{10}$Be and increases again in the $N>6$ region for 
$^{11}$Be and $^{12}$Be reflecting the reduction and enhancement 
with increase of $N$. 
The $N$ dependence of  proton radii is consistent with the experimental data 
reduced from charge radii determined by isotope shift measurements.
The increase of the proton radius in $^{12}$Be 
is an experimental evidence of the $N=8$ shell breaking as pointed out in 
Ref.~\cite{Krieger:2012jx}.
The $N$ dependence of proton radii coincides 
with that of the proton deformation $\beta_p$. 
The minimums of $r_p$ and $\beta_p$ at $N=6$ instead of $N=8$ can be interpreted as
the migration of the neutron magic number from $N=8$ to $N=6$ in Be isotopes.

Also in B isotopes, calculated proton radii are enhanced 
in $^{15}$B and $^{17}$B with the cluster development. 
The $N$ dependence of 
the calculated proton radii is consistent with the experimental proton radii
reduced from charge changing cross sections in the 
$6\le N\le 12$ region. For $^{19}$B, decrease of the proton radius was predicted by 
the AMD+VAP calculation because of the weak clustering. 

In contrast to Be and B isotopes, calculated proton radii of C isotopes 
show a weaker $N$ dependence. It originates in the stability of proton structure 
because of the less clustering in neutron-rich C isotopes. At present, the experimental  
data is not enough to discuss the $N$ dependence of  proton radii in the neutron-rich C isotopes.

\subsection{Radii and deformation of O, Ne and Mg isotopes, and the ground state clustering in Ne
  isotopes} \label{sec:3.2}
As discussed for the Be, B and C isotopes, the proton and neutron radii
reflect various structural characteristics such as the magnitude and pattern of deformation,
the breakdown of the magic number and the formation of neutron skin or halo. In particular, the 
behavior of the proton radii is attributed to the enhancement and reduction of the
clustering in Be and B isotopes. Therefore, it is interesting to extend the survey to heavier mass
isotopes. 

\begin{figure}[h] 
 \begin{center}
  \resizebox{0.7\textwidth}{!}{
  \includegraphics{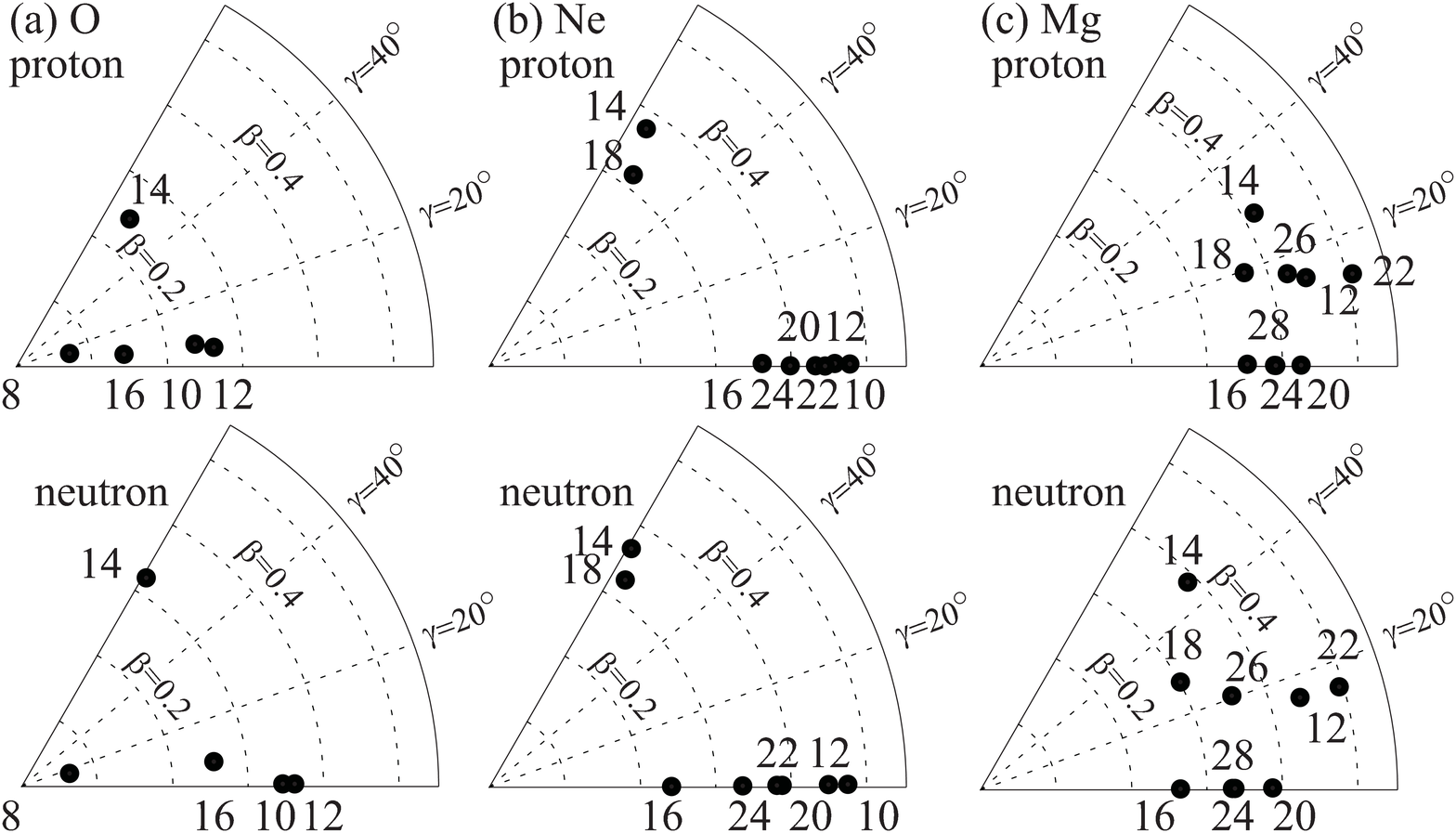}}
  \caption{Intrinsic quadrupole deformation parameters of the ground states of O, Ne and Mg
  isotopes obtained by AMD calculation. Numbers in the figure show neutron number. }
  \label{fig:gs1}       % Give a unique label
 \end{center}
\end{figure}

\begin{figure}[h] 
 \begin{center}
  \resizebox{0.6\textwidth}{!}{
  \includegraphics{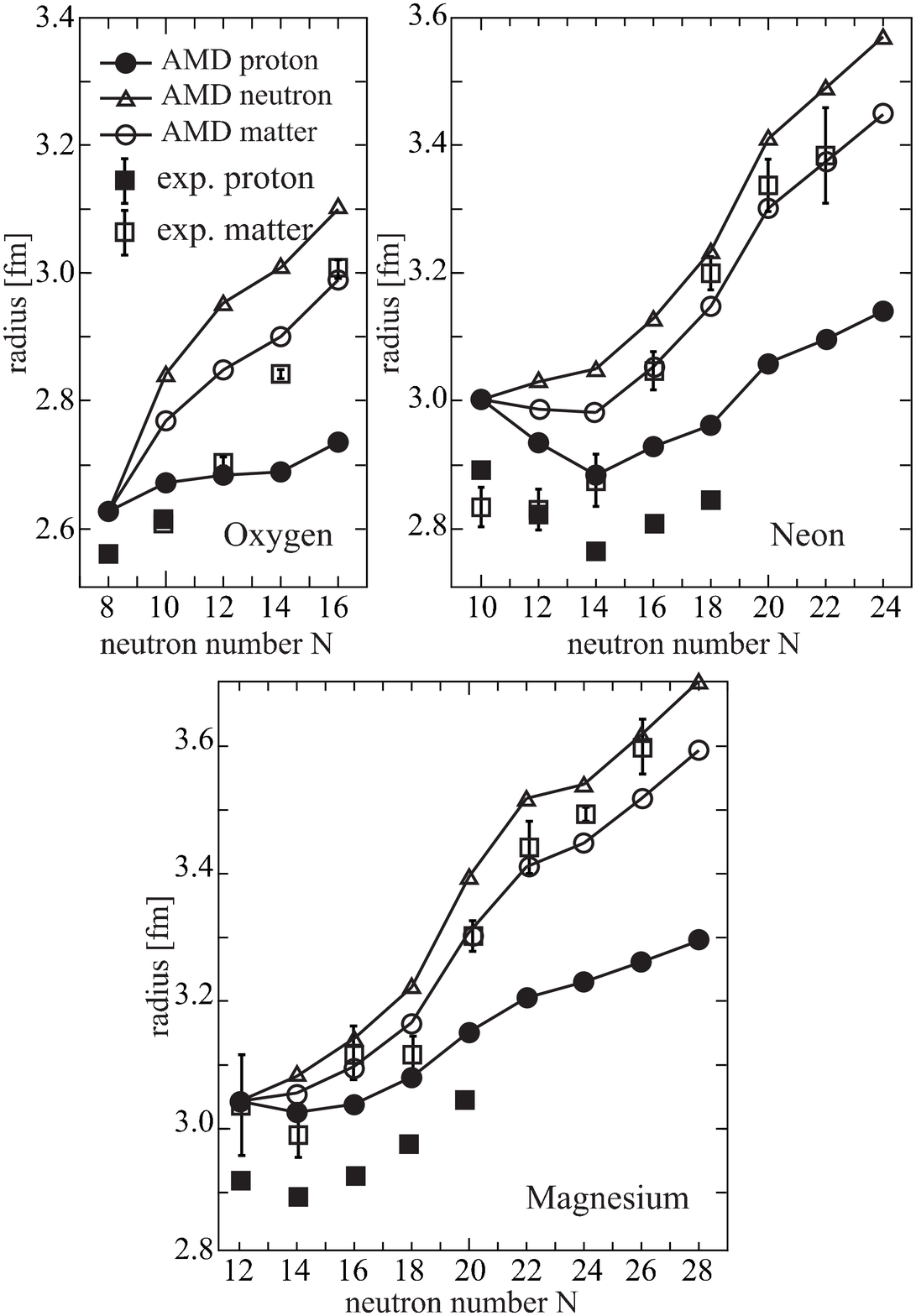}}
  \caption{Calculated and observed proton, neutron and matter radii of O, Ne and Mg 
  isotopes. Experimental data are taken from
  Refs. \cite{Takechi:2012zz,Takechi:2014eza,Marinova:2011zz,Yordanov:2012zz,Angeli:2013a}.} 
  \label{fig:gs2}       % Give a unique label
 \end{center}
\end{figure}

\begin{figure}[h] 
 \begin{center}
  \resizebox{0.7\textwidth}{!}{
  \includegraphics{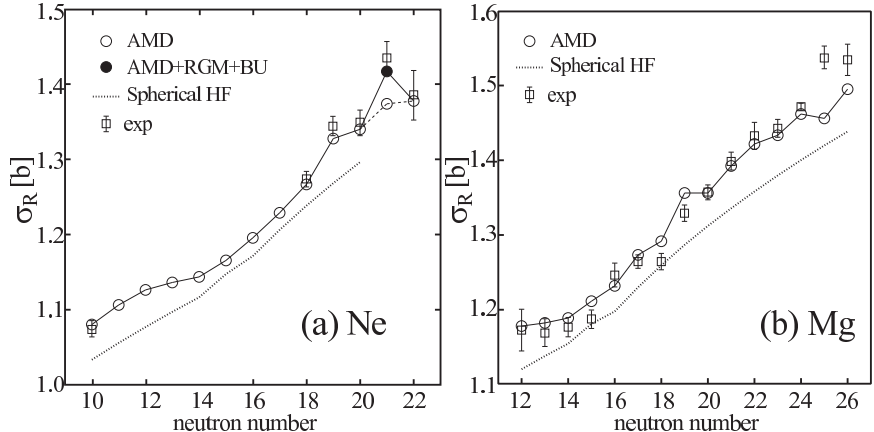}}
  \caption{Observed and calculated reaction cross sections of Ne and Mg isotopes. Open circles
  show the calculated cross sections by the density folding model which use the ground state
  density distribution obtained by AMD \cite{Sumi:2012fr,Watanabe:2014zea}, while the dotted lines
  use the density   distributions obtained by spherical Hartree Fock. Filled circles (AMD+RGM+BU)
  shows the results obtained by using the density distribution of AMD+RGM \cite{Minomo:2011bb}
  which is able to describe one neutron halo structure of  $^{31}{\rm Ne}$ and the breakup effect
  is taken into account. Experimental data are taken from
  Refs. \cite{Takechi:2012zz,Takechi:2014eza}.}  \label{fig:gs3}       % Give a unique label
 \end{center}
\end{figure}

Ne and Mg isotopes have longer isotope chains than Be and B isotopes and are famous for
the breakdown of the $N=20$ magic number 
\cite{Sorlin:2008jg,Thibault:1975zz,Huber:1978zz,Warburton:1990zza}.  
AMD has been also applied to the structure study of the Ne and Mg isotopes 
\cite{Kimura:2002a,Kimura:2004ey,Kimura:2007qx}, 
and we here discuss what kinds of structural information can 
be extracted from the behavior of the proton, neutron and matter radii. We first examine how the
quadrupole deformation of proton and neutron distributions shown in Fig. \ref{fig:gs1} are
correlated to the behavior of the radii and reaction cross sections shown in Figs. \ref{fig:gs2}
and \ref{fig:gs3}.  

The deformations of O isotopes are always small compared with Ne and Mg isotopes owing to the
$Z=8$ proton shell closure.  It is confirmed from Fig. \ref{fig:gs1} (a) that the proton
deformation is always smaller than that of neutron. It is also noted that the overlap between the
ground state and the  spherical state $|\braket{\Phi_{g.s.}|P^{J=0}\Phi(\beta=0)}|^2$ are large
and not less than 0.8 in all O isotopes. This means that all ground states are approximated well
by the spherical state, and hence,  deformation plays only a minor role in O isotopes. As a
result, compared to other isotope chains such as Be, B, Ne and Mg, the proton radii are kept
almost constant except for the drip-line nucleus $^{24}{\rm O}$ despite of the increasing neutron
radii and the formation of the neutron skin toward the neutron drip line.  

\begin{figure}[h] 
 \begin{center}
  \resizebox{0.7\textwidth}{!}{
  \includegraphics{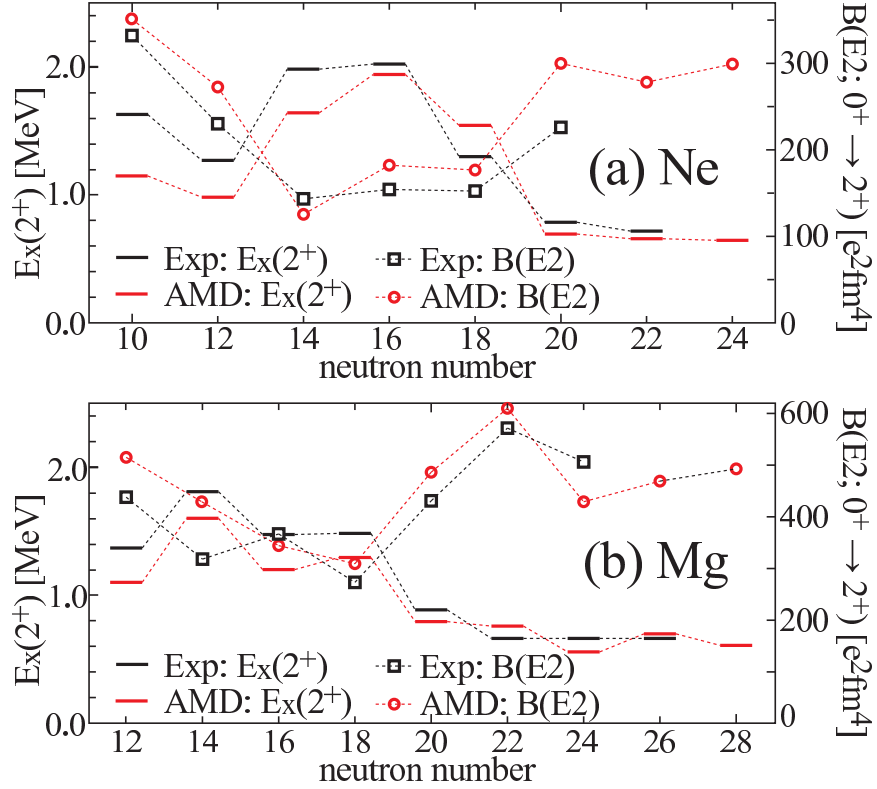}}
  \caption{Observed and calculated excitation energies of the $2^+_1$ states and the reduced
  transition probabilities $B(E2;0^+_1 \rightarrow 2^+_1)$
  in even-mass Ne and Mg isotopes. Experimental data are taken from Refs. 
  \cite{Motobayashi:1995ei,Pritychenko:1999wvv,Yanagisawa:2003nkp,Iwasaki:2005npt,Gibelin:2007wb,Pritychenko:2013gwa,Raman:1201zz,Yoneda:2001rdq,Gade:2007zz}.} 
  \label{fig:gs4}       % Give a unique label
 \end{center}
\end{figure}

In contrast to the O isotopes, deformation plays a major role for the structure evolution of
Ne and Mg isotopes in which the breakdown of the $N=20$ magic number and the resultant large
quadrupole deformation are well known 
\cite{Sorlin:2008jg,Motobayashi:1995ei,Detraz:1979zz,Guillemaud-Mueller:1984ngh}. The stable
nucleus $^{20}{\rm Ne}$ has the largest deformation among the Ne isotopes, which owes to the
$\alpha+{}^{16}{\rm O}$ clustering \cite{Fujiwara-supp,Horiuchi:68a,Dufour:1994zz}. The addition of
valence neutrons reduces deformation and changes the deformation pattern in $^{22,24,26}{\rm Ne}$. 
Note that the $d_{5/2}$ sub-shell closure leads to the oblate deformation of $^{24}{\rm Ne}$ which
is common to other lighter $N=14$ isotones, $^{19}{\rm B}$ and  $^{20}{\rm C}$. The neutron number
$N=18$ is also energetically favors the oblate deformation. As a result, $^{28}{\rm Ne}$ is
another oblate deformed Ne isotope. Further addition of valence neutron changes the trend of
deformation, because the Ne isotopes with $N>18$ are located in the island of inversion. Due to
the breakdown of the neutron magic number $N=20$, the ground state of  $^{30}{\rm Ne}$ ($N=20$) is
dominated by the intruder $2\hbar\omega$ configuration and strongly deformed
\cite{Kimura:2002a,Kimura:2004ey,Pritychenko:1999wvv,Yanagisawa:2003nkp}. The deformation of the
ground state continues until the end of the isotope chain, $^{34}{\rm Ne}$. This onset of the
deformation in $N>18$ isotopes are experimentally confirmed from the sudden increase of $B(E2)$
and the decrease of $2^+_1$ energy (Fig. \ref{fig:gs4})
\cite{Pritychenko:1999wvv,Yanagisawa:2003nkp,Iwasaki:2005npt,Gibelin:2007wb}. 

Mg isotopes have similar deformation pattern to Ne isotopes: The stable nucleus
$^{24}{\rm Mg}$ is strongly deformed 
\cite{Batchelor:1960,Girod:1983zz,Bender:2008zv,Kimura:2012ab}, 
and the addition of valence neutrons reduces the 
deformation of $N=14,16$ and 18 systems ($^{26}{\rm Mg}$, $^{28}{\rm Mg}$ and $^{30}{\rm Mg}$). 
The Mg isotopes with $N>18$ are also located in the island of inversion and their ground states
are strongly deformed owing to the dominance of the intruder $2\hbar\omega$ configurations
\cite{Warburton:1990zza,Kimura:2002a,Motobayashi:1995ei,Pritychenko:1999wvv,Fukunishi:1992ykk}. 
It is  interesting to note the differences of the deformation patterns
between Ne and Mg isotopes. First, because the proton number $Z=12$ energetically favors the
triaxial deformation \cite{Batchelor:1960,Girod:1983zz,Bender:2008zv,Kimura:2012ab}, there are
many isotopes  ($N=12$, 14, 18, 22 and 26) exhibiting 
triaxial deformation. This suggests that Mg isotopes are rather soft against the $\gamma$
deformation \cite{Hinohara:2011wh,Yao:2010at}. Another point  is the breakdown of the $N=28$
magic number. Mg isotopes have longer  isotope chain than Ne isotopes and the present calculation
shows that all isotopes with $N>18$, including the drip-line nucleus $^{40}{\rm Mg}$ ($N=28$), are
deformed. It suggests that the island of inversion (neutron-rich $N\simeq 20$ nuclei) and another
region of nuclear deformation around  $^{42}{\rm Si}$ 
\cite{Sorlin:1993zz,Bastin:2007gm,Force:2010ng,SantiagoGonzalez:2011zz,Takeuchi:2012kf,Crawford:2014iza,Retamosa:1996rz,Li:2011as,Kimura:2013gc,Utsuno:2014psa} where the magic number $N=28$ is
broken are merged. This is consistent with the recent observation
\cite{Doornenbal:2013ina,Doornenbal:2016fcv} which confirmed the systematic deformations of Mg
isotopes toward neutron drip line. 

\begin{figure}[h] 
 \begin{center}
  \resizebox{0.6\textwidth}{!}{
  \includegraphics{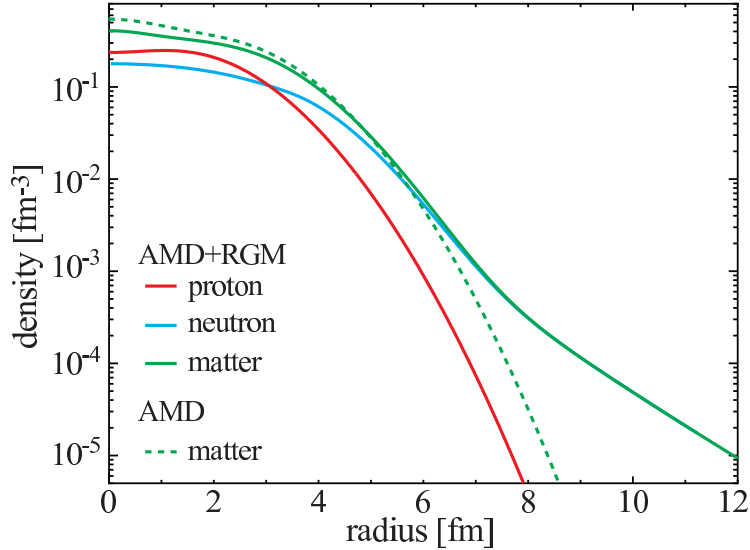}}
  \caption{Proton, neutron and matter (proton+neutron) density distributions of $^{31}{\rm Ne}$
  calculated by AMD+RGM model \cite{Minomo:2011bb}. Dashed line shows the matter density
  distribution obtained by ordinary AMD model. This figure is reconstructed from the data
  presented in Ref. \cite{Minomo:2011bb}.} 
  \label{fig:gs5}       % Give a unique label
 \end{center}
\end{figure}
The above-mentioned characteristics of Ne and Mg isotopes are reflected well to the reaction
cross sections and matter radii. Figure \ref{fig:gs3} compares the observed reaction cross
sections of Ne and Mg isotopes with the calculation by the density-folding model 
\cite{Takechi:2012zz,Takechi:2014eza,Sumi:2012fr,Watanabe:2014zea,Minomo:2011bb,Minomo:2011sj}. To 
clarify the role of the nuclear deformation, the density-folding model employs two
different density distributions of the ground states. The open circles show the results obtained
by using the density distribution of 
the AMD (deformation is unrestricted), while the dotted lines show the results
obtained by the density distribution of the spherical Hartree Fock (HF). It is clear that the AMD
results reasonably agree with the observation indicating the importance of deformation. The
differences between the AMD and spherical HF is large for $N\simeq Z$ nuclei which are strongly
deformed, but reduced for $N=14$, 16 and 18 isotones, and then, it is again enlarged for $N>18$
isotones. In both isotope chains, the large deviation between AMD and spherical HF continues until
the neutron drip line, which  is common to the trend of the deformation pattern shown in
Fig. \ref{fig:gs1}, the $2^+_1$ energies and $B(E2)$ values shown in Fig. \ref{fig:gs4}.
It must be noted that the observed and calculated (AMD) cross sections are 
discontinuously increased from $N=18$ to $19$ isotones, which clearly indicates that the west end
of the island of inversion is at $N=19$ \cite{Kimura:2007qx}. We also note that the AMD results
considerably underestimate the cross sections of $^{31}{\rm Ne}$ and $^{37}{\rm Mg}$
\cite{Watanabe:2014zea,Minomo:2011sj}. This implies the formation of neutron halo in these
nuclei, because the ordinary AMD employs the Gaussian wave function and cannot describe the
long-range part of the neutron halo wave function. To describe the halo structure, an extended
version of AMD denoted by AMD+RGM was developed which combines the AMD with the resonating group
method (RGM). By adopting AMD+RGM \cite{Minomo:2011sj}, the long-tail of the neutron density
distribution of  $^{31}{\rm Ne}$ is properly described as shown in Fig. \ref{fig:gs5} and the huge
cross section of $^{31}{\rm Ne}$ is reasonably reproduced.   

\begin{figure}[h] 
 \begin{center}
  \resizebox{0.7\textwidth}{!}{
  \includegraphics{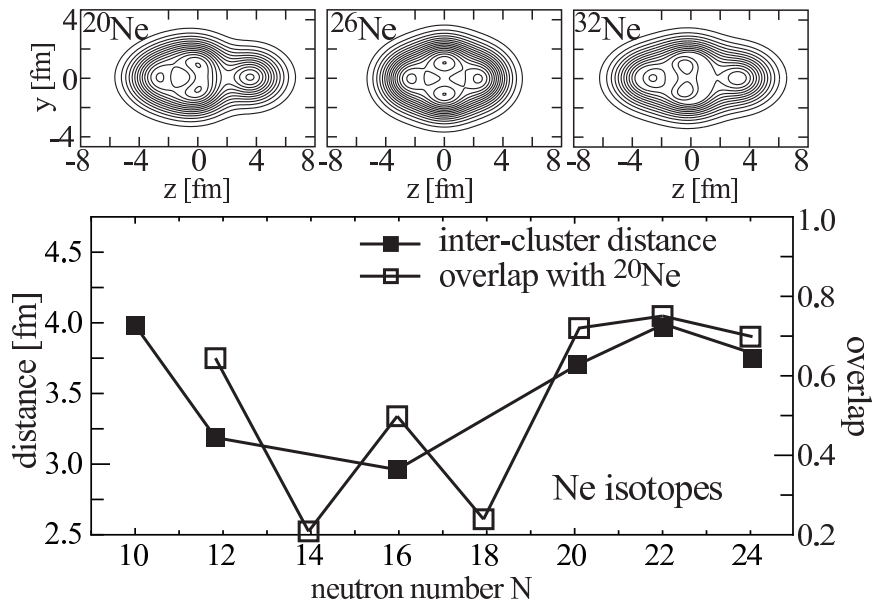}}
  \caption{Upper panels show the intrinsic proton density distributions of $^{20}{\rm Ne}$, 
  $^{26}{\rm Ne}$ and $^{32}{\rm Ne}$. Lower panel show the overlap between the proton wave
  functions of $^{20}{\rm Ne}$ and other Ne isotopes. It also shows an estimation of the distances
  between $\alpha$ and $^{16}{\rm O}$ clusters. The inter-cluster distances in $^{24}{\rm Ne}$ and 
  $^{28}{\rm Ne}$ were not evaluated because of too small overlap.}   
  \label{fig:gs6}       % Give a unique label
 \end{center}
\end{figure}

Now, we examine how the fingerprints of clustering can be seen in the radii. Similar to the Be and
B isotopes, the behavior of proton radii of Ne isotopes can be attributed to the clustering. The
global behavior of proton radii in Ne and Mg isotope chains are similar to each other: They
continue to increase in $N>18$ isotopes because of the large deformation. However, in addition to 
the deformation change, the reduction and growth of the clustering occurs in Ne isotopes. As
confirmed from the density distribution shown in Fig. \ref{fig:gs6}, $^{20}{\rm Ne}$ shows
the $\alpha+{}^{16}{\rm O}$ clustering. The additional neutrons reduce or vanish this
clustering from $^{22}{\rm Ne}$ to $^{28}{\rm Ne}$. But further addition of valence neutrons
revives the clustering in Ne isotopes located in the island of inversion where the strong
deformation of neutron distribution induces the clustering of proton distribution. 
As a result, the proton radius is reduced from $^{22}{\rm Ne}$ to $^{28}{\rm Ne}$ compared to
$^{20}{\rm Ne}$, and  increases in neutron-rich Ne isotopes. Therefore, in Fig. \ref{fig:gs2}, one
sees that the proton radius of $N=10 \sim 24$ isotopes changes more largely  in Ne isotopes than
in Mg isotopes. 

To evaluate this reduction and growth of $\alpha+{}^{16}{\rm O}$ clustering more quantitatively,
we calculated the overlap of the proton wave functions between $^{20}{\rm Ne}$ and other Ne
isotopes \cite{Kimura:2004ey}. Here, the overlap of the proton wave functions are defined and
calculated as follows. First, we choose a single intrinsic wave function $\Phi_{int}$ that has the
largest squared overlap with the GCM wave function of the ground state $\Psi_{gs}(^{A}{\rm Ne})$,
and calculate the single-particle orbits $\phi_i$. Then, we pickup only the proton orbits and
construct the antisymmetrized product of ten proton orbits, 
\begin{align}
\Psi_{p}(^{A}{\rm Ne})=\mathcal A\set{\phi_1\phi_2,...,\phi_{10}}.
\end{align}
 The overlap of proton wave function $\mathcal O$ is defined as the overlap between $^{20}{\rm Ne}$
 and $^{A}{\rm Ne}$;
\begin{align}
 \mathcal O=|\braket{\Psi_{p}(^{20}{\rm Ne})|\Psi_{p}(^{A}{\rm Ne})}|^2.
\end{align}
We also estimated the inter-cluster distance $d$ between $\alpha$ and $^{16}{\rm O}$ clusters,
which is calculated as follows. We first construct Brink-Bloch wave function \cite{Brink:1966a} for
$^{20}{\rm Ne}$  in which $\alpha$ and $^{16}{\rm O}$ are placed with inter-cluster distance $d$;
\begin{align}
 \Psi_{BB}(d) = \mathcal A\set{\phi_\alpha(-\frac{16}{20}\bm d)
 \phi_{^{16}{\rm O}}(\frac{4}{20}\bm d)},  \bm d = (0,0,d), 
\end{align}
where $\phi_\alpha(-\frac{16}{20}\bm d)$ and $\phi_{^{16}{\rm O}}(\frac{4}{20}\bm d)$
are the wave functions of the $\alpha$ and $^{16}{\rm O}$ clusters. They are assumed to have the
double closed-shell configurations described by Harmonic oscillator wave functions and placed at 
$-\frac{16}{20}\bm d$ and $\frac{4}{20}\bm d$, respectively. Using the same procedure explained
above, we construct the antisymmetrized product of proton wave functions from $\Psi_{BB}(d)$, and
calculate the overlap,
\begin{align}
 \mathcal O(d)=|\braket{\Psi_{BB;p}(d)|\Psi_{p}(^{A}{\rm Ne})}|^2.
\end{align}
We regard that the value of $d$ which maximizes the overlap $\mathcal O(d)$ as the inter-cluster
distance. Thus-obtained overlap and the inter-cluster distance are given in
Fig. \ref{fig:gs6}. They clearly show that the $\alpha+{}^{16}{\rm O}$ clustering is reduced in
$N=12$, 14, 16 and 18 isotopes, but enhanced in the island of inversion. Note that this reduction and
enhancement of the clustering is correlated well with the proton radii of Ne
isotopes. Experimentally, charge radii of Ne and Mg isotopes are measured by the isotope shift up
to $^{28}{\rm Ne}$ and $^{32}{\rm Mg}$ \cite{Marinova:2011zz,Yordanov:2012zz}. Although the AMD
calculation systematically overestimates the data, it nicely describes the neutron number
dependence.

\section{Clusters in the excited states}\label{sec:4}
Behind the variation of the ground state clustering discussed in the previous section, the shell
effect of excess neutrons is playing an essential role. It was suggested that a special class of
the shell structure called molecular orbits is formed around the clustered core in Be
isotopes. The AMD calculations proved the existence of the molecular orbits without {\it a priori}
assumptions. 

Theoretical calculations including AMD showed that the molecular orbits naturally explain the
variation of the ground state clustering. In addition to this, they also showed that not only the
ground states but also the excited states are explained and predicted by the molecular
orbits. Thus, the study of neutron-rich Be isotopes revealed a novel type of clustering; the
clustered core with covalent neutrons. In this decade the study is extended to highly excited
states where the atomic orbits are formed and ionic bonding dominates over the covalent
bonding. The studies of Be and B isotopes by AMD are summarized in the section \ref{sec:4.1}. 

This success of the molecular orbits strongly motivated the extention of the concept to other 
nuclei. The highlights of the extention is summarized as follows; (1)
Universality of the concept: It is of interest and importance to investigate if the concept of the
molecular orbit applies to other nuclei universally and if it yields a novel type clustering in a
broader region of the nuclear chart. (2) Molecular orbits in the asymmetric systems: The molecular
orbits formed around the parity asymmetric cluster core should have different nature from those of
Be isotopes.  (3) Extension to the multi-cluster systems: Since the valence neutrons in the
molecular orbits play a glue-like role, we expect that the multi-cluster systems such as $n\alpha$
cluster systems may be stabilized by the addition of the valence neutrons. In the section
\ref{sec:4.2}, we examine the points (1) and (2) by using O, F and Ne isotopes as examples. The
point (3) is discussed in the section \ref{sec:4.3} focusing on the $3\alpha$ cluster states in C
isotopes. 

\subsection{Molecular orbits and di-cluster resonances in Be isotopes}\label{sec:4.1}
%Be-13 paper
\subsubsection{Overview of cluster structures in Be isotopes}
As already discussed in the previous section, the ground states of Be isotopes have
cluster structures, which play an important role in 
the shell breaking mechanism of neutron-rich Be. 
In these decades, cluster structures in the ground and excited states 
of Be isotopes have been intensively investigated 
by many theoretical works
\cite{SEYA,OERTZENa,OERTZENb,Oertzen-rev,KanadaEn'yo:2001qw,KanadaEn'yo:2012bj,Dote:1997zz,KanadaEnyo:1995tb,Arai:1996dq,Ogawa:1998et,KanadaEn'yo:1999ub,Itagaki:1999vm,Itagaki:2000nn,Itagaki:2001az,Descouvemont:2001wek,Descouvemont:2002mnw,KanadaEn'yo:2002ay,KanadaEn'yo:2002rh,KanadaEn'yo:2003ue,Ito:2003px,Neff:2003ib,Arai:2004yf,Ito:2005yy,Ito:2008zza,Suhara:2009jb,Dufour:2010dmf,Ito:2012zza,KanadaEn'yo:2012rm,Ito2014-rev}.
Low-lying states of neutron-rich Be isotopes
are described well by molecular orbit models
assuming valence neutrons in molecular orbits surrounding 
the $2\alpha$ core. In the highly excited states near and above the $\alpha$-decay threshold
energy,   dinuclear-type He+He cluster resonances (di-cluster resonances) have been predicted by
two-body cluster models \cite{Descouvemont:2001wek,Descouvemont:2002mnw,Dufour:2010dmf} and more
generalized  cluster models
\cite{Ito:2003px,Arai:2004yf,Ito:2005yy,Ito:2008zza,Ito:2012zza,Ito2014-rev}. 
Appearance of such  cluster structure in neutron-rich Be isotopes has been theoretically
verified  by the AMD, FMD and no-core
shell model approaches \cite{KanadaEnyo:1995tb,Neff:2003ib,Suhara:2009jb,Yoshida:2014tta}, which 
do not rely on {\it a priori} assumption of 
existence of any clusters. 

\begin{figure}[htb]
\begin{center}
\resizebox{0.6\textwidth}{!}{%
\includegraphics{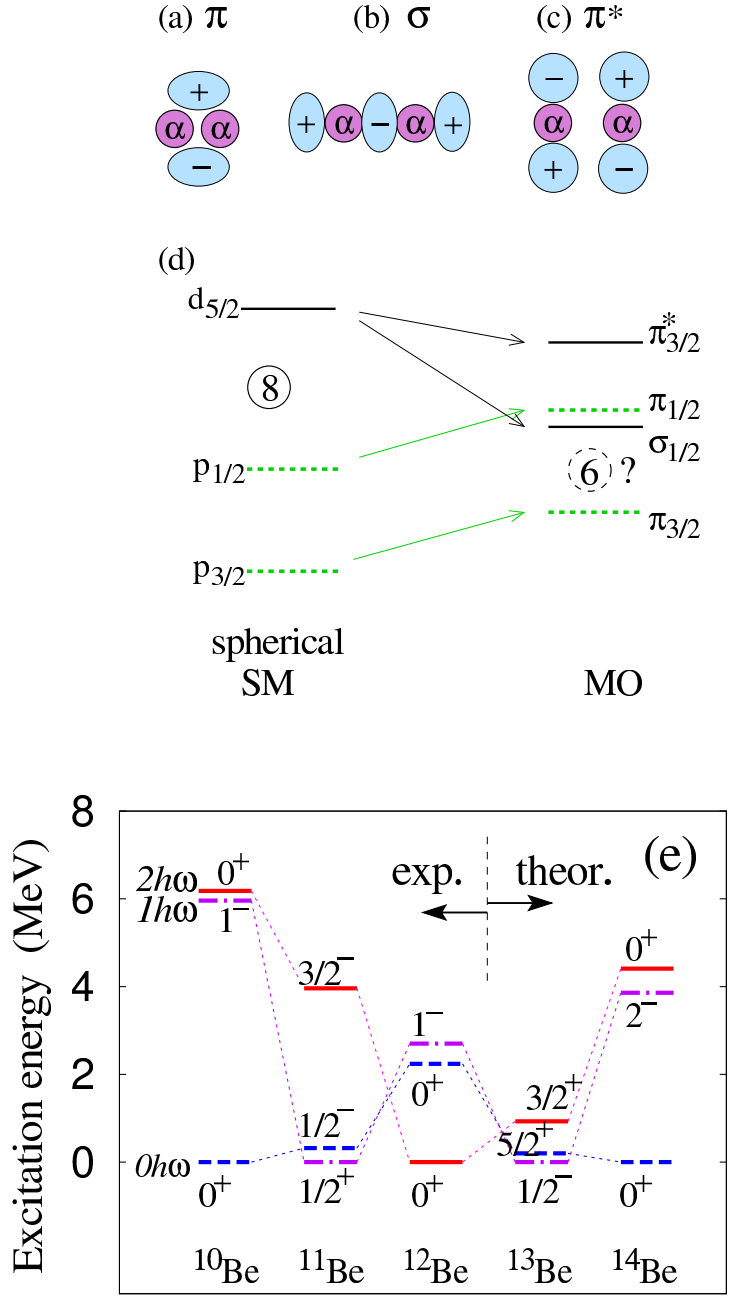} }	
%\includegraphics[width=12.5cm]{beiso-orbital.eps} 	
%\vspace{0.5cm}
  \caption{(a)(b)(c) Schematic figures of
molecular orbits around the $2\alpha$ core in Be isotopes. 
(d) A schematic figure of
evolution of single-particle energies from spherical shell-model orbits to molecular orbits. 
(e) Experimental energy levels assigned to band-head states in 
low-lying spectra
of $^{10}$Be, $^{11}$Be, and $^{12}$Be, and theoretical ones of $^{13}$Be and $^{14}$Be obtained by 
the AMD+VAP calculations using the MV1  ($m=0.65$, $b=h=0$) central and G3RS ($u_1=-u_2=-3700$ MeV) 
spin-orbit interactions \cite{KanadaEn'yo:2002ay,KanadaEn'yo:2012rm}.  
Figures are modified from the original ones in Ref.~\cite{KanadaEn'yo:2012rm}.
\label{fig:beiso-orbital}}
\end{center}
\end{figure}

%be-13
In order to understand cluster features of low-lying states of neutron-rich Be isotopes,
the molecular orbit (MO) is a useful picture, which has been 
applied for $^9$Be \cite{OKABEa,OKABEb,OKABEc} and extended to neutron-rich 
Be isotopes \cite{SEYA,OERTZENa,OERTZENb,Oertzen-rev,Itagaki:1999vm,Itagaki:2000nn}.
As already mentioned, the $2\alpha$ core is formed in neutron-rich Be isotopes as a result of
many-body correlation 
in $A$-nucleon dynamics. In the $2\alpha$ systems with valence neutrons, 
molecular orbits around the $2\alpha$ core are constructed by linear combination of $p$-orbits
around $\alpha$ clusters and are occupied by valence neutrons. 
Schematic figures of molecular orbits are shown in Fig.~\ref{fig:beiso-orbital}. Negative-parity
orbits called "$\pi_{3/2}$-orbit" and "$\pi_{1/2}$-orbit" are the lowest nodal orbits  
with one node in the transverse direction corresponding to $p$-orbits in the spherical shell model limit.
The $\pi_{3/2}$-orbit is an $ls$-favored orbit and the lowest orbit in the $2\alpha$ system.
The positive-parity orbits denoted by "$\sigma_{1/2}$-orbit" and "$\pi^*_{3/2}$-orbit" are 
higher nodal orbits. 
Since the $\sigma_{1/2}$-orbit is a longitudinal orbit having two nodes along the
$\alpha$-$\alpha$ direction,  
its kinetic energy reduces as the $2\alpha$ cluster develops. 
Consequently, valence neutrons in the $\sigma_{1/2}$-orbit push two $\alpha$s 
outward keeping them at a moderate distance, and enhance 
the cluster structure of neutron-rich Be isotopes. 
Moreover, in a developed cluster structure, the lowered $\sigma_{1/2}$-orbit eventually comes down
below the $ls$-unfavored $\pi_{1/2}$-orbit, and induces the breaking of the $N=8$ shell closure.

%%%%%%%%%%%%%%%%%%%%%%%%%%%%%%
%%%%%%%%%%%%%%%%%%%%%%%%%%%%%
 
In a cluster state with a $\sigma_{1/2}$-orbit configuration, $2\alpha$ clusters are tightly 
bonded at a moderate distance by the valence neutrons in the $\sigma_{1/2}$-orbit. 
We call this structure with $\sigma_{1/2}$-orbit neutrons "the MO $\sigma$-bond structure".
However, in the asymptotic region of  large inter-cluster distance,
the system approaches di-cluster states of He+He, 
in which valence neutrons occupy atomic orbits around either of $\alpha$ clusters 
instead of  molecular orbits to gain correlation energy between valence neutrons. 
Owing to inter-cluster excitations in the di-cluster structure, 
He+He cluster resonances (called di-cluster resonances) appear in highly excited region.
For instance, $^6$He+$^4$He 
and $^{6(8)}$He+$^{6(4)}$He cluster resonances have been predicted
in $^{10}$Be and $^{12}$Be, respectively.
It means that two kinds of remarkable cluster structures 
coexist in neutron-rich Be isotopes. 
One is the MO $\sigma$-bond states in the strong coupling regime, and the other is the 
di-cluster resonances in the weak coupling regime. The latter usually appears in relatively 
higher energy region than the MO $\sigma$-bond states.
In order to unify these two kinds of cluster states, the MO $\sigma$-bond and the di-cluster resonance states,  
Ito and his collaborators developed a method of the generalized two-center cluster model (GTCM), 
which successfully describes the cluster structures of $^{10}$Be and $^{12}$Be, and 
showed transition of valence neutron configurations from molecular orbits 
to atomic orbits. 
Motivated by the theoretical predictions, 
many experimental studies have been achieved to search for rotational band members 
of the MO $\sigma$-bond states and di-cluster resonances. 

%%%%%%%%%%%%%%%%%%%%%%%%%%%%%%
\begin{figure}[htb]
\begin{center}
\resizebox{0.6\textwidth}{!}{%
\includegraphics{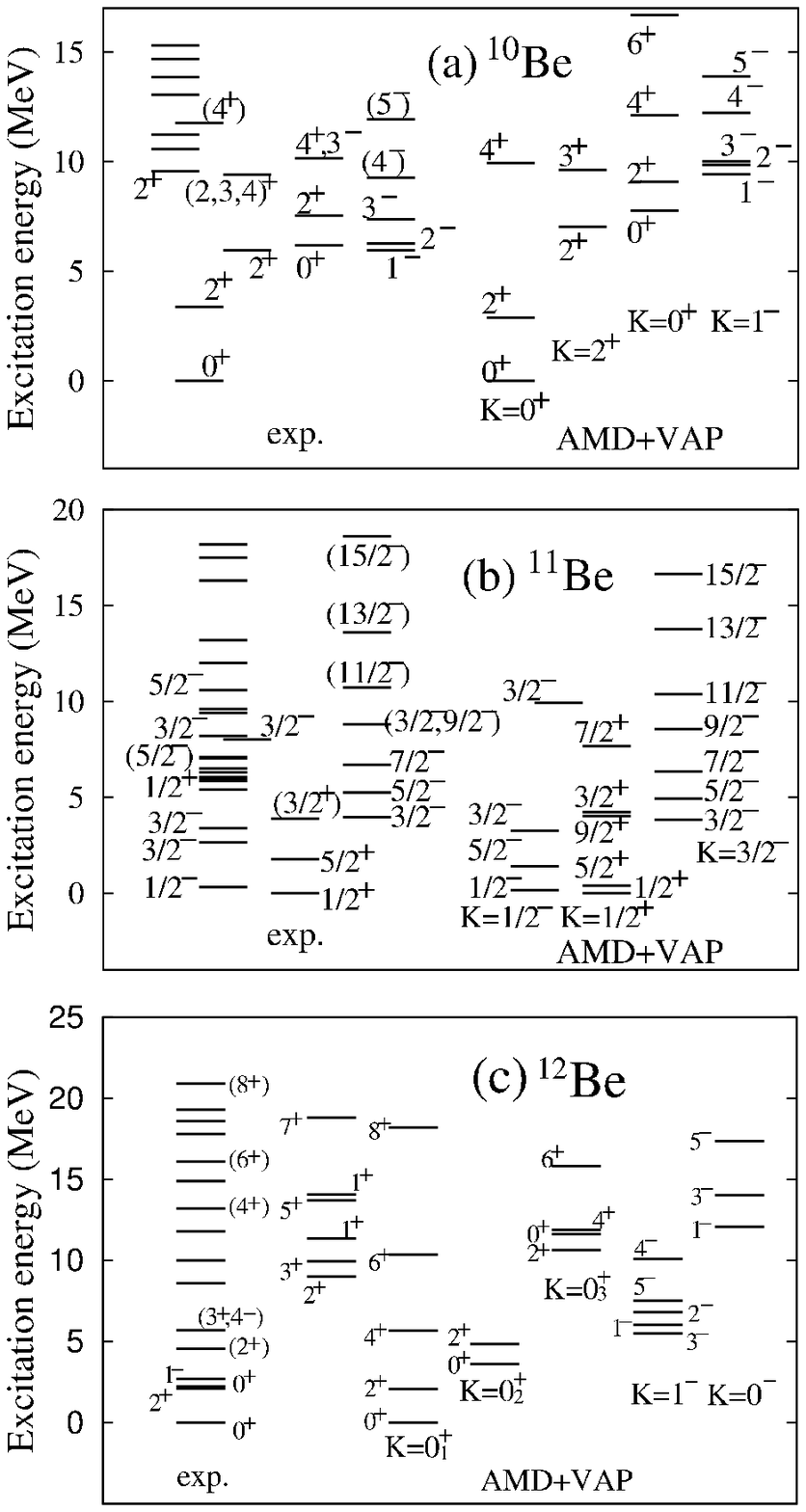} }	
%\includegraphics[width=15cm]{bespe-fig.eps} 	
%\vspace{0.5cm}
  \caption{Energy spectra of $^{10}$Be, $^{11}$Be, and $^{12}$Be.
Theoretical energy levels calculated with the AMD+VAP
using the MV1 central and G3RS spin-orbit interactions
\cite{KanadaEn'yo:1999ub,KanadaEn'yo:2002rh,KanadaEn'yo:2003ue,KanadaEn'yo:2004hy}
and experimental ones are shown.
The interaction parameters, $m=0.65$, $b=h=0$, and $u_1=-u_2=-3700$ MeV, are used 
for $^{11}$Be and $^{12}$Be, and $m=0.62$, $b=h=0$, and $u_1=-u_2=-3000$ MeV
are used for $^{10}$Be.
The experimental data for $^{10}$Be 
are from Ref.~\cite{Tilley:2004zz} and references therein. Those for 
$^{11}$Be are from Refs.~\cite{bohlen98,Kelley:2012qua}, and those for $^{12}$Be
are from Refs.~\cite{Freer:1999zz,Freer:2001ef,Fortune:1994zz,Shimoura:2007zz}.
The original figures are  
in Refs.~\cite{KanadaEn'yo:1999ub,KanadaEn'yo:2002rh,KanadaEn'yo:2003ue}.
\label{fig:bespe}
}
\end{center}
\end{figure}
%%%%%%%%%%%%%%%%%%%%%%%%%%%%%

%10Be spectra :from aris2014 %Beck %9Li
\subsubsection{Band structures of $^{10}$Be, $^{11}$Be, and $^{12}$Be}
Let us discuss the band structures of $^{10}$Be, $^{11}$Be, and $^{12}$Be based on the AMD+VAP
results using the MV1 central and G3RS spin-orbit interactions \cite{KanadaEn'yo:1999ub,KanadaEn'yo:2002rh,KanadaEn'yo:2003ue,KanadaEn'yo:2004hy}. 
Calculated energy levels are shown in Fig.~\ref{fig:bespe} compared with experimental ones.
Energy levels of $^{10}$Be %in Fig.~\ref{fig:bespe}(a)
are classified into four rotational bands, $K^\pi=0^+_1$, $2^+$, $0^+_2$, and $1^-$,  
consisting of band members $J^\pi=$
($0^+_1$, $2^+_1$, $4^+_1$),
($2^+_2$, $3^+_1$),
($0^+_2$, $2^+_3$, $4^+_2$, $6^+_1$), and
($1^-$, $2^-$, $3^-$, $4^-$, $5^-$), respectively.
The band-head states of these bands are assigned to the experimentally observed states, 
$0^+_1$, $2^+_2$(5.96 MeV), $0^+_2$(6.18 MeV), and $1^-_1$(5.96 MeV). 
Various kinds of cluster structures are found in the intrinsic states of these bands.
The ground state of $^{10}$Be 
has the normal neutron configuration ($\pi_{3/2}^2$) with the $2\alpha$ core.
The $K^\pi=2^+$ band is regarded as a side band of the ground band arising from 
two neutron correlation, in 
other words, a triaxial intrinsic structure as discussed in Refs.~\cite{Itagaki:2001az,Suhara:2009jb}. 
The $K^\pi=0^+_2$ band has a MO $\sigma$-bond structure with a remarkably developed 
cluster structure, in which $2\alpha$ clusters are bonded by two valence neutrons in the 
$\sigma_{1/2}$-orbit (the $\sigma^2_{1/2}$ configuration).
The $K^\pi=1^-$ band is a negative-parity band with a $\pi_{3/2}$-orbit neutron and 
a $\sigma_{1/2}$-orbit neutron (the $\pi_{3/2}\sigma_{1/2}$ configuration).
Because of the $\sigma_{1/2}$-orbit neutron, the intrinsic state has a
moderately developed cluster structure and constructs the rotational band.

%%%%%%%%%%%%%%%%%%%%%%%%%%%%%%
\begin{figure}[htb]
\begin{center}
\resizebox{0.6\textwidth}{!}{%
\includegraphics{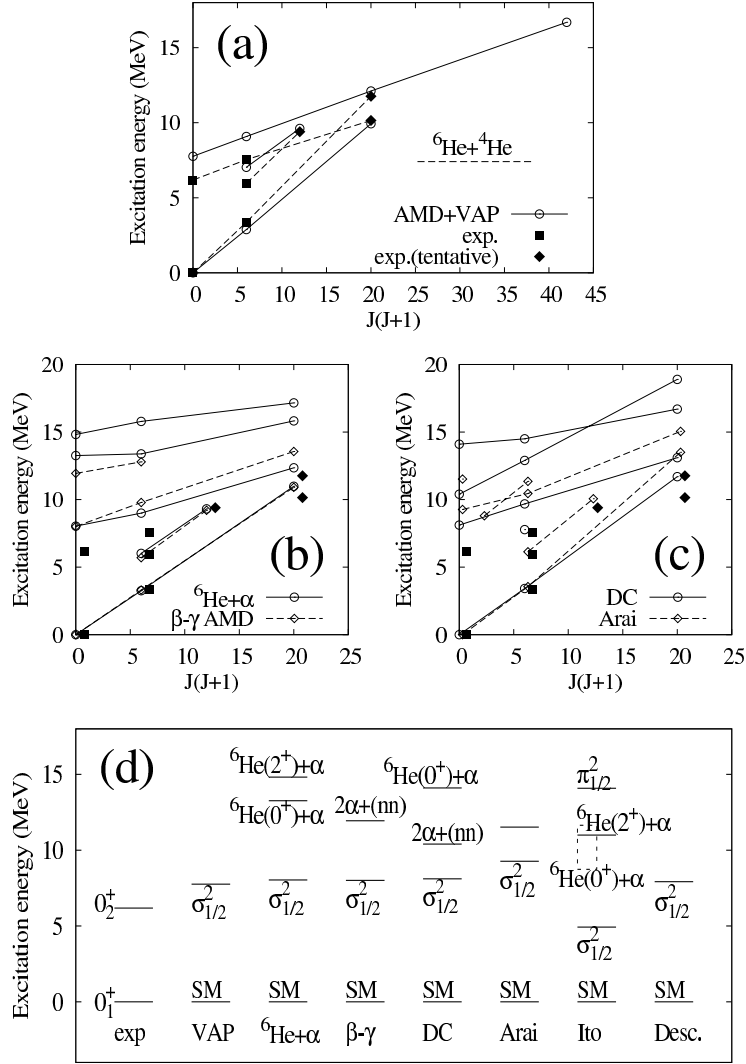} }	
%\includegraphics[width=15cm]{be10rot-fig.eps} 	
%\vspace{0.5cm}
  \caption{(a)(b)(c): Energy spectra of positive parity states in $^{10}$Be. 
(d): excitation energies of $0^+$ states in $^{10}$Be. 
Theoretical energies are those obtained by 
the AMD+VAP using the MV1  ($m=0.62$, $b=h=0$) central and G3RS ($u_1=-u_2=-3000$ MeV) 
spin-orbit interactions \cite{KanadaEn'yo:1999ub}, the
$^6$He+$\alpha$ cluster model 
in Ref.~\cite{KanadaEn'yo:2011nc}, the
$\beta$-$\gamma$ constraint AMD \cite{Suhara:2009jb}, the 
dineutron condensation(DC)+AMD model \cite{Kobayashi:2012di}, 
the stochastic variational method of $2\alpha+2n$ by 
Arai {\it et al.} \cite{Arai:2004yf},
the generator coordinate method of 
$^6$He+$\alpha$ by Descouvemont {\it et al.} \cite{Descouvemont:2002mnw},
and the generalized two-center cluster model 
by Ito {\it et al.} \cite{Ito2014-rev}.
The effective nuclear interactions used in Refs.~\cite{KanadaEn'yo:2011nc,Suhara:2009jb,Kobayashi:2012di}
are the Volkov No.2  ($m=0.6$, $b=h=0.125$) central and G3RS  ($u_1=-u_2=-1600$ MeV)  spin-orbit interactions.
For other theoretical calculations, 
spectra calculated by the Minnesota central interaction \cite{Thompson:1977zz} supplemented by spin-orbit interaction of 
Refs.~\cite{Arai:2004yf,Descouvemont:2002mnw} and 
those using the Volkov No.2  ($m=0.643$, $b=h=0.125$) central and G3RS  ($u_1=-2000$ MeV, $u_2=3000$ MeV)  spin-orbit interactions of Ref.~\cite{Ito2014-rev} are shown.
The experimental data are those from Ref.~\cite{Tilley:2004zz} and references therein. 
\label{fig:be10rot}
}
\end{center}
\end{figure}
%%%%Dufour10%%%%%%%%%%%%%%%%%%%%%%%%%

Many experiments have been performed
to discover new states and confirm band structures, in particular, coexistence of 
the ground $K^\pi=0^+_1$ band and the MO $\sigma$-bond $K^\pi=0^+_2$ band
\cite{Hamada:1994zz,Soic:1995av,Curtis:2001sd,Fletcher:2003wk,Curtis:2004wr,Ahmed:2004th,Millin05,Freer:2006zz,Bohlen:2007qx,Suzuki:2013mga}. 
In Fig.~\ref{fig:be10rot}, 
the observed positive-parity energy levels of $^{10}$Be 
are plotted as functions of $J(J+1)$  compared with theoretical
energy spectra of the AMD+VAP \cite{KanadaEn'yo:1999ub}, $\beta$-$\gamma$ AMD \cite{Suhara:2009jb}, 
$^6$He+$\alpha$ cluster model \cite{KanadaEn'yo:2011nc}, AMD+DC \cite{Kobayashi:2012di}, 
and 4-body $2\alpha+2n$ calculations \cite{Arai:2004yf}.

In the AMD+VAP result, the MO $\sigma$-bond structure having the enhanced cluster structure 
constructs the $K^\pi=0^+_2$ band up to $J=6$ with
 a large moment of inertia.
Experimental studies have revealed a
$2^+$ state at 7.54 MeV and a $4^+$ state at 10.2 MeV, which likely posses a
$^6$He+$\alpha$ cluster structure \cite{Millin05,Freer:2006zz}.
These $2^+$ and $4^+$ states are candidates of the 
$K^\pi=0^+_2$ band members starting from the $0^+_2$(6.18 MeV).
This assignment is 
consistent with the energy slope of the calculated $K^\pi=0^+_2$ band.
Other theoretical calculations, the $\beta$-$\gamma$ AMD \cite{Suhara:2009jb}, 
$^6$He+$\alpha$ cluster model \cite{KanadaEn'yo:2011nc}, AMD+DC \cite{Kobayashi:2012di}, 
and 4-body $2\alpha+2n$ calculations \cite{Arai:2004yf}, 
as well as the $^6$He+$\alpha$ GCM \cite{Descouvemont:2002mnw} give almost consistent results 
for these bands with the AMD+VAP result.

Another interesting problem in $^{10}$Be is whether di-cluster resonances of $^6$He+$^4$He 
exist. 
Theoretical calculations predicted $0^+$ states
a few MeV higher than the $0^+_2$ state
as shown in Fig.~\ref{fig:be10rot}. The AMD+DC
and $^6$He+$\alpha$ cluster models predicted $^6$He($0^+$)+$^4$He cluster resonances, 
whereas the GTCM predicted the 
$^6$He($0^+$)+$^4$He state as a broad continuum state (the dashed line in Fig.~\ref{fig:be10rot}(d)). Moreover, 
$^6$He($2^+$)+$^4$He cluster resonances and a three-cluster resonance of $2\alpha+nn$ have been 
theoretically predicted. 
%There is no experimentally confirmed $0^+$ states above the $0^+_2$ state
%in $^{10}$Be. 
A very broad $0^+$ state of the $^6$Li($T=1$)+$\alpha$ resonance, 
which was recently observed at 11 MeV in $^{10}$B \cite{Kuchera:2011ax},
is a candidate of the isobaric analog state of a $^6$He+$\alpha$ cluster resonance. 

%11Be spectra:von-Oetzen02-rev, borlen02
For $^{11}$Be (Fig.~\ref{fig:bespe}(b)), 
we obtain three rotational bands, the $K^\pi=1/2^+$,  $K^\pi=1/2^-$, and $K^\pi=3/2^-$ bands.
In the AMD+VAP calculation, the effective nuclear interaction is adjusted
to reproduce the parity inversion of the $1/2^-_1$ and $1/2^+_1$ states 
in $^{11}$Be \cite{KanadaEn'yo:2002rh}. 
The $K^\pi=1/2^-$ is the normal state with the $\pi^2_{3/2}\pi_{1/2}$ configuration 
corresponding to the $0\hbar\omega$ $p$-shell configuration. 
The $K^\pi=1/2^+$ is a MO $\sigma$-bond state with two neutrons in the $\pi_{3/2}$-orbit and the 
last neutron in the $\sigma_{1/2}$-orbit (the $\pi_{3/2}^2\sigma_{1/2}$ configuration).
This corresponds to the intruder $1\hbar\omega$ state with one particle 
in the higher shell ($sd$-shell) in terms of the spherical shell model. 
The ground state is the band-head 
$1/2^+$ state of the $K^\pi=1/2^+$ band constructed by the MO $\sigma$-bond structure. 
The abnormal spin-parity $1/2^+$ of the ground state in $^{11}$Be has been 
known as the breaking of the $N=8$ magic number. 
Because of one neutron in the $\sigma_{1/2}$-orbit, the $K^\pi=1/2^+$ band 
has the moderately developed cluster structure. 
In the calculation, another type of MO $\sigma$-bond structure with 
two $\sigma_{1/2}$-orbit neutrons (the $\pi_{3/2}\sigma_{1/2}^2$ configuration)  is obtained as 
an excited band. 
Because of two neutrons in the $\sigma_{1/2}$-orbit, the cluster structure is developed further 
and it constructs the $K^\pi=3/2^-$ band up to high spin states with small level spacing, 
{\it i.e.}, a large moment of inertia.
The existence of the MO $\sigma$-bond band with the $\sigma_{1/2}^2$ configuration 
has been suggested by von Oertzen {\it et al.} (Ref.~\cite{oertzen03-rev} and references therein).
Candidates for $K^\pi=3/2^-$ band members
have been experimentally observed by two-neutron transfer reactions 
\cite{oertzen03-rev,bohlen98,bohlen02}.  The theoretical energy spectra of 
the $K^\pi=3/2^-$ band obtained by  
the AMD+VAP is consistent with the experimental assignment of
the $K^\pi=3/2^-$ band starting from $3/2^-$ state at 3.96 MeV. 

%12Be spectra:
%Beck-rev

The ground and excited states of $^{12}$Be have been calculated using the same effective nuclear
interaction as that used for  $^{11}$Be.
The calculated energy levels of $^{12}$Be 
are classified into three positive-parity bands ($K^\pi=0^+_1, 0^+_2$, and $0^+_3$)
and  two negative-parity bands ($K^\pi=1^-$ and $0^-$) (Fig.~\ref{fig:bespe}(c)). 
The low-lying $K^\pi=0^+_1$, $0^+_2$, and $1^-$ bands are understood 
as molecular $\pi$- and $\sigma$-orbit configurations, 
whereas the $K^\pi=0^+_3$ and 
$0^-$ bands in the high-energy region are He+He di-cluster resonances.

The ground band ($K^\pi=0^+_1$) has a MO $\sigma$-bond structure 
with two $\sigma_{1/2}$-orbit neutrons (the $\pi^2_{3/2}\sigma^2_{1/2}$
configuration), which corresponds to the intruder $2\hbar\omega$ configuration 
(2 particles in the $sd$ shell) indicating the breaking of the $N=8$ shell. 
The $K^\pi=0^+_2$ band is dominated by the $\pi^2_{3/2}\pi^2_{1/2}$ 
configuration corresponding to a normal $0\hbar\omega$ configuration with the 
neutron $p$-shell closure. The
band-head state of the $K^\pi=0^+_2$ band is assigned to the experimental 
$0^+_2$ state at 2.1 MeV. 
As a result of the inversion between the normal and 
intruder configurations, the ground state of  $^{12}$Be has a large deformation
with the cluster structure enhanced by two $\sigma_{1/2}$-orbit neutrons
even though $^{12}$Be is an $N=8$ nucleus.
The intruder configuration and large deformation of the ground state 
have been experimentally supported by weak $\beta$ decays to $^{12}$B \cite{Barker76,Suzuki:1997zza},
strong $E2$ transitions in the ground band \cite{Iwasaki:2000gh,Iwasaki:2000gp,Imai:2009zza}, 
and other experiments \cite{Alburger:1978zza,Fortune:1994zz,shimoura03,Meharchand:2012zz},
and more directly evidenced by 1$n$-knockout reactions \cite{Navin:2000zz,Pain:2005xw}.
The $K^\pi=1^-$ band is constructed by a MO $\sigma$-bond structure with 
the $\pi^2_{3/2}\pi_{1/2}\sigma_{1/2}$ configuration. This band 
has a moderately enhanced cluster structure with  a $\sigma_{1/2}$-orbit neutron.

%%%%%%%%%%%%%%%%%%%%%%%%%%%%%%
\begin{figure}[htb]
\begin{center}
\resizebox{0.5\textwidth}{!}{%
\includegraphics{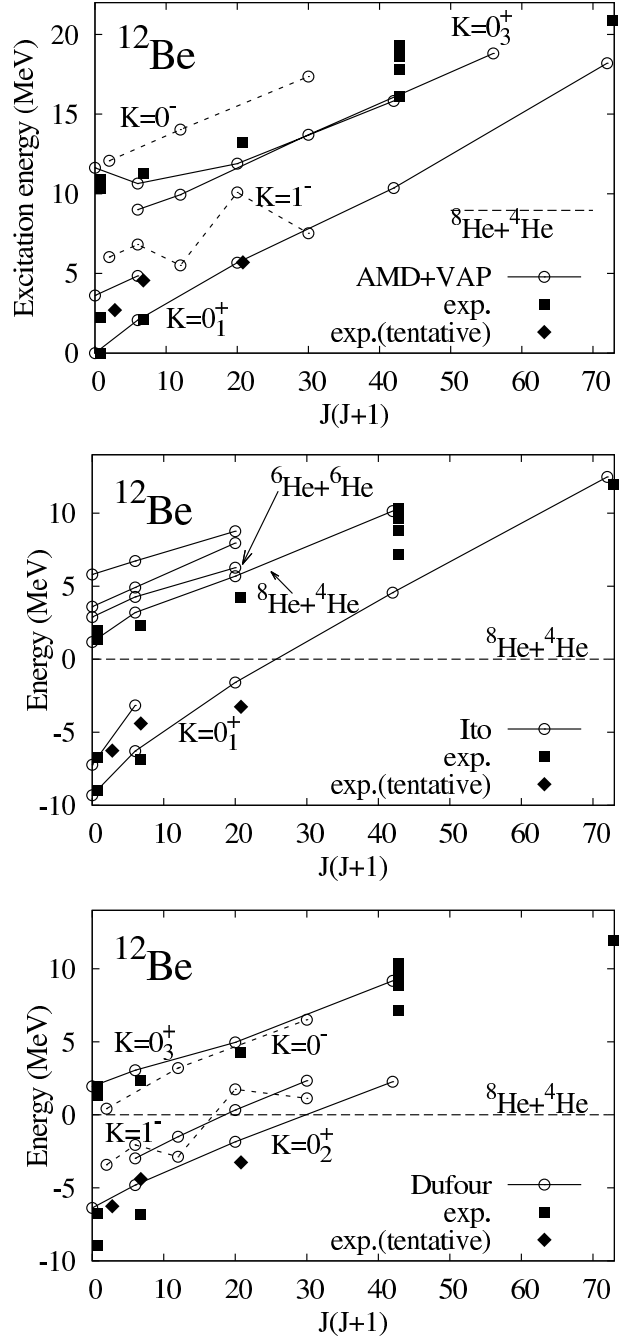} }	
%\includegraphics[width=15cm]{be12rot-fig.eps} 	
%\vspace{0.5cm}
  \caption{Energy spectra of positive- and negative-parity rotational bands in $^{12}$Be. 
Excitation energies obtained by the AMD+VAP
using the MV1  ($m=0.65$, $b=h=0$) central and G3RS ($u_1=-u_2=-3700$ MeV) 
spin-orbit interactions
\cite{KanadaEn'yo:2003ue,KanadaEn'yo:2004hy} are shown in the top panel.
Energies from the $\alpha$-decay threshold 
calculated with the generalized two-center cluster model
by Ito {\it et al.} \cite{Ito:2008zza}, and those with the 
GCM of $^6$He+$^6$He and $^8$He+$\alpha$ 
by Dufour {\it et al.} \cite{Dufour:2010dmf} are shown in the middle and
bottom panels, respectively. The Volkov No.2 central interactions and the spin-orbit interactions
are used in those calculations of Refs.~\cite{Ito:2008zza,Dufour:2010dmf}.
Experimental data of
spin-assigned energy levels from Refs.~\cite{Freer:1999zz,Freer:2001ef,SAITO04,Yang:2014kxa,Fortune:1994zz,Ajzenberg90,Shimoura:2007zz}. 
are shown by filled symbols. Energy levels with tentative spin assignment 
are shown by filled diamonds.
\label{fig:be12rot}
}
\end{center}
\end{figure}
%%%%%%%%%%%%%%%%%%%%%%%%%%%%%

In the progress of experimental and theoretical investigations of $^{12}$Be in the recent past years, 
He+He di-cluster resonances 
in highly excited states are being revealed 
\cite{Oertzen-rev,Descouvemont:2001wek,KanadaEn'yo:2003ue,Ito:2008zza,Dufour:2010dmf,Ito:2012zza,Ito2014-rev,bohlen02,korsheninnikov95,Freer:1999zz,Freer:2001ef,SAITO04,Yang:2014kxa}.
Many excited states have been observed above the He+He threshold energies
by $^6$He+$^6$He and $^8$He+$^4$He break-up 
reactions \cite{Freer:1999zz,Freer:2001ef,SAITO04,Yang:2014kxa}. They are  
considered to be He+He di-cluster resonances.
For the theoretical side, He+He resonances
have been predicted near and above the threshold energies
by the  GCM \cite{Descouvemont:2001wek,Dufour:2010dmf}, AMD+VAP \cite{KanadaEn'yo:2003ue,KanadaEn'yo:2004hy},
and GTCM calculations \cite{Ito:2008zza,Ito:2012zza,Ito2014-rev}. 
The theoretical rotational band structures of $^{12}$Be are shown in Fig.~\ref{fig:be12rot}. 
The results of low-lying levels for the
$K^\pi=0^+_1$ and $K^\pi=0^+_2$ bands obtained by the AMD+VAP and GTCM
are qualitatively consistent with each other, and they reproduce well the experimental 
low-energy spectra. In the energy region above the He+He thresholds,  
the GTCM predicts the $^6$He+$^6$He di-cluster resonance band and also the $^8$He+$^4$He
di-cluster  resonance band. The di-cluster resonances have been also predicted by the GCM
calculations,  however, the $^6$He+$^6$He and $^8$He+$^4$He components are mixed in the
$K^\pi=0^+_3$ rotational band, that is, 
the $0^+$ state has the dominant $^6$He+$^6$He component while the  $2^+$ and $4^+$ states 
contain mixed components of two channels. 
The AMD+VAP calculation shows a di-cluster resonance feature in the $K^\pi=0^+_3$ band, in which 
the structure changes as the increase of $J$ from the $^6$He+$^6$He structure 
to the strong coupling cluster structure, 
probably containing mixed components of $^6$He+$^6$He and $^8$He+$^4$He. The side-band,
$K^\pi=2^+$ band associated with the $K^\pi=0^+_3$ band, is also predicted.

In addition to the positive-parity bands, the GCM calculations
\cite{Descouvemont:2001wek,Dufour:2010dmf} and the 
AMD+VAP calculation \cite{KanadaEn'yo:2004hy} predict
a negative-parity band $K^\pi=0^-$ of $^8$He+$^4$He di-cluster resonances consisting of 
$1^-$, $3^-$, and $5^-$ states.  
Considering that a reflection asymmetric intrinsic structure of the ground $K^\pi=0^+_1$ band 
having $^8$He+$^4$He clustering, 
the $K^\pi=0^-$ band can be 
interpreted as the parity partner of the $K^\pi=0^+_1$ band 
caused by the negative-parity excitation of $^8$He-$^4$He motion, and is 
associated with the well-known parity doublet $K^\pi=0^-$ band in $^{20}$Ne.
It means that negative-parity excitations in the two bands,
$K^\pi=1^-$ and $K^\pi=0^-$, of $^{12}$Be are 
different.
The former is a single-particle excitation between $\pi_{1/2}$- and $\sigma_{1/2}$-orbits,
and the latter is an excitation of the inter-cluster motion. 
It should be pointed out that the energy cost for the single-particle excitation in the MO configuration 
is small because of  vanishing of the $N=8$ shell gap 
in the developed cluster system. This is a reason why the $K^\pi=1^-$ band appears in the low-energy 
region. Intense experimental efforts are being made to establish 
 band structures of $^{12}$Be including the low-lying MO $\sigma$-bond states 
and the high-lying di-cluster resonances.

%Vanishing of N=8 in 11Be,12Be,13Be from MO picture: from aris2014
\subsubsection{Systematics of low-lying states of Be isotopes from molecular orbit picture}

As discussed previously, the molecular orbits play an important role in cluster structures of 
low-lying states of neutron-rich Be isotopes. In particular, $\sigma_{1/2}$-orbit neutrons
enhance the cluster structure to form the MO $\sigma$-bond structure.  
%in which 2$\alpha$ clusters is bonded by the $\sigma_{1/2}$-orbit neutrons at the moderate distance. 
Moreover,  the lowering mechanism of the $\sigma_{1/2}$-orbit in developed cluster structures is 
essential for the vanishing of the $N=8$ magic number in $^{11}$Be and $^{12}$Be. 
Looking into systematics of low-lying states in the series of 
Be isotopes, we here discuss evolution of the clustering in relation to 
the vanishing of the $N=8$ magic number from the molecular orbit picture.

Let us briefly review again the molecular orbits proposed by Seya {\it et al.} and von Oertzen
{\it et al.} 
In the molecular orbit model for 2$\alpha$ cluster systems, molecular orbits 
given by the linear combinations of $p$-orbits around $\alpha$ clusters are considered for valence
neutron configurations. As shown in Refs.~\cite{OERTZENa,OERTZENb,Oertzen-rev,TC-SM}, 
single-particle energy levels are smoothly connected from the one-center to the two-center limits
as functions of $\alpha$-$\alpha$ distance. The orbits 
in a moderate distance region correspond to molecular orbits. 
We call negative-parity orbits constructed by $p$-orbits perpendicular to 
the $\alpha$-$\alpha$ direction ``$\pi$-orbits'', and
a positive-parity orbit from $p$-orbits parallel to 
the $\alpha$-$\alpha$ direction ``$\sigma$-orbit''  in analogy to covalent orbits of electrons in molecules
(see Figs.~\ref{fig:beiso-orbital} (a) and (b)). 
We call the other positive-party orbit given by $p$-orbits perpendicular to 
the $\alpha$-$\alpha$ direction ``$\pi^*$-orbit'' (Fig.~\ref{fig:beiso-orbital}(c)). 
In addition to the spatial configurations ($\pi$, $\sigma$, and $\pi^*$),  
molecular orbits are specified by the angular momentum $\Omega\equiv j_z$ 
projected on to the symmetry axis $z$. With the label $\Omega$, 
we use the notations, $\pi_{3/2,1/2}$, $\sigma_{1/2}$, and $\pi^*_{3/2,1/2}$. 
Here, the $\pi_{3/2(1/2)}$- and  $\pi^*_{3/2(1/2)}$-orbits 
are $ls$-favored (unfavored) orbits.
Note that the present notations,  $\pi_{3/2}$, $\pi_{1/2}$,  $\sigma_{1/2}$, and  $\pi^*_{3/2}$ correspond to 
the labels $\pi 3/2^-(g)$, $\sigma 1/2^-(g)$, $\sigma 1/2^+(u)$, and $\pi 3/2^+(u)$ in Ref.~\cite{Oertzen-rev}, and 
the labels $(3u,1)$, $(1u,2)$, $(1g,2)$, and $(1g,2)$ in Ref.~\cite{SEYA}, respectively.
In the spherical shell model limit, the $\pi_{3/2}$- and $\pi_{1/2}$-orbits 
become the $p_{3/2}$- and $p_{1/2}$-orbits, respectively, whereas the $\sigma_{1/2}$- and $\pi^*_{3/2}$-orbits become
the $d_{5/2}$-orbits (see Fig.~\ref{fig:beiso-orbital}(d)).
As the $\alpha$-$\alpha$ distance increases, 
the energy of the $\sigma_{1/2}$($\pi^*_{3/2,1/2}$)-orbit with two nodes (one node) along
the longitudinal ($z$) axis decreases because of the kinetic energy reduction, 
whereas the energies of the $\pi_{3/2,1/2}$-orbits with no node rise up. Consequently, 
the inversion of the $\pi_{1/2}$- and $\sigma_{1/2}$-orbits occurs in the developed cluster system. 
It is also important that $\sigma_{1/2}$- and  $\pi^*_{3/2}$-orbit neutrons enhance 
the cluster structure because of the lowering mechanism, however, $\pi_{3/2,1/2}$-orbit
neutrons tend to suppress the cluster structure to gain the potential energy.

%%%%%%%%%%%%%%%%%%%%%%%%%%%%%%
\begin{table}[ht]
\caption{\label{tab:beiso}
Molecular orbit configurations of band-head states 
of low-lying states in Be isotopes.
Neutron configurations around the $2\alpha$ core are listed.
The $0\hbar\omega$, $1\hbar\omega$, 
and $2\hbar\omega$ excitations of the shell model (SM) configurations  
are also shown based on the correspondence of negative-parity orbits ($\pi_{3/2}$,$\pi_{1/2}$)
and positive-parity orbits ($\sigma_{1/2},\pi^*_{3/2}$) to  
the $p$- and $sd$-orbits in the spherical limit, respectively.
The numbers ($n_\pi,n_\sigma,n_{\pi *}$) of $\pi$-, $\sigma$-, $\pi^*$-orbit neutrons are also listed.
The assignment of the SM and MO configurations are based 
on Ref.~\cite{KanadaEn'yo:2012rm}.
(The configuration for $^{13}$Be($5/2^+$) is tentative.)
}
\begin{center}
\begin{tabular}{cccccc}
\hline
\hline
 & SM & MO config.  & $n_\pi, n_\sigma+n_{\pi *}$\\
%\hline
$^{10}$Be($0^+_1$) &  $0\hbar\omega$ & $\pi^2_{3/2}$ & 2,0 \\
$^{10}$Be($1^-$) &  $1\hbar\omega$ & $\pi_{3/2}\sigma_{1/2}$& 1,1 \\
$^{10}$Be($0^+_2$) &  $2\hbar\omega$ & $\sigma^2_{1/2}$ & 0,2 \\
%\hline
$^{11}$Be($1/2^+$) &  $1\hbar\omega$ & $\pi^2_{3/2}\sigma_{1/2}$ & 3,0 \\
$^{11}$Be($1/2^-$) &  $0\hbar\omega$ & $\pi^2_{3/2}\pi_{1/2}$ & 2,1 \\
$^{11}$Be($3/2^-$) &  $2\hbar\omega$ & $\pi_{3/2}\sigma^2_{1/2}$ & 1,2 \\
%\hline
$^{12}$Be($0^+_1$) &  $2\hbar\omega$ & $\pi^2_{3/2}\sigma^2_{1/2}$ & 2,2 \\
$^{12}$Be($0^+_2$) &  $0\hbar\omega$ & $\pi^2_{3/2}\pi^2_{1/2}$ & 4,0 \\
$^{12}$Be($1^-$) &  $1\hbar\omega$ & $\pi^2_{3/2}\pi_{1/2}\sigma_{1/2}$ &3,1 \\
%\hline
$^{13}$Be($1/2^-$) &  $1\hbar\omega$ & $\pi^2_{3/2}\pi_{1/2}\sigma^2_{1/2}$ & 3,2 \\
$^{13}$Be($5/2^+$) &  $0\hbar\omega$ & $(\pi^2_{3/2}\pi^2_{1/2}\sigma_{1/2})$  & 4,1\\
$^{13}$Be($3/2^+$) &  $2\hbar\omega$ & $\pi^2_{3/2}\sigma^2_{1/2}\pi^*_{3/2}$ & 2,2+1\\
%\hline
$^{14}$Be($0^+_1$) &  $0\hbar\omega$ & $\pi^2_{3/2}\pi^2_{1/2}\sigma^2_{1/2}$ & 4,2\\
$^{14}$Be($2^-$) &  $1\hbar\omega$ & $\pi^2_{3/2}\pi_{1/2}\sigma^2_{1/2}\pi^{*}_{3/2}$ & 3,2+1\\
$^{14}$Be($0^+_2$) &  $2\hbar\omega$ & $\pi^2_{3/2}\sigma^2_{1/2}\pi^{*2}_{3/2}$ &2,2+1 \\
\hline
\hline
\end{tabular}
\end{center}
\end{table}
%%%%%%%%%%%%%%%%%%%%%%%%%%%%%%%%%

%%%%%%%%%%%%%%%%%%%%%%%%%%%%%%
\begin{figure}[htb]
\begin{center}
\resizebox{0.6\textwidth}{!}{%
\includegraphics{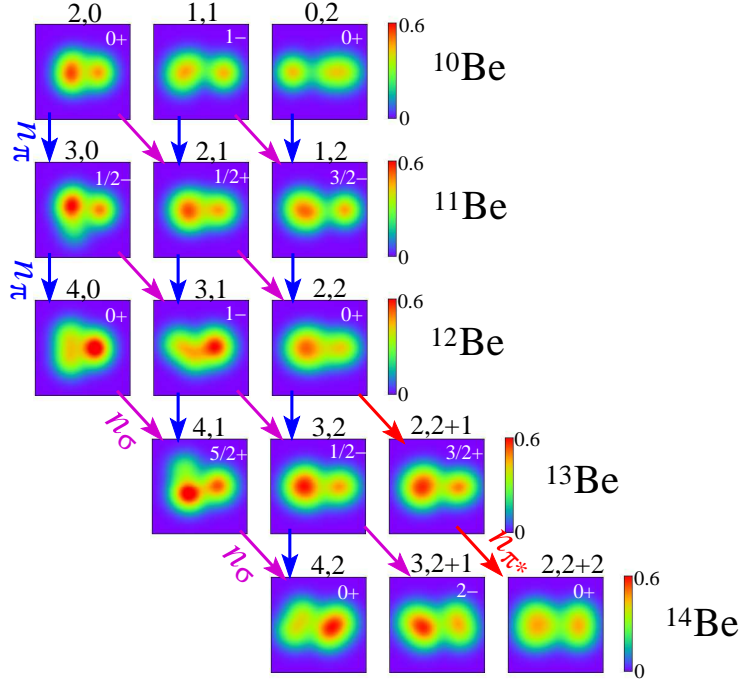} }	
%\includegraphics[width=9.0cm]{beiso-bandhead-dense-fig.eps} 	
%\vspace{0.5cm}
  \caption{Matter density distributions in the band-head states 
of Be isotopes obtained by the AMD+VAP 
using the MV1  ($m=0.65$, $b=h=0$) central and G3RS ($u_1=-u_2=-3700$ MeV) 
spin-orbit interactions 
\cite{KanadaEn'yo:1999ub,KanadaEn'yo:2002rh,KanadaEn'yo:2003ue,KanadaEn'yo:2002ay,KanadaEn'yo:2012rm}.
For $^{10}$Be, the results obtained using the MV1  ($m=0.62$, $b=h=0$) central and G3RS ($u_1=-u_2=-3000$ MeV) 
spin-orbit interactions are shown.
Panels for the states with 
$\sigma_{1/2}^0$, $\sigma_{1/2}^1$, $\sigma_{1/2}^2$,
$\sigma_{1/2}^2\pi^*_{3/2}$, and $\sigma_{1/2}^2(\pi^*_{3/2})^2$
configurations are aligned in the first, second, third, fourth, and fifth columns
from the left based on the dominant MO configurations listed in Table \ref{tab:beiso}. 
The neutron numbers ($n_\pi,n_\sigma+n_{\pi *}$) in the $\pi_{3/2,1/2}$-, $\sigma_{1/2}$-, $\pi^*_{3/2}$-orbits are shown above each panel.
Down arrows indicate the increase of $\pi_{3/2,1/2}$-orbit neutrons,
and down-right arrows stand for the increase of the $\sigma_{1/2}$- and
$\pi^*_{3/2}$-orbit neutrons. 
The densities of intrinsic states are integrated with respect to
the $z$ axis and plotted on the $x$-$y$ plane in the unit of fm$^{-2}$. 
The axes of the intrinsic frame are chosen so as to be
$\langle x^2\rangle \ge \langle y^2\rangle \ge \langle z^2\rangle$.  
\label{fig:beiso-bh-dense}}
\end{center}
\end{figure}
%%%%%%%%%%%%%%%%%%%%%%%%%%%%%

%%%%%%%%%%%%%%%%%%%%%%%%%%%%%%
\begin{figure}[htb]
\begin{center}
\resizebox{0.55\textwidth}{!}{%
\includegraphics{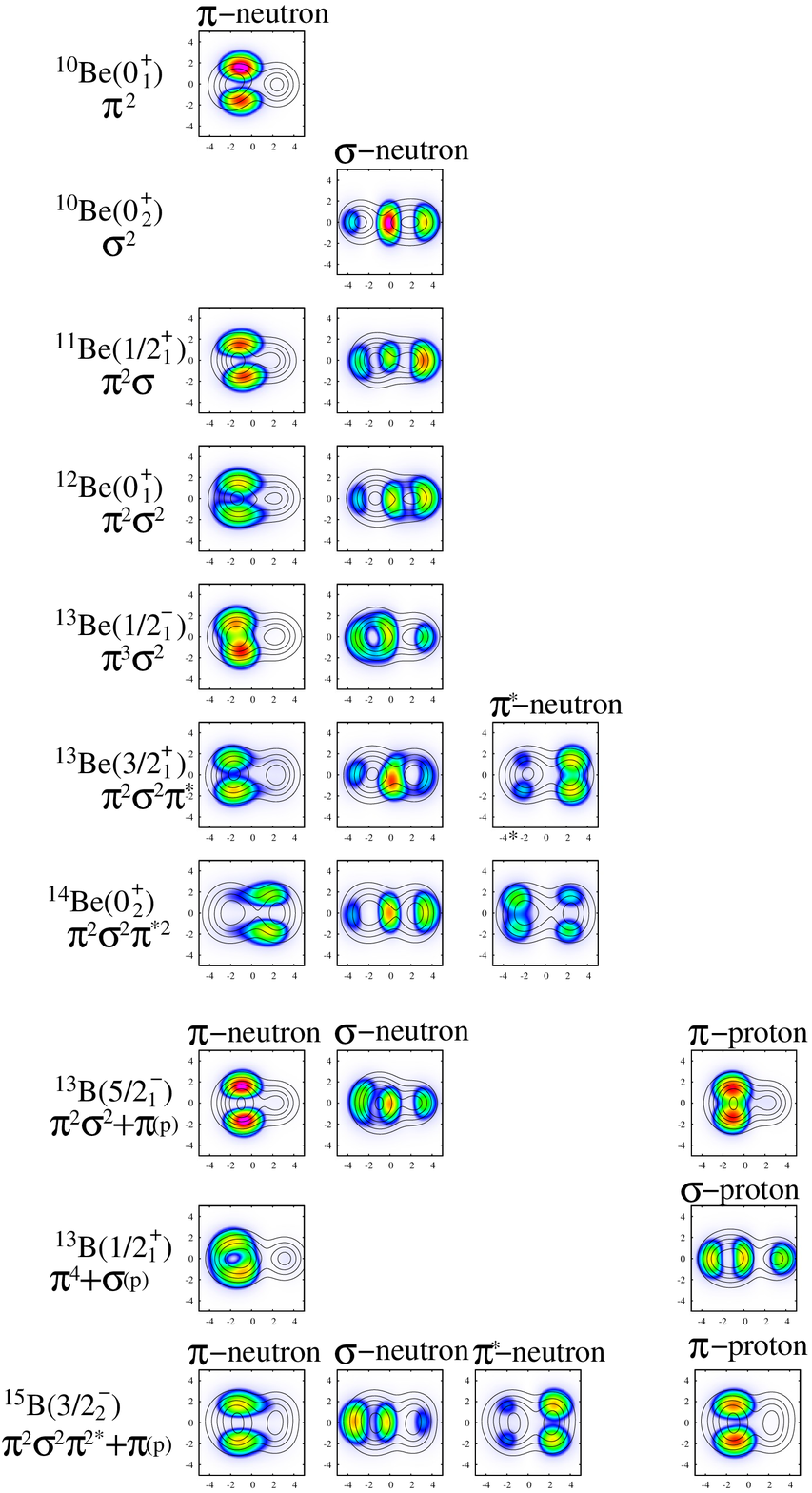}}	
%\includegraphics[width=15.0cm]{beiso-single.eps} 	
%\vspace{0.5cm}
  \caption{Single-particle wave functions of valence neutrons 
in the MO $\sigma$-bond structures in Be and B isotopes as well as those in the $^{10}$Be($0^+_1$)
obtained by the AMD+VAP using the MV1  ($m=0.65$, $b=h=0$) central and G3RS 
($u_1=-u_2=-3700$ MeV) spin-orbit interactions
\cite{KanadaEn'yo:1999ub,KanadaEn'yo:2002rh,KanadaEn'yo:2003ue,KanadaEn'yo:2002ay,KanadaEn'yo:2012rm,enyo-b13}.  
For $^{10}$Be, the results obtained using the MV1  ($m=0.62$, $b=h=0$) central 
and G3RS ($u_1=-u_2=-3000$ MeV) spin-orbit interactions are shown.
Single-particle wave functions in the band-head states are shown
(except for $^{13}$B($5/2^-_1$)). 
The $^{13}$B($5/2^-_1$) state is the band member of $K^\pi=3/2^-$ starting from the 
$3/2^-_2$ state. 
The densities of intrinsic states are integrated with respect to
the $z$ axis and plotted on the $x$-$y$ plane.
The single-particle densities of the molecular orbits are displayed in color maps.
The total matter densities are shown by contour lines.
\label{fig:beiso-single}}
\end{center}
\end{figure}
%%%%%%%%%%%%%%%%%%%%%%%%%%%%%

We summarize the molecular orbit configurations of the band-head states of Be isotopes
in Table \ref{tab:beiso}. The matter density distributions of the intrinsic states are shown in 
Fig.~\ref{fig:beiso-bh-dense}, and the single-particle densities 
are shown in Fig.~\ref{fig:beiso-single}.
Strictly speaking, realistic Be states do not 
have pure molecular orbit configurations but they are given by mixed configurations.
 As shown in Figs.~\ref{fig:beiso-bh-dense} and \ref{fig:beiso-single}, 
most of the intrinsic states of Be obtained by the AMD+VAP show reflection 
asymmetry indicating that  
single-particle wave functions somewhat deviate from the ideal 
molecular orbits, which should be reflection symmetric.
The asymmetry is caused by many-body correlations 
between valence neutrons. In the strong correlation limit in the asymptotic region, 
valence neutrons are distributed around 
either of 2 $\alpha$s and molecular orbit configurations change into atomic orbit configurations as discussed by Ito 
{\it et al.} \cite{Ito:2003px,Ito:2008zza,Ito2014-rev}. In a transient region,
the molecular orbit structures somewhat couple with two-body He+He di-cluster components.
The assignment of molecular orbit  configurations in Table \ref{tab:beiso} is based on 
dominant features such as 
nodal structures and percentages of positive- and negative-parity components of single-particle wave 
functions obtained by the AMD+VAP
calculations \cite{KanadaEn'yo:1999ub,KanadaEn'yo:2002ay,KanadaEn'yo:2002rh,KanadaEn'yo:2003ue,KanadaEn'yo:2012rm}.
As already described, 
$^{10}$Be($0^+_1$), $^{10}$Be($1^-$), and $^{10}$Be($0^+_2$) are understood as the 
$\pi_{3/2}^2$,  $\pi_{3/2}\sigma_{1/2}$, and $\sigma_{1/2}^2$, respectively.  
In a similar way,  
$^{11}$Be($1/2^+$),$^{11}$Be($1/2^-$), and $^{11}$Be($3/2^-_2$) states
are described by the $\pi_{3/2}^2\sigma_{1/2}$,  $\pi_{3/2}^2\pi_{1/2}$  and $\pi_{3/2}\sigma_{1/2}^2$ configurations,
and  $^{12}$Be($0^+_1$), $^{12}$Be($0^+_2$), and $^{12}$Be($1^-_1$) states
correspond to $\pi_{3/2}^2\sigma_{1/2}^2$,  $\pi_{3/2}^2\pi_{1/2}^2$  and $\pi_{3/2}^2\pi_{1/2}\sigma_{1/2}^1$, respectively.
These assignments are consistent with the results of molecular orbit models 
\cite{Oertzen-rev,Itagaki:1999vm} and those of 
cluster models \cite{Ito:2003px,Arai:2004yf,Ito:2005yy,Ito:2008zza,Ito2014-rev}.
For $^{13}$Be and $^{14}$Be, excited bands with developed cluster structures 
have been theoretically predicted by the AMD+VAP calculations  \cite{KanadaEn'yo:2002ay,KanadaEn'yo:2012rm}. The band-head states, 
$^{13}$Be($1/2^-$),  $^{13}$Be($3/2^+$), and $^{13}$Be($5/2^+$), are 
regarded as $\pi_{3/2}^2\pi_{1/2}\sigma_{1/2}^2$, $\pi_{3/2}^2\sigma_{1/2}^2\pi^{*}_{3/2}$
and $\pi^2_{3/2}\pi^2_{1/2}\sigma_{1/2}$. And $^{14}$Be($0^+_1$), $^{14}$Be$(2^-)$, and
$^{14}$Be($0^+_2$) are described by the  $\pi_{3/2}^2\pi_{1/2}^2\sigma_{1/2}^2$,
$\pi_{3/2}^2\pi_{1/2}\sigma_{1/2}^2\pi^{*}_{3/2}$, and $\pi_{3/2}^2\sigma_{1/2}^2\pi^{*2}_{3/2}$
configurations, respectively.  Experimental data are not enough to assign band structures of
$^{13}$Be and $^{14}$Be. 

To see how the cluster structure is enhanced (suppressed) by 
$\sigma$($\pi$)-orbit neutrons, we show density distributions of the intrinsic states in 
Fig.~\ref{fig:beiso-bh-dense}. The numbers ($n_\pi,n_\sigma,n_{\pi *}$) of $\pi$-, $\sigma$-, $\pi^*$-orbit
neutrons are shown in the figure corresponding to the configurations listed in Table \ref{tab:beiso}.
As clearly seen, $\sigma$-orbit and $\pi^*$-orbit neutrons enhance the 
cluster structure, whereas $\pi$-orbit neutrons suppress the cluster structure:
the cluster structure is enhanced 
by $n_\sigma$ increasing and enhanced further by $n_{\pi}^*$ 
increasing, whereas it is suppressed as $n_\pi$ increases.

Finally, we discuss energy levels of the band-head states labeled by the molecular orbit  configurations.
Figure \ref{fig:beiso-orbital} shows experimental excitation energies of $^{10-12}$Be and
theoretical values of $^{13}$Be and $^{14}$Be.
In the spherical shell model limit, $0\hbar\omega$ configurations  
are the lowest, whereas
$1\hbar\omega$ and $2\hbar\omega$ are excited configurations.
The inversion between $0\hbar\omega$, $1\hbar\omega$, and $2\hbar\omega$ occurs in $^{11-13}$Be, and 
the energy spectra seem to be out of the normal ordering of the spherical shell model configurations.
However, in the molecular orbit  configurations in the developed cluster structure, the $\sigma_{1/2}$- and 
$\pi_{1/2}$-orbits almost degenerate (see Fig.~\ref{fig:beiso-orbital}(d)) and they compose
the major shell on the top of the $N=6$ shell gap. It means that 
$^{10}$Be is regarded as a closed MO-shell ($\pi_{3/2}$-orbit) nucleus, whereas, 
$^{11-13}$Be are open MO-shell ($\pi_{1/2}$- and $\sigma_{1/2}$-orbits)
nuclei, in which 
the neutron Fermi level exists at the major $\sigma_{1/2}$-$\pi_{1/2}$ shell.
In the molecular orbit  picture, 
$^{10}$Be$(0^+_1)$, $^{11}$Be$(1/2^+)$, $^{11}$Be$(1/2^-)$, $^{12}$Be$(0^+_1)$, 
$^{12}$Be$(0^+_2)$, $^{12}$Be$(1^-) $, $^{13}$Be$(1/2^-)$,  and $^{14}$Be$(0^+_1)$ 
are ``normal'' states described by major MO-shell configurations, 
and therefore, it is not surprising that they appear in low-energy regions. 
On the other hand, 
$\pi_{3/2}$-orbit holes and $\pi^*_{3/2}$-orbit particles are interpreted as
particle-hole excited states in the molecular orbit  configurations. 
For instance, 
$^{10}$Be$(1^-)$, $^{11}$Be$(3/2^-)$, $^{14}$Be$(2^-)$ and $^{14}$Be$(0^+_2)$ have
excited molecular orbit  configurations and they exist in 
relatively higher energy regions ($E_x=4\sim 6$ MeV).
The $^{13}$Be($3/2^+$) state is the exception that the excited molecular orbit  configuration
is obtained in the low-energy region, probably, because of many-body correlations.

For $^{13}$Be, which is an unbound nucleus, 
experimental spectra and spin-parity assignments 
in the low-energy region are still controversial \cite{korsheninnikov95,Kondo:2010zza,ostrowski92,belozerov98,Thoennessen:2001yj,Lecouey:2003tg,simon04,simon07,Falou:2010if,Fortune:2010zz,Randisi:2013urw}. An experimental report of a low-lying resonance with the abnormal spin-parity $J^\pi=1/2^-$
supports  the breaking of the $N=8$ magic number \cite{Kondo:2010zza}.

\subsubsection{Analogous molecular orbit  states in B isotopes}
In excited states of B isotopes, 
MO $\sigma$-bond structures analogous to those of Be isotopes have been predicted.
In B isotopes, the MO $\sigma$-bond structures are not so favorable 
as Be isotopes because of an additional proton, and therefore, they appear in excited states.  For instance, AMD+VAP calculations predicted $K^\pi=3/2^-$ bands in highly excited states of $^{13}$B and $^{15}$B  \cite{KanadaEn'yo:2002ay,enyo-b13}. 
These states are dominated by $\pi^2\sigma^2$ and $\pi^2\sigma^2\pi^{*2}$ neutron 
configurations, respectively, and  regarded
as analogous states of $^{12}$Be($0^+_1$) and $^{14}$Be($0^+_2$) with an additional proton in the $\pi$-orbit.  
Moreover, $^{13}$B($1/2^+$) with an exotic cluster structure 
having a proton in the $\sigma$-orbit has been predicted.
It is surprising that 
the molecular $\sigma$-orbit appears also in the proton configuration despite that protons are 
rather deeply bound in the neutron-rich systems. $^{13}$B($1/2^+$) has a remarkably 
large deformation with $^8$He and $\alpha$ clusters bonded by the $\sigma$-orbit proton, 
and it constructs the $K^\pi=1/2^+$ rotational band with a large moment of inertia. 
Intrinsic structures of these excited states
of B isotopes are shown in Fig.~\ref{fig:beiso-single}. In the figure, 
the density distributions of single-particle wave functions are illustrated together with the total matter density.
Also for $^{11}$B, cluster structures in excited states can be understood by the 
molecular orbit picture
as discussed in Ref.~\cite{Suhara:2012zr}.

%It is concluded that 
%the lowering mechanism of the $\sigma$-orbit derives the enhancement of the
%cluster structure resulting in the breaking of the neutron magic number 
%in $^{11}$Be, $^{12}$Be, and $^{13}$Be.

\subsection{Molecular orbitals in O, Ne and F isotopes}\label{sec:4.2}
The success of the molecular orbit in Be isotopes strongly motivated the extention of this
concept to other nuclei. One of the motivation of the extension is the universality of the
molecular orbits. It is of interest and importance to investigate if the  molecular orbit 
appears in other nuclei universally.

O, F and Ne isotopes are the good candidates of the molecular orbit, because the clustering of
their stable nuclei $^{16}{\rm O}$, $^{19}{\rm F}$ and $^{20}{\rm Ne}$ are well known
\cite{Fujiwara-supp,Furutani:1980a,Horiuchi:68a,Dufour:1994zz,Roth:1960a,Arima:1967tck,Suzuki:1976zz}. 
Different from Be isotopes, these isotopes can have the parity asymmetric cores, 
$\alpha+{}^{12}{\rm C}$, $\alpha+{}^{15}{\rm N}$ and $\alpha+{}^{16}{\rm O}$,
respectively. Therefore, they should have new aspects caused by this asymmetry. In addition to
this, these isotopes have much longer isotope chain than Be isotopes, and hence, have more
valence neutrons. Here, after a short introduction of the clustering in stable 
$^{16}{\rm O}$, $^{19}{\rm F}$ and $^{20}{\rm Ne}$, we summarize the AMD studies for neutron-rich
O, Ne and F isotopes.

\subsubsection{$\alpha$ clustering in $^{16}$O, $^{20}$Ne and $^{19}$F}
For a better understanding of the clustering in O, Ne and F isotopes, it is useful to review the 
clustering of their stable isotopes $^{16}{\rm O}$, $^{20}{\rm Ne}$ and $^{19}{\rm F}$. 

\begin{figure}[h] 
 \begin{center}
  \resizebox{0.6\textwidth}{!}{
  \includegraphics{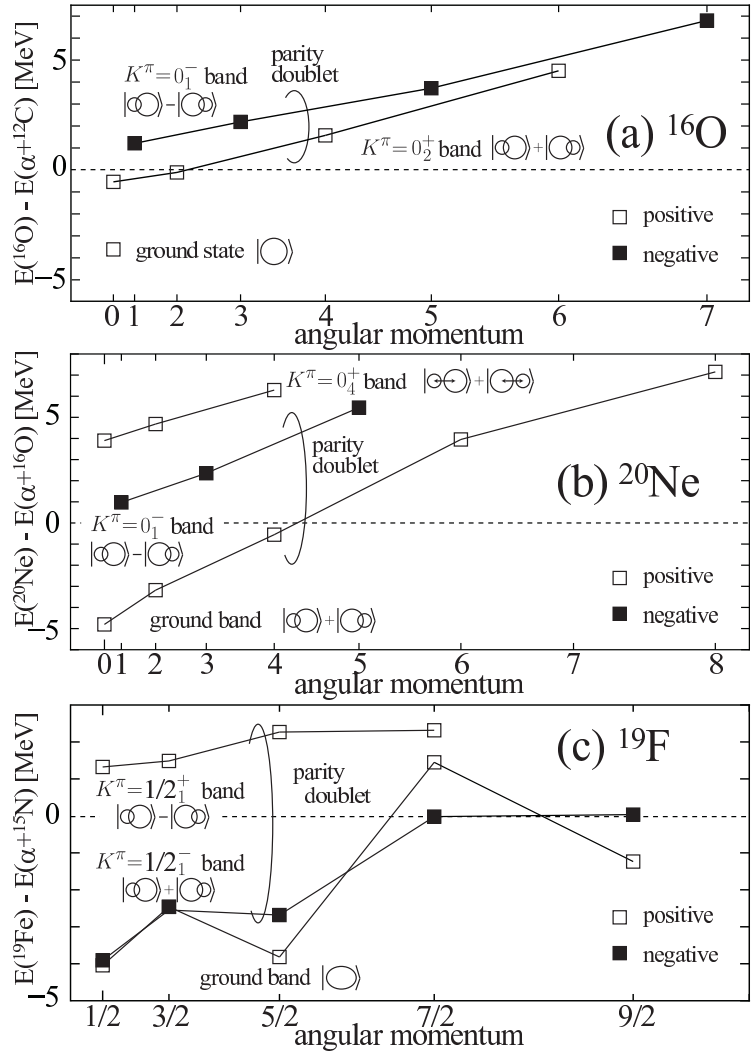}}
  \caption{Observed cluster bands in $^{16}{\rm O}$, $^{20}{\rm Ne}$ and $^{19}{\rm F}$.
  Experimental data are taken from Ref. \cite{Tilley:1993zz,Tilley:1995zz,Tilley:1998wli}.}  
  \label{fig:ex1}       % Give a unique label
 \end{center}
\end{figure}

$^{16}{\rm O}$ and $^{20}{\rm Ne}$ are famous for their asymmetric cluster structures of
$\alpha+{}^{12}{\rm C}$ and $\alpha+{}^{16}{\rm O}$. Because of the parity
asymmetric  configuration, they must constitute the parity doublet (a pair of the positive- and
negative-parity bands) \cite{Horiuchi:68a}.  Many theoretical and experimental studies have been
devoted to study the clustering in $^{16}{\rm O}$ and $^{20}{\rm Ne}$, and the cluster band
assignment is established well as  summarized in Fig. \ref{fig:ex1}
\cite{Tilley:1993zz,Tilley:1995zz,Tilley:1998wli}.  
The ground state of $^{16}{\rm O}$ has the doubly closed shell configuration, but the $0^+_2$
and $1^-_2$ states at 6.0 and 9.6 MeV have prominent $\alpha+{}^{12}{\rm C}(0^+_1)$ cluster
structure. Two rotational bands with positive- and negative-parity  ($K^\pi=0^+_2$ and
$0^-_1$ bands) are built on these states and constitute the parity doublet. In the case of
$^{20}{\rm Ne}$, the ground band and the $K^\pi=0^-_1$ band built on the $1^-$ state at 5.8 MeV
are regarded as the parity doublet. In addition to them, another cluster band
($K^\pi=0^+_4$ band), in which the relative motion between clusters is excited by $2\hbar\omega$,
is built on the $0^+_4$ state around 8.7 MeV.  

\begin{figure}[h] 
 \begin{center}
  \resizebox{0.6\textwidth}{!}{
  \includegraphics{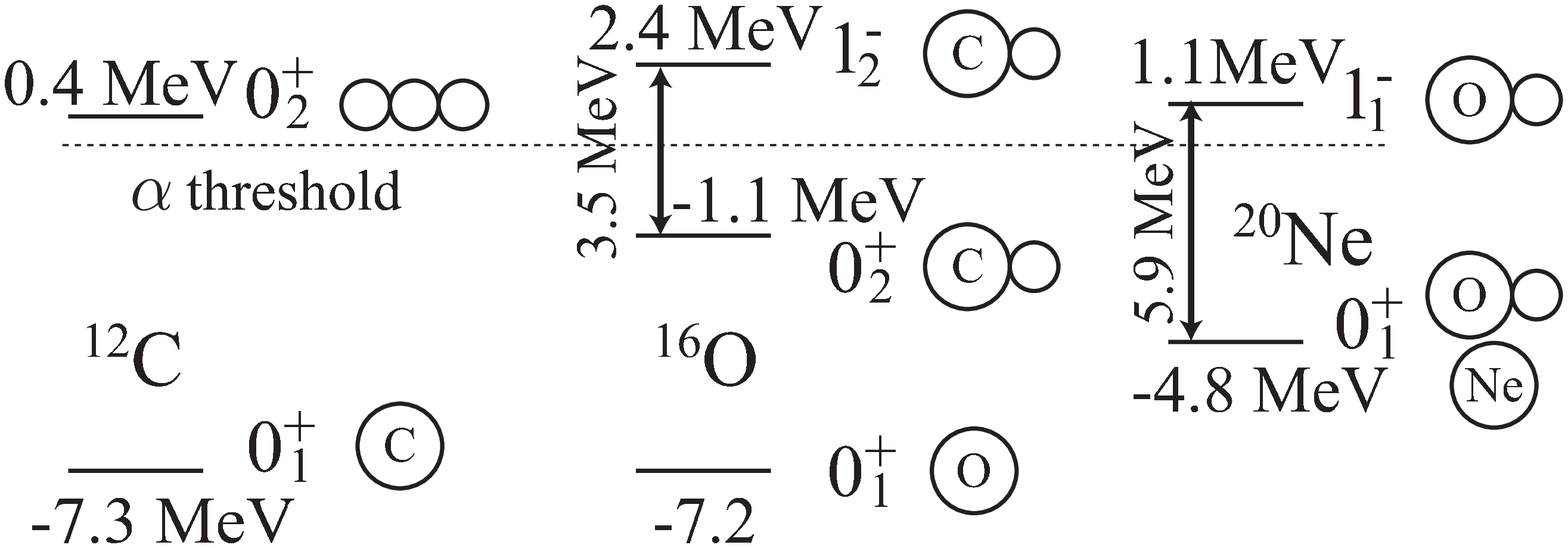}}
  \caption{Systematics of $\alpha$ cluster states in $^{12}{\rm C}$, $^{16}{\rm O}$ and
  $^{20}{\rm  Ne}$. Numbers in the figure show the energy measured from the $\alpha$
  threshold. }   
  \label{fig:ex2}       % Give a unique label
 \end{center}
\end{figure}

Here, let us note the difference between $^{20}{\rm Ne}$ and lighter nuclei such as $^{12}{\rm C}$
and $^{16}{\rm O}$. To understand the difference, the energies of the $0^+_2$ state of 
$^{12}{\rm C}$ ($3\alpha$ cluster state),  the parity doublets of $^{16}{\rm O}$ and 
$^{20}{\rm Ne}$ and the ground states of those nuclei are summarized in Fig. \ref{fig:ex2}. 
In this figure, since the ground state of $^{20}{\rm Ne}$ is assigned as the cluster state, one 
sees that the $0^+$ state of $^{20}{\rm Ne}$ having shell structure illustrated in
Fig. \ref{fig:int2} looks missing in $^{20}{\rm Ne}$, while it exists as the ground states in
$^{12}{\rm C}$ and $^{16}{\rm O}$ (compare Fig. \ref{fig:ex2} with Fig. \ref{fig:int2}). This
suggests the transient nature of  the ground state of $^{20}{\rm Ne}$. Namely, the ground state of
$^{20}{\rm Ne}$ is not a pure $\alpha$ cluster state but a mixture of the shell and $\alpha$
cluster configurations. This transient nature is confirmed in another way. The energy of the
$\alpha$ cluster states of $^{12}{\rm C}$ ($+0.4$ MeV) and $^{16}{\rm O}$ ($-1.1$ MeV) are close
to the $\alpha$ threshold energy, but the ground state of $^{20}{\rm Ne}$ ($-4.8$ MeV) is much
deeper than the $\alpha$ threshold indicating stronger attraction between $\alpha$ and 
$^{16}{\rm O}$ clusters. This stronger attraction induces the distortion of clusters 
leading to the mixing between cluster and shell configurations. Yet another evidence of the
transient nature is the energy splitting of the parity doublets. As explained in
Ref. \cite{Horiuchi:68a}, the energy splitting becomes smaller for enhanced cluster state, while it is
enlarged for modest cluster state. It is clear that the splitting is enlarged in $^{20}{\rm Ne}$
(5.9 MeV) compared with that in  $^{16}{\rm O}$ (3.5 MeV) indicating the reduction of clustering
in the $^{20}{\rm Ne}$ ground state. The highlight of the discussion on $^{22}{\rm Ne}$ made in
the following section is how this transient nature of the $\alpha+{}^{16}{\rm O}$ clustering is 
affected and modified by the excess neutrons. It will alter the systematics of the $\alpha$
clustering and change the energy splitting  of the parity doublet. 

\begin{figure}[h] 
 \begin{center}
  \resizebox{0.6\textwidth}{!}{
  \includegraphics{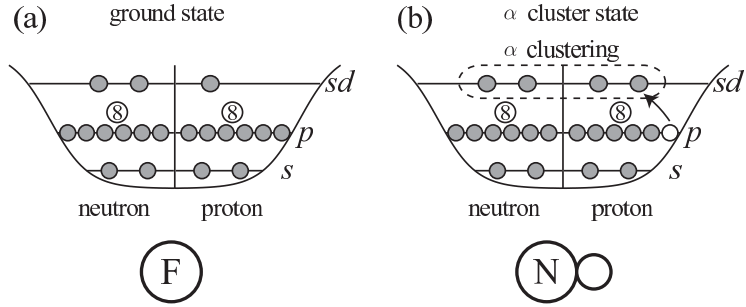}}
  \caption{Particle-hole configurations in the ground and the first excited states of 
  $^{19}{\rm  F}$.}   \label{fig:ex3}       % Give a unique label
 \end{center}
\end{figure}

$^{19}{\rm F}$ can be regarded as the nucleus in which a proton hole is coupled to 
$^{20}{\rm Ne}$. In the ground state, this proton hole occupies the $sd$-shell, {\it i.e.} three
nucleons occupy $sd$-shell on top of $^{16}{\rm O}$ core as illustrated in Fig. \ref{fig:ex3} (a). 
Since the valence nucleons cannot form $\alpha$ particle, the cluster correlation in the ground
state is rather weak and the shell model gives reasonable description \cite{Arima:1967tck},
although the contribution from the $t+{}^{16}{\rm O}$ is not negligible.  If a
proton hole occupies $p$-shell as illustrated in Fig. \ref{fig:ex3} (b), in other words, a proton is
excited from $p$- to $sd$-shell across $Z=8$ shell gap, valence nucleons in $sd$-shell form an
$\alpha$ particle to generate the $\alpha+{}^{15}{\rm N}$ cluster state
\cite{Furutani:1980a}. Since the energy gain owing to the $\alpha$ clustering is large enough to
compensate the energy loss due to the 
proton excitation, the energy of this proton hole configuration is comparable with the ground
state. As a result, the excitation energy of the first excited state having this configuration
($1/2^-_1$ state) is as small as 110 keV. The $\alpha$ cluster bands in $^{19}{\rm F}$ have been
also studied in detail and assigned as shown in the lower panel of Fig. \ref{fig:ex1}
\cite{Tilley:1995zz}. Thus, it is reminded that a proton excitation across $Z=8$ shell gap
triggers the clustering in the case of F isotopes.   

\subsubsection{$\alpha$ clustering in $^{22}$Ne}
\begin{figure}[h] 
 \begin{center}
  \resizebox{0.6\textwidth}{!}{
  \includegraphics{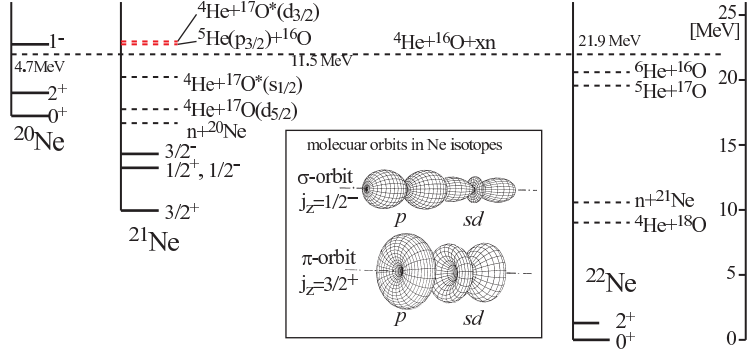}}
  \caption{Solid lines show the ground and excited states of $^{20}{\rm Ne}$,   $^{21}{\rm Ne}$
  and $^{22}{\rm Ne}$, while dashed lines show threshold energies. Figure in the inset shows the
  molecular orbits that are composed from the $p_{3/2}$ orbit around $^{4}{\rm He}$ and the
  $d_{3/2}$ orbit around $^{16}{\rm O}$. This figure is reconstructed from Ref. \cite{oertzen-ne}.}   
  \label{fig:ex4}       % Give a unique label
 \end{center}
\end{figure}

One of the earliest extention of the molecular-orbit picture to the nuclei other than Be
isotopes was suggested for Ne isotopes by von Oertzen \cite{oertzen-ne,Oertzen-rev}. His idea is
based on the threshold energies in Ne isotopes shown in Fig. \ref{fig:ex4}.  
In $^{21}{\rm Ne}$, one sees that the $^{5}{\rm He^*(3/2^-)}+{}^{16}{\rm O}$ and 
$^{4}{\rm He}+{}^{17}{\rm O^*(3/2^+)}$ thresholds energies are almost degenerated at approximately
1 MeV above the $^{4}{\rm He}+{}^{16}{\rm O}+n$ threshold. This implies that the valence neutron
occupying $sd$-orbit around $^{16}{\rm O}$ cluster and that occupying $p$-orbit around 
$^{4}{\rm He}$ cluster can be shared by these two clusters forming the molecular orbit. One also
sees similar near degeneracy of the $^{5}{\rm He}+{}^{17}{\rm O}$ and 
$^{4}{\rm He}+{}^{18}{\rm O}$ thresholds in $^{22}{\rm Ne}$. Note that this matching of the
threshold energy is inevitable for symmetric clusters such as Be isotopes, but not for the
asymmetric clusters. In an analogous way to Be isotopes, the parallel and transverse alignment
of these two orbits generate $\sigma$- and $\pi$-orbits as illustrated in the inset of
Fig. \ref{fig:ex4}. By considering the combinations of these molecular orbits, he suggested
a possible assignment of cluster bands in $^{21}{\rm Ne}$ and $^{22}{\rm Ne}$
\cite{oertzen-ne,Wheldon:2005a}.  

\begin{figure}[h] 
 \begin{center}
  \resizebox{0.6\textwidth}{!}{
  \includegraphics{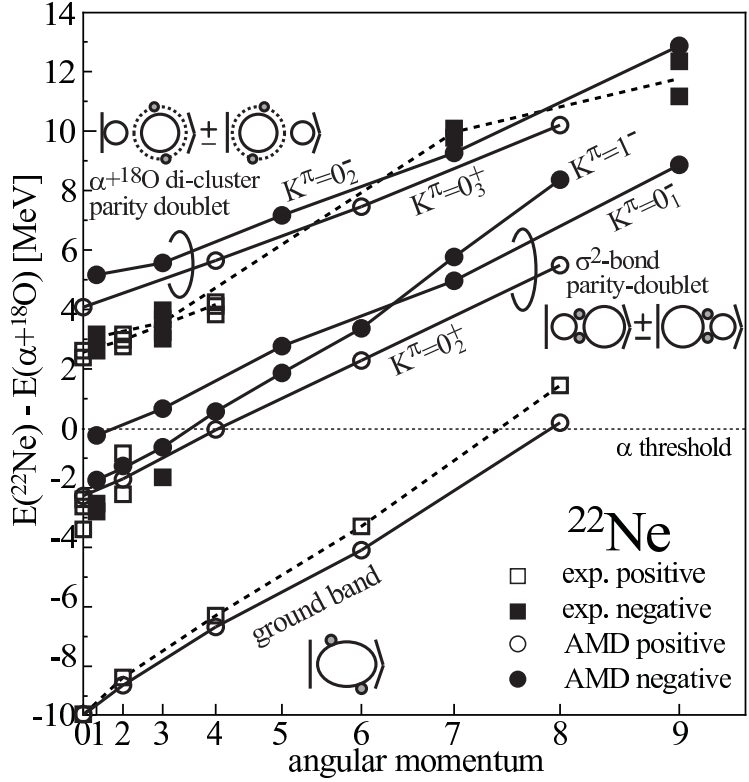}}
  \caption{Calculated and observed cluster states in $^{22}{\rm Ne}$. Boxes show the
  experimental data
  \cite{Scholz:1969a,Scholz:1972zz,Rogachev:2001ti,Curtis:2002mg,Goldberg:2004yk}, 
  while the circles show the AMD result. This figure is reconstructed from the data presented in
  Ref. \cite{Kimura:2007kz}.}   
  \label{fig:ex5}       % Give a unique label
 \end{center}
\end{figure}

To test whether the suggested molecular orbits are really formed in Ne isotopes and to identify
the cluster bands, AMD calculation was performed for $^{22}{\rm Ne}$ \cite{Kimura:2007kz} using
the Gogny D1S effective interaction. The band assignment by the AMD calculation is summarized in  
Fig. \ref{fig:ex5} together with the observed candidates of the cluster bands. The result is
summarized as follows. (1) In the ground band, the valence neutrons reduces the 
$\alpha+{}^{16}{\rm O}$ clustering and make the structure more like shell model state.  (2)
On the other hand, in the excited state, valence neutrons occupy the $\sigma$-orbit and enhance
the $\alpha+{}^{16}{\rm O}$ clustering. Three rotational bands around 
the $\alpha+{}^{18}{\rm O}$ threshold are regarded as the cluster bands with covalent bonding, 
which we denote by $\sigma$- and $\sigma^2$-bond bands. (3) Another group of cluster bands appears
at approximately 5 MeV above the  $\alpha+{}^{18}{\rm O}$ threshold. In those bands the valence
neutrons orbit only around $^{16}{\rm O}$, and hence, regarded as the ordinary two cluster system
having  $\alpha+{}^{18}{\rm O}$ di-cluster configuration. We denote them by di-cluster
bands. Because of the parity asymmetric intrinsic structure,  the $\sigma$-, $\sigma^2$-bond bands
and di-cluster bands constitute the parity doublet. 

\begin{figure}[h] 
 \begin{center}
  \resizebox{0.6\textwidth}{!}{
  \includegraphics{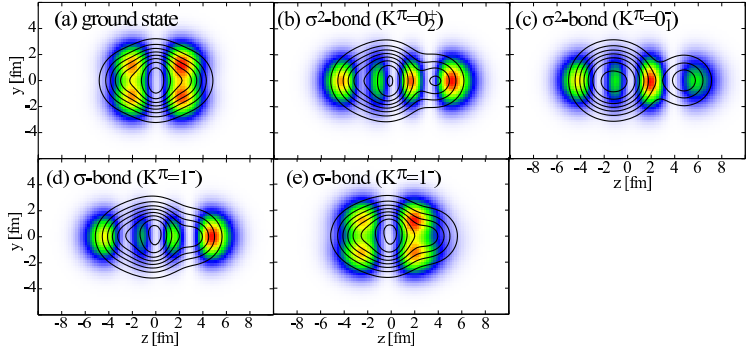}}
  \caption{Intrinsic density distributions of $^{22}{\rm Ne}$ for the band head states of (a) the
  ground band, (b) $K^\pi=0^+_2$ $\sigma^2$-bond band, (c) $K^\pi=0^-_1$  $\sigma^2$-bond band, and
  (d)(e) $K^\pi=1^-$ $\sigma$-bond band. Solid lines show the density distribution of the 
  $^{20}{\rm  Ne}$ core, while the color plots show that of valence neutrons.
  This figure is reconstructed from the data presented in Ref. \cite{Kimura:2007kz}.}   
  \label{fig:ex6}       % Give a unique label
 \end{center}
\end{figure}

The structure of the ground band and $\sigma$-, $\sigma^2$-bands become clear by looking at the
core and valence neutron densities shown in Fig. \ref{fig:ex6}.
Here, we define the most weakly bound two neutrons as valence neutrons and the remaining 20
nucleons as the $^{20}{\rm Ne}$ core which are plotted by color plot and contour lines,
respectively. In the ground state, the valence neutrons occupy the orbit which have a node along
the symmetry axis (Fig. \ref{fig:ex6} (a)). This orbit has $sd$-shell nature and may correspond to
the $\pi$-orbit illustrated in Fig. \ref{fig:ex4}. However, the system has the  parity symmetric
shape and the $\alpha$ clustering of $^{20}{\rm Ne}$ core is lost indicating the deformed
mean-field nature of the ground state. Indeed, the calculated properties of this valence neutron
orbit listed in Tab. \ref{tab:ex1} is in good accordance with the Nilsson orbit with the
asymptotic quantum number of $[Nn_zm_l\Omega^\pi] = [211\ 3/2^+]$. Here, the amount of the
positive-parity component $p^+$ and the angular momenta of the valence neutron orbit
$\phi_i$ are defined 
as
\begin{align}
 p^+ &= |\langle \phi_i|\frac{1+P_x}{2}| \phi_i\rangle|^2,
 \label{eq:sp1}\\ 
 j(j+1)&= \langle \phi_i|\hat{j}^2| \phi_i\rangle, \quad
 \Omega = \sqrt{\langle \phi_i|\hat{j}_z^2| \phi_i\rangle},\label{eq:sp2}\\
 l(l+1)&= \langle \phi_i|\hat{l}^2| \phi_i\rangle, \quad
 m_l = \sqrt{\langle \phi_i|\hat{l}_z^2| \phi_i\rangle}.\label{eq:sp3}
\end{align}
In other words, when the valence neutrons occupy $\pi$-orbit, they diminish the $\alpha$
clustering of the core and the system exhibit the deformed mean-field nature. Because of this
symmetric shape, we have not obtained the negative-parity partner of the ground band that
constitutes the parity doublet.  As already discussed in the section \ref{sec:3.2}, the reduction
of $\alpha$ clustering in the ground state can also be observed as the reduction of the $B(E2)$
and the proton radius. 

\begin{table}[h]
\begin{center}
\caption{Properties of the valence neutron orbits shown in Fig. \ref{fig:ex6} (a)-(e). The
 single-particle energy $\varepsilon$ is given in MeV, and other  quantities are defined by
 Eq. (\ref{eq:sp1})-(\ref{eq:sp3}). The Nilsson asymptotic quantum numbers   $[Nn_zm_l\Omega^\pi]$
 deduced from these properties are also given. The data is taken from Ref. \cite{Kimura:2007kz}.} 
\label{tab:ex1}       % Give a unique label
% For LaTeX tables use
\begin{tabular}{lccccccc}
\hline\noalign{\smallskip}
orbit & $\varepsilon$ & $p^+$ & $j$ & $l$  & $m_l$ & $\Omega$ & $[Nn_zm_l\Omega^\pi]$\\
 \noalign{\smallskip}\hline\noalign{\smallskip}
(a) & $-8.9$ & 0.94 & 2.5 & 2.0  & 1.1  & 1.5  & $[211\ 3/2^+]$\\
(b) & $-5.6$ & 0.36 & 2.9 & 2.4  & 0.2  & 0.5  & \\
(c) & $-6.5$ & 0.49 & 2.8 & 2.3  & 0.1  & 0.5  & \\
(d) & $-3.4$ & 0.21 & 2.8 & 2.4  & 0.2  & 0.5  & \\
(e) & $-7.9$ & 0.90 & 2.3 & 2.1  & 1.0  & 1.6  & $[211\ 3/2^+]$\\
\noalign{\smallskip}\hline
\end{tabular}
\end{center}
\end{table}

When the valence neutrons are excited, the structure is drastically changed. In the
$\sigma^2$-bond bands with positive parity (Fig. \ref{fig:ex6} (b)), two valence neutrons occupy 
the orbit having three nodes along the symmetry axis which exhibits $pf$-shell nature. This
valence neutron orbit 
induces strong $\alpha$ clustering as clearly observed in the core density shown in
Fig. \ref{fig:ex6} (b). Interestingly, because of the strong asymmetry of the core, this valence
neutron orbit is a mixture of the positive- and negative-parity components (Tab. \ref{tab:ex1}) and
cannot be interpreted as a single Nilsson orbit. It is noteworthy that the molecular-orbit picture 
($\sigma$-orbit shown in Fig. \ref{fig:ex4}) \cite{oertzen-ne} gives a natural and reasonable
interpretation of this orbit. Because of the parity asymmetry, the negative-parity 
states ($K^\pi=0^-_1$ band) shown in Fig. \ref{fig:ex6} (c) accompany with the positive-parity
states to constitute the $\sigma^2$-bond parity doublet. It is interesting to note that the
valence neutron densitiy shown in Fig. \ref{fig:ex6} (c) indicates that the probability is
largest in between $^{4}{\rm He}$ and $^{16}{\rm O}$ clusters to bond them, namely
covalency. Thus, we are able to conclude that the valence neutrons in $\pi$-orbit ($sd$-shell)
diminish the  clustering, while those in $\sigma$-orbit induce it. This is confirmed from the
structure of the $K^\pi=1^-$ band with $\sigma$-bond shown in Fig. \ref{fig:ex6} (d) and (e) in 
which a valence neutron occupies the $\pi$-orbit and the other occupies $\sigma$-orbit. A
valence neutron in $\sigma$-orbit induces the moderate clustering of the core but not as prominent
as the $\sigma^2$-bond ($K^\pi=0^\pm$) bands because another valence  neutron in $\pi$-orbit
reduces it. Above these molecular-orbit bands, we also obtained the atomic-orbit bands
($K^\pi=0^+_3$ and $0^-_2$) as similar to the $\alpha$+$^{8}{\rm He}$ di-cluster states in
$^{12}{\rm Be}$. In these bands, two valence neutrons orbit only around the $^{16}{\rm O}$ cluster
analogous to the ionic bonding of  atomic molecule.

We also note the energy splitting of the parity doublets (the energy difference between the
positive- and netgative-parity states which constitute the doublet).
In $^{22}{\rm Ne}$, the splitting of the $\sigma^2$-bond bands is approximately 2 MeV and that of
the $\alpha+{}^{18}{\rm O}$ di-cluster bands is appriximately 1 MeV, which are much smaller than
the 6 MeV splitting in $^{20}{\rm Ne}$. The small splitting of the $\alpha+{}^{18}{\rm O}$
di-cluster bands is due to their much enhanced clustering compared to the $^{20}{\rm Ne}$
doublet. In the case of the $\sigma^2$-bond bands, the valence neutrons in $\sigma$-orbit prevent
the reduction of the inter-cluster distance between $\alpha$ and $^{16}{\rm O}$ clusters, because
it enlarges the kinetic energy of the valence neutrons. As a result, the splitting of the
$\sigma^2$-bond bands is also kept small. 

As shown in Fig \ref{fig:ex5}, the candidates of these cluster bands are reported by the
experiments 
\cite{Scholz:1969a,Scholz:1972zz,Rogachev:2001ti,Curtis:2002mg,Goldberg:2004yk,Neogy:1972zz,Torilov:2011a}. 
Several states below the $\alpha$ threshold energy are reported as the
candidates of the $\alpha$ cluster states by the $\alpha$ transfer reaction on
$^{18}$O  \cite{Scholz:1969a,Scholz:1972zz}. 
They have tens times larger $\alpha$ reduced width amplitude 
($\alpha$RWA) than the ground state. Since the $\alpha$RWAs of the molecular-orbit states with
$\sigma$- and $\sigma^2$-bond by AMD show the qualitative agreement with the observation, the
observed $\alpha$ cluster states should correspond to the molecular-orbit states with $\sigma$-
and $\sigma^2$-bond discussed above. They will also have large $^{6}$He 
reduced width amplitude to be confirmed by the experiment, because of covalent nature of 
the $\sigma$ orbit. Above these $\sigma$-bond bands, the di-cluster bands are confirmed by
the $\alpha$ resonant scattering. AMD results and also other theoretical calculations 
\cite{Descouvemont:1988zz,Dufour:2003kgu,Dufour:2004xei} show  good agreement with the observed
energies, moment of inertia and $\alpha$RWA
\cite{Rogachev:2001ti,Curtis:2002mg,Goldberg:2004yk,Torilov:2011a}.  

\subsubsection{$\alpha$ clustering in O isotopes}
\begin{figure}[h] 
 \begin{center}
  \resizebox{0.6\textwidth}{!}{
  \includegraphics{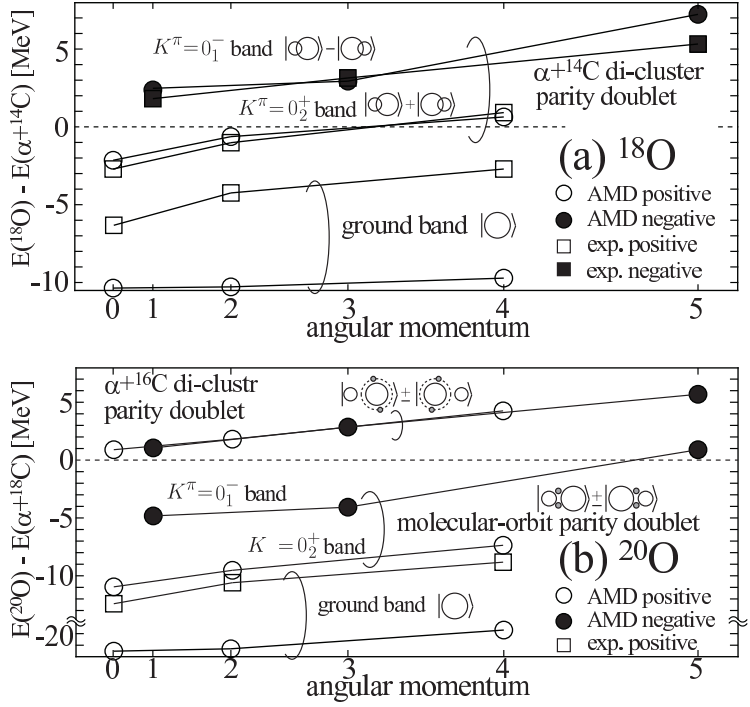}}
  \caption{Calculated and observed cluster states in $^{18}{\rm O}$ and $^{20}{\rm O}$. Boxes show
  the experimental data, while the circles show the AMD result. Energy is measured from the
  $\alpha$ threshold. This figure is reconstructed from the data presented in
  Ref. \cite{Furutachi:2007vz}.} 
  \label{fig:ex7}       % Give a unique label
 \end{center}
\end{figure}

\begin{figure}[h] 
 \begin{center}
  \resizebox{0.60\textwidth}{!}{
  \includegraphics{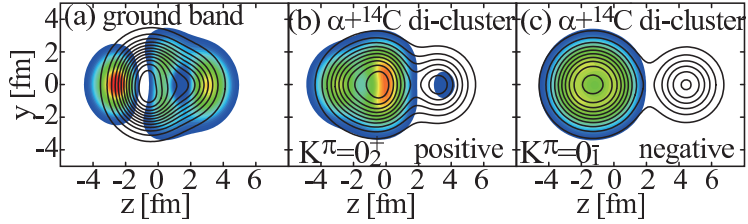}}
  \caption{Intrinsic density distributions of $^{18}{\rm O}$ for the band head states of (a) the
  ground band, (b) the $K^\pi=0^+_2$ band with $\alpha+{}^{14}{\rm C}$ clustering and (c) the
  $K^\pi=0^-_1$ band that is paired with the $K^\pi=0^+_2$ band.
  Solid lines show the density distribution of the  $^{16}{\rm O}$ core, while the color plots
  show that of valence neutrons. This figure is reconstructed from the data presented in
  Ref. \cite{Furutachi:2007vz}.}    
  \label{fig:ex8}       % Give a unique label
 \end{center}
\end{figure}

The structure of neutron-rich O isotope gives us a good insight to the relationship between
the molecular-orbit and atomic-orbit bands. Figure \ref{fig:ex7} summarizes the assignment of the
cluster bands in $^{18}{\rm O}$ and $^{20}{\rm O}$ obtained by the AMD study \cite{Furutachi:2007vz}. 
In $^{18}{\rm O}$, a pair of the positive- and negative-parity cluster bands are built on the
$0^+_2$ and $1^-_1$ states 2.2 MeV below and 2.4 MeV above the $\alpha$ threshold,
respectively. The intrinsic density distribution of these band-head states are shown in
Fig. \ref{fig:ex8} (b) and (c). 
Compared to the ground state (Fig. \ref{fig:ex8} (a)), it is clear that those states have
pronounced $\alpha$ cluster structure, and positive- and negative-parity states have similar
intrinsic structure to constitute the parity doublet. It must be noted that the valence
neutrons are well confined around $^{12}{\rm C}$ cluster differently from the $\sigma$-orbit 
of $^{22}{\rm Ne}$ shown in Fig. \ref{fig:ex6} (c). In other words, these states are regarded as
the atomic orbit state or an ordinary di-cluster state with $\alpha$+$^{14}{\rm C}$
configuration. In Ref. \cite{Furutachi:2007vz}, it was also shown that energy of the
molecular-orbit-like configuration is higher than  the $\alpha$+$^{14}{\rm C}$ configuration and
located at approximately 10 MeV above the $\alpha$ threshold. Thus, the order of the
molecular-orbit and atomic-orbit bands are inverted in $^{18}{\rm O}$. The origin of this
inversion is attributed to the shell effect of $^{14}{\rm C}$ cluster. Because of the magicity of
the neutron number $N=8$, the last two neutrons in $^{14}{\rm C}$ are tightly bound and are hardly
picked out from  $^{14}{\rm C}$ cluster. As a result, the $\alpha$+$^{14}{\rm C}$ cluster states
appear at smaller excitation energies than the molecular-orbit configurations.  

\begin{figure}[h] 
 \begin{center}
  \resizebox{0.650\textwidth}{!}{
  \includegraphics{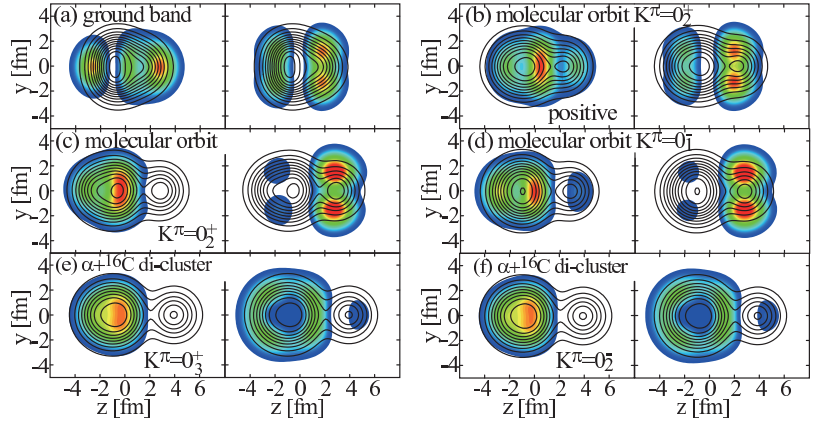}}
  \caption{Intrinsic density distributions of $^{20}{\rm O}$ for the band head states of (a) the
  ground band, (b)(c) the $K^\pi=0^+_2$ molecular orbit band, (d) the $K^\pi=0^-_1$ molecular orbit
  band that is paired with the $K^\pi=0^+_2$ band, and a pair of the atomic orbit bands (e)
  $K^\pi=0^+_3$ and (f) $K^\pi=0^-_2$. Solid lines show the density distribution of the 
  $^{16}{\rm O}$ core, while the color plots show that of valence neutrons. In each panel, the
  left side shows the density distribution of the second weakly bound two neutrons and the right
  side shows the most weakly bound two neutrons. This figure is reconstructed from the data
  presented in Ref. \cite{Furutachi:2007vz}. }   
  \label{fig:ex9}       % Give a unique label
 \end{center}
\end{figure}

Above explanation is verified by investigating the clustering systematics of $^{20}{\rm O}$. Since
the last two neutrons in $^{16}{\rm C}$ are much weakly bound than those of $^{14}{\rm C}$, we
expect that the order of molecular-orbit and atomic-orbit bands is reverted and becomes the same
order with the $^{22}{\rm Ne}$ case. This expectation is confirmed in the results for
$^{20}{\rm O}$ summarized in Fig. \ref{fig:ex7} (b) and Fig. \ref{fig:ex9}.  
$^{20}{\rm O}$ has a parity doublet of molecular-orbit configuration of 
$\alpha+{}^{14}{\rm C}+2n$ whose band-head energies are approximately 10 and 5 MeV below the
$\alpha$ threshold, respectively. The positive-parity states are the admixture of the
molecular-orbit and atomic-orbit configurations which are respectively shown in Fig. \ref{fig:ex9}
(b) and (c). The configuration (b) does not have prominent clustering of the core
($\alpha+{}^{14}{\rm C}$), but the last two valence neutrons (right panel of Fig. \ref{fig:ex9}
(b)) are distributed to the entire system showing molecular-orbit-like bonding. Another
configuration shown in Fig. \ref{fig:ex9} (c) has more developed cluster core. The last two
neutrons are localized around the $\alpha$ cluster and another two are around the $^{12}{\rm C}$ 
cluster, 
hence it appears more like  $^{6}{\rm He}+{}^{14}{\rm C}$  di-cluster system. The negative-parity
partner of molecular-orbit band denoted by $K^\pi=0^-_1$ band in Fig. \ref{fig:ex7} is easily
identified, because it is dominated by the configuration shown in the panel (d) which looks almost
identical to the panel (c).  

Above these molecular-orbit-like doublet, the atomic-orbit doublet having $\alpha+{}^{16}{\rm C}$
configuration appears in the vicinity of the $\alpha$ threshold. The intrinsic configurations of
the positive- and negative-parity states shown in the panels (e) and (f) look almost identical to
each other. Four valence nucleons are almost confined in the atomic orbit around the 
$^{12}{\rm C}$ cluster indicating the formation of the $\alpha+{}^{16}{\rm C}$ configuration. This 
energetical order of molecular-orbit and atomic-orbit bands is common to those of $^{22}{\rm Ne}$
and Be isotopes showing that the underlying shell effect influences the formation of the
molecular- and atomic-orbits and clustering systematics. Experimentally, the cluster states of
$^{18}{\rm O}$ have been 
investigated by many authors based on the $\alpha$-transfer, resonant scattering and enhanced $E1$
transitions 
\cite{Torilov:2011a,oertzen-o18,Gai:1983zz,Gai:1987zz,Gai:1991a,Goldberg:2005a,Ashwood:2006sb,Yildiz:2006xc,Johnson:2009kj,Avila:2014zwa}.  
Fig. \ref{fig:ex7} (a) shows that present results reasonably agree with these
experimental candidates, supporting the above-mentioned clustering systematics. Unfortunately, the
experimental information for $^{20}{\rm O}$ is rather obscure 
\cite{LaFrance:1979zz,Wiedeking:2004xz,Bohlen:2011a,Hoffman:2012zz}, and we expect further
experimental studies will reveal the clustering systematics and the relationship between the
molecular-orbit and atomic-orbit states.  

\subsubsection{$\alpha$ clustering in neutron-rich F isotopes}
The neutron drip line of F and Ne isotopes are extended beyond $N=20$ shell gap where the magicity
of neutron number $N=20$ is lost and the shell gap between $sd$ and $pf$-shells is
quenched. Since the 2$\alpha$ clustering in neutron-rich Be isotopes is also closely related to
the breakdown of the $N=8$ magic number, it is very interesting to investigate how the clustering
in F and Ne 
isotopes are affected by the breakdown of the neutron magic number $N=20$. In the section
\ref{sec:3.2}, it was already discussed that the ground state clustering in neutron-rich Ne isotopes
are enhanced owing to the breakdown of the $N=20$ magic number. Here, we discuss how the breakdown
of the $N=20$ magic number affects the excited states of F isotopes. 
\begin{figure}[h] 
 \begin{center}
  \resizebox{0.60\textwidth}{!}{
  \includegraphics{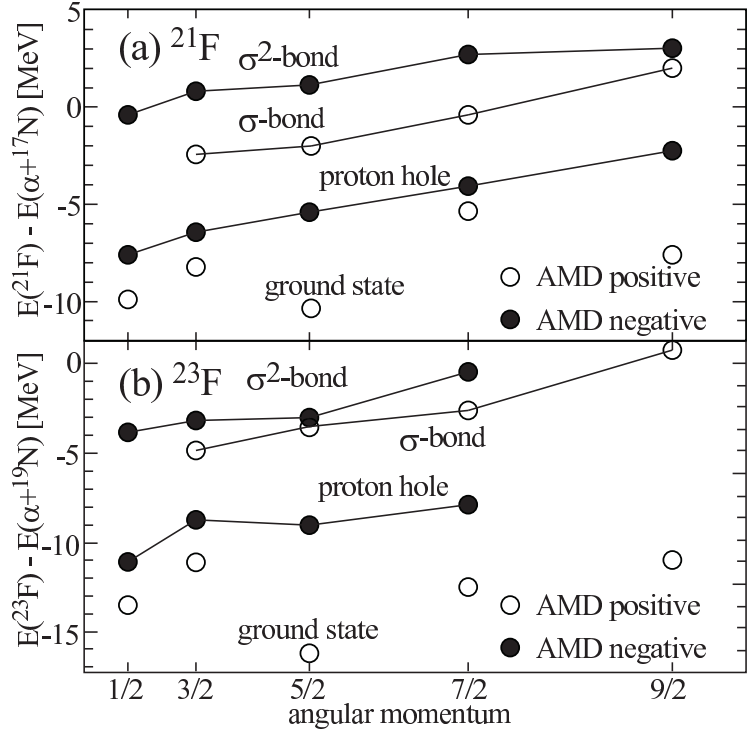}}
  \caption{Calculated partial level scheme of $^{21}{\rm F}$ and $^{23}{\rm F}$. Disconnected
  circles show the states in which a proton and all valence neutrons are occupying the
  $sd$-shell. Connected circles show the rotational band denoted by ``proton hole'',
  ``$\sigma$-bond''  and ``$\sigma^2$-bond'' in which a proton is excited from $p$-shell to
  $sd$-shell. This figure is reconstructed from the data presented in
  Ref. \cite{kimura:2011zza}.}   \label{fig:ex10}       % Give a unique label
 \end{center}
\end{figure}
\begin{figure}[h] 
 \begin{center}
  \resizebox{0.60\textwidth}{!}{
  \includegraphics{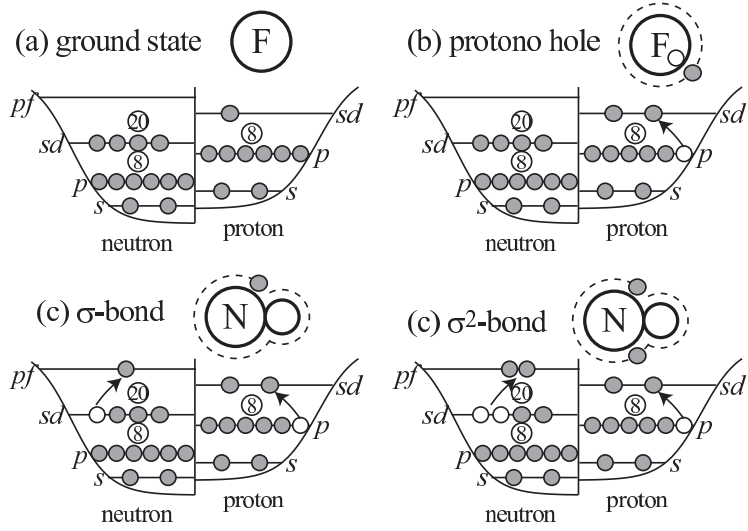}}
  \caption{Configurations of the ground state, proton hole, $\sigma$-bond and $\sigma^2$-bond
  rotational bands of $^{21}{\rm F}$ in terms of the single particle orbit.}   
  \label{fig:ex11}       % Give a unique label
 \end{center}
\end{figure}
\begin{figure}[h] 
 \begin{center}
  \resizebox{0.70\textwidth}{!}{
  \includegraphics{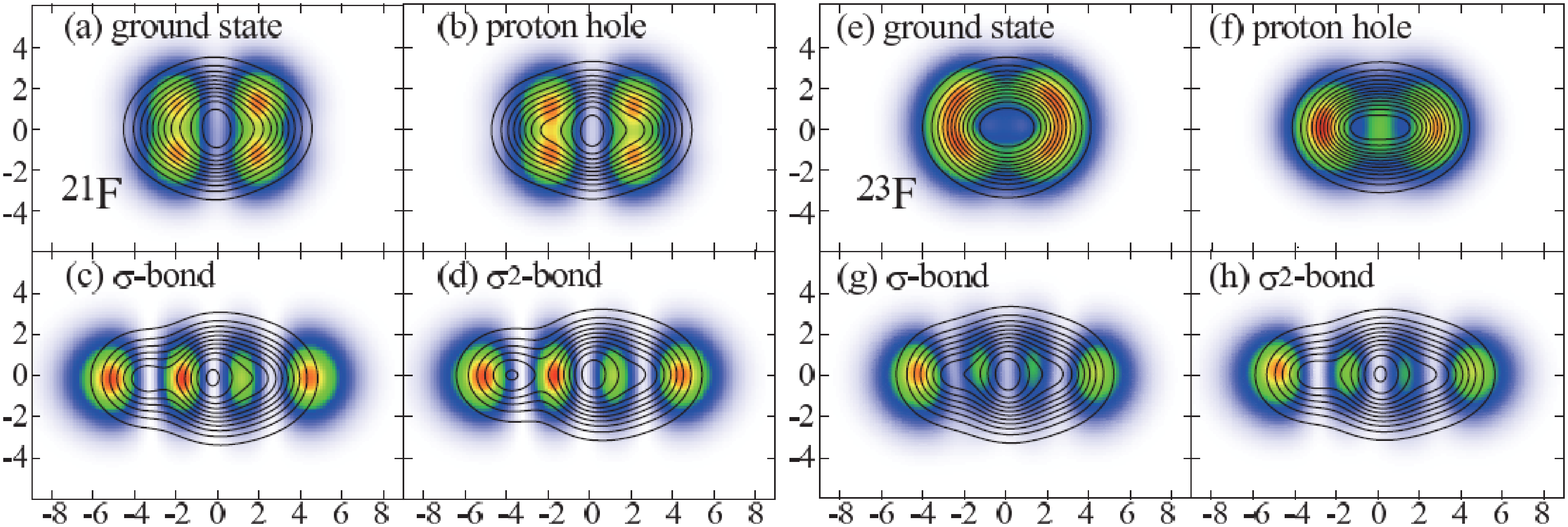}}
  \caption{Density distributions of the $^{19}{\rm F}$ core (contour plot) and those of the last
  valence neutron (color plot) in the ground state, proton hole, $\sigma$-bond and $\sigma^2$-bond
  rotational bands of $^{21}{\rm F}$ and $^{23}{\rm F}$. 
  This figure is reconstructed from the data presented in   Ref. \cite{kimura:2011zza}.}  
  \label{fig:ex12}       % Give a unique label
 \end{center}
\end{figure}

As already explained above, the $\alpha$ cluster state of $^{19}{\rm F}$ must have a proton hole
in $p$-shell. Therefore, we here discuss how the proton hole states of neutron-rich F isotopes
evolve  as function of neutron number \cite{kimura:2011zza}. Figure \ref{fig:ex10} shows the
calculated partial level scheme of (a) $^{21}{\rm F}$ and (b) $^{23}{\rm F}$ where the
disconnected circles show the states without proton hole, while the connected circles show the
proton hole states which are classified into three rotational bands denoted by ``proton hole'',
``$\sigma$-bond'' and ``$\sigma^2$-bond''. For the understanding of their structures,
Fig. \ref{fig:ex11} schematically shows 
the single-particle configurations of $^{21}{\rm F}$, and Fig. \ref{fig:ex12} shows their core and
valence-neutron density distributions. While the ground state of $^{21}{\rm F}$ has no proton hole
in $p$-shell, the rotational band denoted by ``proton hole'' has it as illustrated in
Fig. \ref{fig:ex11} (a) and (b). Different from the case of $^{19}{\rm F}$, the proton excitation
from $p$ to $sd$-shell does not induce the clustering as seen from the density distribution shown
in Fig. \ref{fig:ex12} (b). This is due to the two valence neutrons occupying the $sd$-shell which
energetically unfavor the deformation caused by the clustering. This role of valence neutrons is
the same as the cases of the ground states of $^{10}{\rm Be}$ and $^{22}{\rm Ne}$. When one or two
valence neutrons are promoted into $pf$-shell together with the proton excitation, the result is
very different. The $\sigma$-bond and $\sigma^2$-bond rotational bands have one and two valence
neutrons 
in $pf$-shell (Fig. \ref{fig:ex11} (c) and (d)), respectively. As clearly seen in their density 
distributions (Fig. \ref{fig:ex12} (c) and (d)), these neutron excitation induce the clustering
of the $^{19}{\rm F}$ core. It is also noted that the clustering is more enhanced as the number of
valence neutron in this orbit increases. Hence, we recognize that this valence neutron
orbit is quite analogous to the $\sigma$-orbit in $^{22}{\rm Ne}$; They have similar density
distribution and 
induce the clustering.  The situation is common to  $^{23}{\rm F}$, where the proton hole,
$\sigma$- and $\sigma^2$-bond bands also appear, and the clustering of the core is also induced as
seen in their density distributions shown in Fig. \ref{fig:ex12} (e)-(h). 

\begin{figure}[h] 
 \begin{center}
  \resizebox{0.70\textwidth}{!}{
  \includegraphics{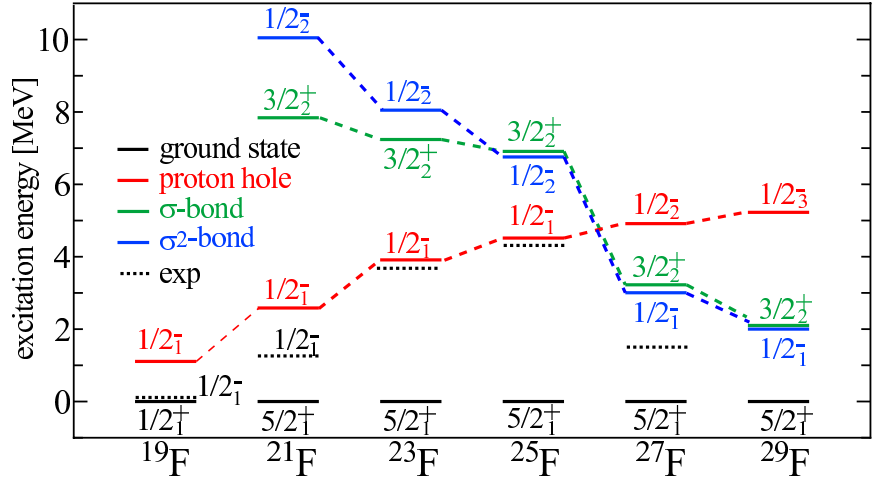}}
  \caption{Excitation energies of the band head states of the proton hole, $\sigma$-bond and
  $\sigma^2$-bond rotational bands in neutron-rich F isotopes as function of the neutron number.
  This figure is reconstructed from the data presented in Ref. \cite{kimura:2011zza}.} 
  \label{fig:ex13}       % Give a unique label
 \end{center}
\end{figure}

Similar calculations were performed up to $^{29}{\rm F}$ and it was found that the proton hole,
$\sigma$- and $\sigma^2$-bond bands always appear in the isotopes. Figure \ref{fig:ex13} shows the
energies of the band-head states of these bands. It is interesting that the energies of the
$\sigma$- and $\sigma^2$-bond bands dramatically decrease toward neutron drip line, while those of
the  proton hole band gradually increase. This characteristic behavior of the molecular-orbit
bands is closely related to the breakdown of $N=20$ magic number explained in the previous
section, and 
summarized as follows. (1) As one approaches to the neutron-drip line, the neutron Fermi level
naturally goes up, which reduces the excitation energy that costs to promote neutron from $sd$ to
$pf$-shell or $\sigma$-orbit. (2) In addition to this, in the vicinity of the island of inversion,
the $N=20$ shell 
gap is quenched or disappears, that further reduces the energy needed for neutron excitation. (3)
Once the neutrons are promoted into $pf$-shell or $\sigma$-orbit, it induces the clustering of
the core.  This brings about the additional binding energy which compensates the energy used for
neutron excitation. At the same time, the clustering brings about the large deformation that
further reduces the energy of the $pf$-shell. Owing to these cooperative effects, the calculation
predicts the great energy reduction of the molecular-orbit states near the neutron drip line of F 
isotopes. 

Up to now, the experimental data is rather obscure for F isotopes. However, there are several
possible candidates are reported. The proton hole $1/2^-$ states in $^{19}{\rm F}$ and 
$^{21}{\rm F}$  are well identified at 0.1 and 1.1 MeV, respectively
\cite{Kaschl:1970trf,Mairle:1981ecs}. By several experiments, the candidates of the
$1/2^-$ states are reported at 3.4 MeV in  
$^{23}{\rm F}$ \cite{Michimasa:2005df,Shimoura:2005ht} and 4.2 MeV in $^{25}{\rm F}$
\cite{Frank:2011zzb,Smith:2012qf}. Those excitation energies are gradually going up as 
neutron number increases and agree with the trend of the calculated proton hole states. An
experiment \cite{Elekes:2004olv} reports the possible low-lying negative-parity state in 
$^{27}{\rm F}$ at 1.3 MeV. The energies of the $1/2^-$ candidates look discontinuously dropping 
at $^{27}{\rm F}$ which agrees with the level crossing between the proton hole band and
$\sigma^2$-bond band predicted by the present calculation.  Although the present data does not
enough to identify the molecular-orbit bands, it is enough to motivate further theoretical and 
experimental studies.  For example, the proton knockout reaction from $^{30}{\rm Ne}$ is very
interesting and useful to identify the molecular states of $^{29}{\rm F}$. Since the dominance of
the neutron $2\hbar\omega$ configuration (two neutrons in $pf$-shell) is experimentally established
well, we can expect that a proton knockout from $p$-shell will strongly populates the
$\sigma^2$-bond band of $^{29}{\rm F}$. 

\subsection{Three center systems}\label{sec:4.3}
\label{Three-center}
In this section, we discuss the linear-chain states in C isotopes, which are the most famous
exotic structures expected in three cluster systems. 

About 60 years ago, Morinaga suggested that the Hoyle state (the $0^{+}_2$ state of $^{12}$C) has
the linear-chain structure, in which three $\alpha$ clusters are linearly aligned
\cite{Morinaga:1956zza,Morinaga:1966}.  However, the cluster model studies 
\cite{Horiuchi:1974,Horiuchi:1975,Uegaki:1977,Kamimura:1981oxj} revealed that the Hoyle
state is not a linear-chain state but a dilute gas-like state, where three $\alpha$ clusters are
very weakly interacting and do not have definite spatial configuration. In addition to this, 
the AMD \cite{KanadaEn'yo:2006ze} and FMD \cite{Neff:2003ib} calculations showed that the
linear-chain of $3\alpha$ particles is unstable against the bending motion (deviation from the
linear alignment). As a result, they found that the highly excited $0^+$ state above the Hoyle
state has a bent-armed shape. Although the recent study based on the concept of the non-localized
clustering \cite{Suhara:2013csa} shed  a new insight into the linear-chain, up to now, no positive 
evidence was observed for the linear-chain formation in $^{12}$C. 

In these decades, the success of the molecular-orbit in Be isotopes strongly motivated the search
for the linear chain in neutron-rich C isotopes \cite{Oertzen:1997}, because the glue-like role of
excess neutrons may stabilize the linear chain. Itagaki {\it et al.} \cite{Itagaki:2001mb}
discussed the stability of linear-chain structures based on the molecular-orbit model. They found
that the combination of the excess neutrons in the $\pi$ and $\sigma$ orbits stabilizes the
linear-chain structure against both of the breathing and bending motions, and concluded that the
$^{16}{\rm C}^*$ with  $\pi_{3/2}^{2} \sigma_{1/2}^{2}$ configuration is the most promising
candidate of the linear-chain state. Also, in other neutron-rich C isotopes, many theoretical
studies predict the linear-chain formation. $^{13}$C was investigated using $3\alpha$+$n$ cluster
model and a bent linear-chain band was suggested
\cite{Itagaki:2006ic,Furutachi:2011zz}. Relativistic Hartree and non-relativistic Hartree-Fock
calculations were also performed for $^{15,16}$C and $^{20}$C \cite{Maruhn:2010dtc,Zhao:2014vfa}
which confirmed the stability of the linear chain within the mean-field approximation. 

Compared to these theoretical models, AMD has several advantages. It can investigate the
linear-chain formation without any {\it a priori} assumption on the clustering. It enables
quantitative discussions, because it can provide a reliable predictions for the excitation energies
and the $\alpha$ decay widths. Up to now,  the AMD calculations have been
performed for $^{14}{\rm C}$ \cite{Suhara:2010ww,Suhara:2011cc,Baba:2016sbi} and for 
$^{16}{\rm C}$ \cite{Baba:2014lsa}.  Recently, rather promising data that imply the formation of
linear-chain states in $^{14}{\rm C}$
\cite{Freer:2014gza,Fritsch:2016vcq,Yamaguchi:2016,Tian:2016vvb} and 
$^{16}{\rm C}$ \cite{Dell'Aquila:2016gyj,Koyama:2016a} were reported.  They show good agreement
with the prediction of AMD studies. Here, we mainly focus on $^{14}{\rm C}$ and introduce the
AMD calculation in Ref. \cite{Suhara:2010ww}. 

\begin{figure*}[t!h!b!]
 \centering
 \resizebox{0.9\textwidth}{!}{\includegraphics{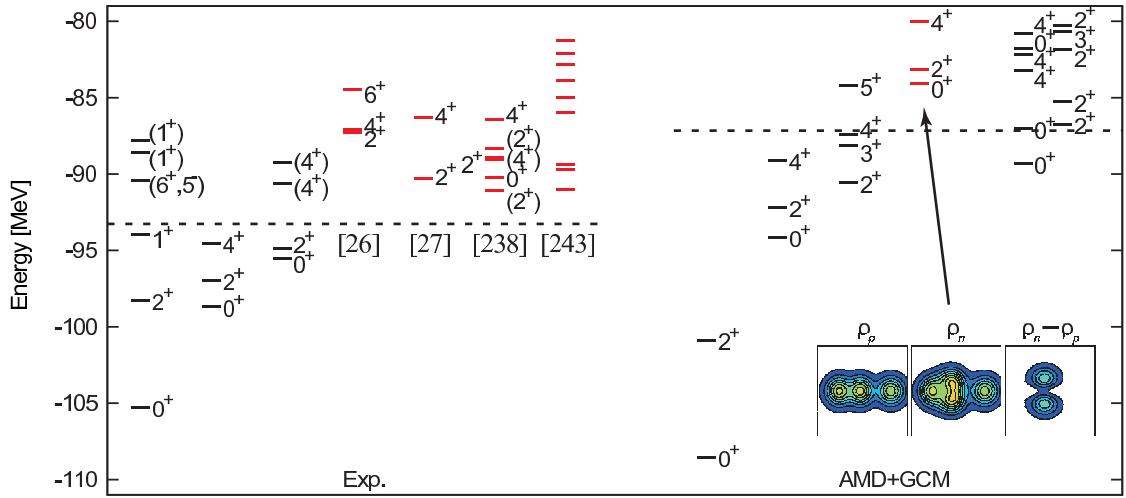}}
 \caption{Energy levels of the positive-parity states in $^{14}$C. 
 The experimental data including the recently observed linear-chain candidates 
 \cite{Freer:2014gza,Fritsch:2016vcq,Yamaguchi:2016,Haigh:2008zz}
 are shown on the
 left side, while the AMD result is shown on the right  side. The observed and calculated
 linear-chain states (candidates) are shown by the red lines.  Dotted lines show the experimental
 and theoretical $^{10}$Be+$\alpha$ threshold energies,  respectively. This figure is
 reconstructed from the data presented in Ref. \cite{Suhara:2010ww}.}  \label{Energy_level_14C_+}
\end{figure*}

In Ref.~\cite{Suhara:2010ww},  $^{14}$C is investigated by using AMD combined with the
two-dimensional GCM using quadrupole deformation parameters $\beta$ and $\gamma$ as the generator
coordinates.  
In this calculation, we used the Volkov No.~2 interaction combined with 
the spin-orbit part of the G3RS interaction.
The adopted parameters are $m=0.6$, $b=h=0.125$, and $u_1=-u_2=-1600$ MeV,
which reproduce the binding energies of deuteron and $^{12}$C.
Fig.~\ref{Energy_level_14C_+} compares the observed positive-parity states with the
calculation. In the left side of the figure, the black lines show the observed positive-parity
states which are 
known long before \cite{AjzenbergSelove:1991zz}, while the red lines show the candidates of the
linear chain. Three different groups independently reported these candidates based on the
$^{10}$Be+$\alpha$ resonant scattering  \cite{Freer:2014gza,Fritsch:2016vcq,Yamaguchi:2016}. In
the same energy region, another group  \cite{Haigh:2008zz} also reported the candidates
based on the $^{10}$Be+$\alpha$ breakup reaction, but the spin-parity assignment was not given.
In the right hand side of the figure, we classified the calculated states into five groups by
analyzing the intrinsic structures and $E2$ transition strengths. The left most band shows
the ground band and the next two bands are the $K^\pi=0^+$ and $2^+$ bands with triangular
configuration of $3\alpha$ clusters.
The calculated linear-chain states are shown by the red lines. We also show the intrinsic proton
$\rho_{p}$, neutron densities $\rho_{n}$ and their difference $\rho_{n}-\rho_{p}$ for the 
linear-chain state. From the proton density, we clearly see the linear-chain formation; three
$\alpha$ clusters develop well and they are linearly aligned. It is also noted that the
inter-cluster distance between $3\alpha$ clusters are not equal showing somewhat asymmetric
$2\alpha$+$\alpha$ correlation. Corresponding to this asymmetry, the excess neutrons distribute
around two of the three $\alpha$ clusters, which have shorter inter-cluster distance. In short,
these indicate $^{10}$Be+$\alpha$ correlation in the linear-chain states. 

When the excitation energies are measured from the $\alpha$ threshold, the agreement
between the calculated and observed linear-chain candidates looks reasonable. 
In particular, the spin-parity assignment and the large moment of inertia of $\hbar^{2}/2I = 0.19$
MeV reported in Ref.~\cite{Yamaguchi:2016} show very good agreement with the AMD.
In addition to this, they also reported the large $\alpha$ decay widths of these linear-chain
candidates which  are the same order of magnitude to the present result estimated by using
the method give in the Ref. \cite{Kanada-En'yo:2014nla}. Hence, we conclude that the linear-chain
formation in $^{14}{\rm C}$ is rather convincing.

\begin{figure}[tb]
 \centering
 \resizebox{0.45\textwidth}{!}{\includegraphics{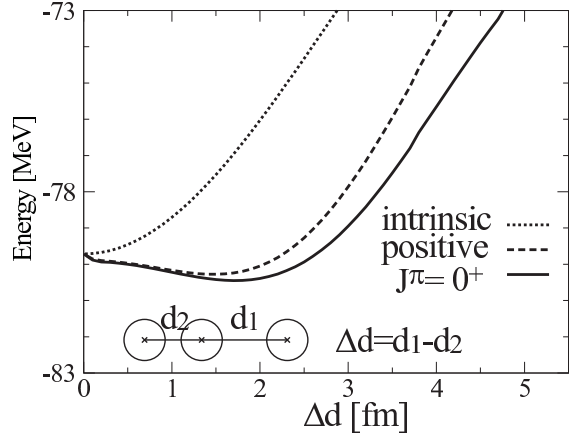}}
 \caption{Energy curves for the intrinsic, positive parity, and $J^\pi=0^{+}$ wave functions of
 $^{14}$C with the linear-chain configuration. The distance between two $\alpha$  clusters
 at both end is fixed to 5.2 fm. The horizontal axis is the difference of the 
 $\alpha$-cluster intervals, $\Delta d$. The intrinsic and positive parity energy curves are
 shifted by $-6.9$ MeV to adjust with the $J^\pi=0^{+}$ energy curve at $\Delta d = 0$ fm.
 This figure is reconstructed from the date presented in Ref. \cite{Suhara:2011cc}.} 
 \label{Energy_curves_14C_+}
\end{figure}

Now, we discuss why the $^{10}$Be+$\alpha$ correlation appears in the linear-chain states by using
a generalized molecular orbital model \cite{Suhara:2011cc}. Fig.~\ref{Energy_curves_14C_+} shows
the calculated energy curves for the intrinsic, positive parity, and $J^\pi=0^{+}$ wave functions
of $^{14}$C with the linear-chain configuration. 
The intrinsic energy curve has the energy minimum at $\Delta d = 0$ fm, while the positive-parity
and the $0^{+}$ energy curves have energy minima at $\Delta d > 0$ fm. 
This indicates that the correlation between clusters is described by the parity projection and it
induces $2\alpha+\alpha$ correlation to gain more binding energy. 
%With the analysis of the motion
%of excess neutrons and their effect in the linear chain structure, we found the excess neutrons
%concentrate around the correlated $2 \alpha$ clusters. As a result, we concludes that the
%linear-chain states in $^{14}$C have $^{10}$Be+$\alpha$ correlation. 

\begin{figure}[tb]
	\centering
	\resizebox{0.45\textwidth}{!}{\includegraphics{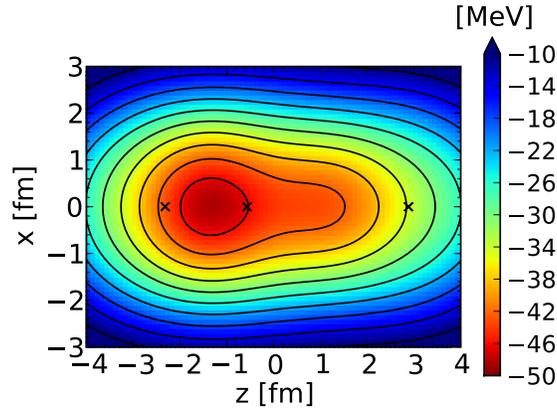}}
	\caption{The contour of the single-folding potential of three $\alpha$ clusters with the
 linear-chain structure. The crosses show the positions of three $\alpha$ clusters.
 This figure is taken from Ref. \cite{Suhara:2011cc}.}
	\label{12C_folding_potential}
\end{figure}

Then, we consider the motion of the excess neutrons around this asymmetric core. 
As already mentioned, the excess neutrons concentrate around the correlated $2 \alpha$ clusters. 
It is different from a simple molecular orbital picture, where the excess neutrons are distributed
widely surrounding all $\alpha$ clusters to reduce the kinetic energy.  
This can be understood by a mean-field potential generated by the aligned three $\alpha$ clusters. 
Fig.~\ref{12C_folding_potential} shows a contour of the valence neutron potential which is
obtained by the single-folding of three $\alpha$ clusters with asymmetric linear alignment.
The potential is deepest between the correlated $\alpha$ clusters and becomes
shallow around the isolated $\alpha$ cluster. Therefore, the excess neutrons tend to gather around
the correlated two  $\alpha$ clusters to gain potential energy. The detailed balance between 
the potential energy gain and kinetic energy loss determines the distribution of the excess
neutrons. As a result, the $^{10}$Be+$\alpha$ correlation appears in the linear-chain states of
$^{14}$C.   

\begin{figure}[tb]
	\centering
	\resizebox{0.4\textwidth}{!}{\includegraphics{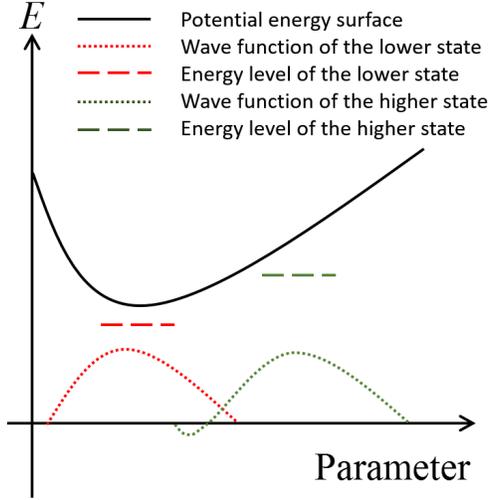}}
	\caption{A schematic figure for the stabilizing mechanism by the orthogonality.}
	\label{Schematic_figure_orthogonality}
\end{figure}

Finally, we discuss the reason why the linear-chain states are stabilized in $^{14}$C in
contradiction to the result of $3\alpha$+$xn$ cluster model by Itagaki \textit{et al.}
\cite{Itagaki:2001mb} which reported the instability of the linear-chain in $^{14}$C against the
bending motion. From the uncertainty principle, if $\alpha$ clusters are fixed to the linear
alignment, the kinetic energy must be increased. As a result, the bending motion which fluctuates
the position of $\alpha$ clusters and reduces the kinetic energy must be energetically favored.  
Therefore, for the stability of the linear-chain, an additional stabilizing mechanism is necessary.
That is the orthogonality of the wave function to the lower energy states required in 
quantum mechanics. A schematic figure shown in Fig.~\ref{Schematic_figure_orthogonality} explains
it.  When the Schr\"odinger equation is solved on the 
potential surface as shown in the figure, the wave function of the lowest energy state
concentrates around the energy minimum. To satisfy the orthogonal condition to the lower state,
the wave function of the excited state cannot have the amplitude near the energy minimum, and
therefore, concentrates in the outer regions. In the case of $^{14}$C, the linear-chain states
must be orthogonal to the ground band and $K^\pi=0^+$ and $2^+$ bands with triangular
configuration of $3\alpha$ clusters. Note that the bending motion of linear chain makes the
overlap between (bent-)linear-chain state and triangular state large that is forbidden by the
orthogonality condition.  Thus, behind the stability of the linear chain, the orthogonal condition
in quantum mechanics plays a crucial role. 

A similar mechanism was also discussed in $^{12}$C \cite{Suhara:2014wua}.  The $0^{+}_{1}$ state
have a compact triangle structure of three $\alpha$ clusters with the significant mixing of the
$\alpha$ cluster breaking component.  The $0^{+}_{2}$ state is described by the superposition of
various triangle configurations of three $\alpha$ clusters. The compact triangle component in the
$0^{+}_{2}$ is hindered due to orthogonality to the $0^{+}_{1}$ state. The details of the
structure of the $0^{+}_{2}$ is affected by the magnitude of the mixing of the $\alpha$ cluster
breaking component in the $0^{+}_{1}$ state. The $0^{+}_{3}$ is dominated by the open triangle
configuration; therefore this state is considered to be a bent linear-chain states. The compact
triangle and acute triangle configurations in the $0^{+}_{3}$ is hindered due to orthogonality to
the $0^{+}_{1}$ and $0^{+}_{2}$ states. The structure of the $0^{+}_{3}$ is significantly affected
by the magnitude of the mixing of the $\alpha$ cluster breaking component in the $0^{+}_{1}$
state. As same as $^{14}$C, to the appearance of the linear-chain state, the orthogonal condition
plays key role in $^{12}$C. It is very interesting to extend the study to other C isotopes, in
particular, to more neutron-rich isotopes. Because the number of excess neutrons is much larger,
they can have different stabilization mechanism.  

\subsection{Cluster aspects in neutron-rich nuclei}\label{sec:4.4}

As discussed in previous sections, a variety of clustering phenomena such as 
the cluster formation, molecular orbit structures, 
and di-cluster resonances are seen in neutron-rich nuclei.
Be and Ne isotopes are the typical examples that show remarkable clustering phenomena.
We here give more general discussions of cluster aspects in neutron-rich nuclei.
In the clustering phenomena in nuclear system, 
many-body correlations as well as single-particle
motions play essential roles. 
For instance, 
cluster formation originates from strong spatial correlations between nucleons, 
whereas molecular orbits are characterized by independent single-particle motion of 
valence neutrons around a two-center cluster core.

\subsubsection{Molecular orbits v.s. Nilsson orbits}
%%%%%%%%%%%%%%%%%%%%%%%%%%%%%%
\begin{figure}[htb]
\begin{center}
\resizebox{0.7\textwidth}{!}{%
\includegraphics{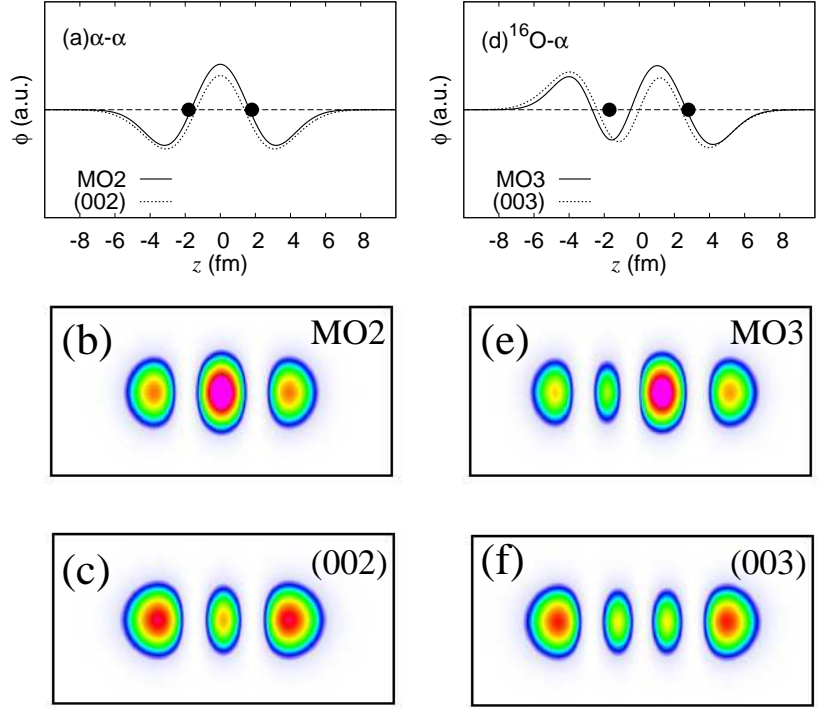} }	
%\includegraphics[width=12.0cm]{nilsson-mo-fig.eps} 	
%\vspace{0.5cm}
  \caption{Molecular $\sigma$-orbits and deformed shell-model orbits. 
(a-c) the $\sigma$-orbit (MO2) in the $\alpha$+$\alpha$ system compared with the $(002)_{\rm def}$ orbit. 
(d-f) the $\sigma$-orbit (MO3) in the $^{16}$O+$\alpha$ system compared with the $(003)_{\rm def}$ orbit. 
Upper panels: the single-particle wave functions on the $z$ axis at $x=y=0$. Middle and lower
 panels: the contour map of the density projected on the $z$-$y$ plane. 
The size parameter $b$ is chosen to be $b=1.5$ fm for the $\sigma$-orbits, 
$b_\perp=1.5$ fm and $b_z=1.3b_\perp$ 
for the $(002)_{\rm def}$ and $(003)_{\rm def}$ orbits.
The cluster positions are shown by filled circles. 
\label{fig:nilsson-mo}}
\end{center}
\end{figure}
%%%%%%%%%%%%%%%%%%%%%%%%%%%%%
Molecular orbits are sometimes associated with deformed shell-model (Nilsson) orbits because 
they are specified by quantum numbers similar to those of deformed shell-model  orbits.
For instance, the total node number $N$, the longitudinal node number $n_z$, 
and the $z$-component $\Omega=j_z$ of the angular momentum are good quantum numbers
for molecular orbits as well as deformed shell-model orbits because of 
the axial symmetry of systems.
Indeed, in the case of a small inter-cluster distance, 
molecular orbits approximately correspond to deformed shell-model orbits in the normal deformation.
However, molecular orbits are not necessarily equivalent to deformed shell-model orbits 
except for the small inter-cluster distance case. 
For example, the $\sigma$-orbits around the $2\alpha$ and $^{16}$O+$\alpha$ cores have the quantum numbers same as those of
$(n_xn_yn_z)=(002)_{\rm def}$ and $(003)_{\rm def}$ orbits in the deformed harmonic oscillator
(h.o.) potential, however, they are quantitatively different from the  
deformed shell-model orbits. The
$\sigma$-orbits typically have largest amplitudes in between clusters, whereas the $(002)_{\rm def}$ and $(003)_{\rm def}$
orbits have maximum amplitudes at the surface region and relatively smaller 
amplitudes in the inner region. 
In Fig.~\ref{fig:nilsson-mo}, we give a comparison between 
the $\sigma$-orbits and deformed shell-model orbits.
Here the molecular $\sigma$-orbit (MO2) around the $2\alpha$ is simply assumed to be a linear combination of 
the spherical h.o. orbit $(001)$ around 
the left and right $\alpha$ clusters ($\alpha_1$ and $\alpha_2$) as
\begin{equation}
 \phi_{\rm MO2}=n_0 \hat P_{\rm orth}\left\{(001)_{\alpha_1}-(001)_{\alpha_2}\right\},
\end{equation}
where  $n_0$ is the normalization factor and $\hat P_{\rm orth}$ is the projection operator 
for the orthogonal condition to the occupied orbits $(000)_{\alpha_1}$ and 
$(000)_{\alpha_2}$.
In a similar way, we also assume the molecular $\sigma$-orbit (MO3) around the $^{16}$O+$\alpha$ 
with $(002)_{^{16}{\rm O}}$ and $(001)_{\alpha}$ as 
\begin{equation}
 \phi_{\rm MO3}=n_0 \hat P_{\rm orth}\left\{(002)_{^{16}{\rm O}}-(001)_{\alpha}\right\},
\end{equation}
where the $\hat P_{\rm orth}$ is defined so as to satisfy the 
orthogonal condition to the occupied orbits 
$(000)_{^{16}{\rm O}}$, $(001)_{^{16}{\rm O}}$ and $(000)_{\alpha}$.
(General molecular orbits in realistic systems are given by linear combination 
of any atomic orbits with coefficients optimized for 
each system, but here we simply consider a specific case of orbits 
and coefficients.)  
As shown in Fig.~\ref{fig:nilsson-mo}, 
amplitudes of the $\sigma$-orbits are enhanced in between two clusters differently from the 
$(002)_{\rm def}$ and $(003)_{\rm def}$ orbits, in which the inner amplitudes are remarkably 
suppressed. It means that, the probability of 
valence neutrons in the $\sigma$ orbits is enhanced
in the region between two clusters, because,  
neutrons feel attractive potentials from both clusters 
and less Pauli repulsion from core nucleons in this low-density region.
As a result of the enhanced amplitude in between clusters, $\sigma$-orbit neutrons push 
outward the clusters bonding them at a moderate 
distance to form the MO $\sigma$-bond structure like a covalent bond in a molecule. 
This is one of the important differences of the molecular orbits 
in the two-center potential 
from the deformed shell-model orbits in the one-center potential. 
 
\subsubsection{Molecular orbits v.s. atomic orbits (strong coupling v.s. weak coupling clustering)}

In the coexistence of the molecular orbit structures and di-cluster
resonances in neutron-rich Be and Ne isotopes, valence neutrons play important roles. 
Valence neutrons in the molecular orbits move around whole the system 
connecting two cluster cores, whereas those in the di-cluster resonances are localized to one side 
(around one of the two cluster cores). 
The former is analogous to covalent orbits and the latter is associated with  
atomic orbits in ionic bondng. 
In a system consisting of two cluster cores and valence 
neutrons, appearance of the molecular orbit structures is not obvious. 
The first condition for the molecular orbit structure is formation of 
single-particle orbits surrounding both the clusters. 
The second condition is that the independent particle feature of valence neutrons 
dominates over many-body correlations which gathers neutrons to
 one of two clusters. 

In the case of identical two clusters such as $\alpha$+$\alpha$, 
single-particle orbits are reflection symmetric because of the 
symmetry of the mean potential from two clusters. 
As a result, at a moderate inter-cluster distance,  
molecular orbits are favored to gain the kinetic energy provided that
many-body correlations between valence neutrons are minor. 
%In the system with the molecular orbit structure, 
%low-lying spectra can be understood by single-particle excitations in the molecular orbit configurations
%because of the independent particle feature of the valence neutrons in the molecular orbits.
However, in the asymptotic region of a large inter-cluster distance, valence neutrons localize 
to one side to gain the correlation energy, and the molecular orbits change to the atomic orbits.
In such the case, the system separates into two clusters to form a di-nuclear structure, in which 
inter-cluster motion is decoupled from internal degrees of freedom of clusters,
and then, di-cluster resonances arise from excitation of the inter-cluster motion.
By means of the GTCM, 
Ito {\it et al.} solved valence neutron wave functions in $2\alpha+2n$ with a fixed inter-cluster distance and 
demonstrated smooth transition of valence neutron orbits 
from the molecular(covalent) orbits at a moderate distance to 
atomic orbits at a large distance \cite{Ito:2003px,Ito:2008zza,Ito2014-rev}.

It is also important that different types of low-energy excitation modes 
appear in the molecular orbit and  di-cluster structures.
The low-energy modes in the molecular orbit structure are based on 
single-particle excitations in the molecular orbit configurations, whereas those 
in the di-nuclear structure originate from the
inter-cluster excitation as well as internal excitations of clusters.
In Be isotopes, the molecular orbits are favored at 
a moderate inter-cluster distance. As discussed previously, 
the low-energy excitations are described by single-particle excitations in the molecular orbit configurations, 
whereas the di-cluster resonances in the relatively higher energy region arise from the inter-cluster excitations. 
For instance, in $^{10}$Be, the $K^\pi=0^+_2$ and $K^\pi=1^-$ bands are understood by 
the single-particle excitations in the molecular orbit configurations, whereas
the  $^6$He($2^+$)+$\alpha$ and $^6$He($0^+$)+$\alpha$ cluster resonances are expected
in highly excited $0^+$ states
 from the inter-cluster excitations with and without the internal excitation of the $^6$He cluster. 
The existence of two excitation modes are more clearly seen in the calculated
negative-parity spectra of $^{12}$Be. The $K^\pi=1^-$ band arises from the 
single-particle excitation in the 
molecular orbit configurations, whereas the $K^\pi=0^-$ band appears from the negative-parity inter-cluster excitation between $^8$He 
and $\alpha$ clusters. 

It should be noted that, with the increase of valence neutrons, the effect of many-body correlations becomes 
no longer minor even at the moderate distance, and it increases asymmetry of the total system. 
In other words, the reflection symmetry is spontaneously 
broken by many-body correlations between valence neutrons as seen in the
asymmetric intrinsic densities
of Be isotopes as already shown in Fig.~\ref{fig:beiso-bh-dense}. 
Because of the asymmetry, the single-particle orbits somewhat 
deviate from ideal molecular orbits. 
It means mixing of di-cluster components,  
even though the low-lying states of Be isotopes can be qualitatively 
understood by dominant molecular orbit components.

Let us next consider the case that two cluster cores are not identical.
In such the case of asymmetric two clusters with valence neutrons, 
reflection symmetry of the mean potential is explicitly broken, and hence, 
the formation of molecular orbits is not obvious. 
%%%%%%%%%%%%%%%%%%
%An example is neutron-rich O isotopes. As discussed previously, XXXX
%
%As another example, we consider $^9$Li as an $\alpha+t$ core ...
As an example, we consider 
$^9$Li as an $\alpha+t$ core with two valence neutrons in analogy to $^{10}$Be having the $2\alpha$ core with two valence neutrons. 
In the $\alpha$+$t$ system at a moderate inter-cluster distance,  
valence neutrons feel stronger attraction from the $\alpha$ cluster 
than that from the $t$ cluster and 
tend to be localized around the $\alpha$ cluster to form a $^6$He cluster. 
Moreover, the $t$ cluster feels a weak attraction from the $^6$He cluster and
therefore the system favors a di-cluster structure of $^6$He+$t$.
It meant that the MO $\sigma$-bond structure is not favored in $^9$Li.
Indeed, in the theoretical calculation of $^9$Li in Ref.~\cite{KanadaEn'yo:2011nc}, 
$^6$He+$t$ cluster resonances are obtained 
but the MO $\sigma$-bond structure is not obtained in excited states of $^9$Li.

Neutron-rich B is another example, in which 
the molecular orbit picture is useful to understand the cluster structures 
as discussed in Refs.~\cite{enyo-b13,Suhara:2012zr}. 
In order to understand the formation of molecular orbits
in $^{13}$B, it is better to consider 
the $2\alpha$ core surrounded by valence nucleons (a proton and 4 neutrons) in 
the molecular orbits rather than a Li+$\alpha$ core.
 Then, we can understand the appearance of 
the molecular orbits around the $2\alpha$ core in $^{13}$B 
similarly to Be isotopes.

Now, a question is why 
molecular orbit structures appear in such $sd$-shell nuclei as 
F and Ne isotopes, which have further asymmetric two clusters. 
In the case of Ne isotopes, 
the $\sigma$-orbit is constructed by $sd$ orbits around the $^{16}$O cluster 
and a $p$ orbit around the $\alpha$ cluster \cite{oertzen-ne}. 
Even though the potential from the $^{16}$O cluster
is stronger than that from the $\alpha$ cluster, the lowest allowed orbits are 
$sd$ orbits because of Pauli blocking from core nucleons.
Consequently, 
valence neutrons in the higher-shell ($sd$) orbits feel effectively weak attraction
and can contribute to form the molecular $\sigma$-orbit together with 
the $p$ orbit around the $\alpha$ cluster. 
The question to be answered is whether or not the formation of  
molecular orbits in 
the asymmetric cluster system is caused by accidental matching of the single-particle
orbits around
different clusters, probably, relating to matching condition of single-particle energies. 
The origin of molecular orbits is also discussed by von Oertzen from the view point of 
threshold energies \cite{oertzen-ne,OERTZENa,Oertzen-rev}.
To answer this question, further examples of molecular-orbit
structures in various nuclei are required. 

\subsubsection{Di-cluster resonances}
%B10-he4
Di-cluster resonances have been well known in stable nuclei such as 
$^{16}$O+$\alpha$ cluster states in $^{20}$Ne and $^{12}$C+$\alpha$ cluster 
states in $^{16}$O 
(see Ref.~\cite{Fujiwara-supp} and references therein).
In unstable nuclei, di-cluster resonances
have been discovered by recent experimental and theoretical studies, as discussed previously for He+He cluster resonances in neutron-rich Be isotopes.
Also in other unstable nuclei, di-cluster resonances containing an $\alpha$ cluster 
are expected to appear 
near $\alpha$-decay threshold energies. Examples are
$^{10}$Be+$\alpha$ states in $^{14}$C \cite{Suhara:2010ww,Freer:2014gza,Fritsch:2016vcq,Haigh:2008zz,Soic:2003yg,oertzen04,Price:2007mm,Malcolm:2012jw}, 
$^9$Li+$\alpha$ states in $^{13}$B \cite{enyo-b13},  
$^{14}$C+$\alpha$  states in $^{18}$O and their mirror states 
\cite{Curtis:2002mg,Furutachi:2007vz,oertzen-o18,Gai:1983zz,Gai:1987zz,Descouvemont:1985zz,Fu:2008zzf,Ashwood:2006sb,Yildiz:2006xc,Johnson:2009kj}, 
$^{18}$O+$\alpha$ states in $^{22}$Ne 
\cite{Scholz:1972zz,Rogachev:2001ti,Curtis:2002mg,Goldberg:2004yk,Kimura:2007kz,Descouvemont:1988zz,Ashwood:2006sb,Yildiz:2006xc}.
These facts may imply that di-cluster resonances arising from $\alpha$-cluster excitations can be 
general phenomena in stable and unstable nuclei. 

Experimental and theoretical efforts of searching for di-cluster resonances 
are being made. It is also an challenging problem to discover more exotic di-cluster resonances 
comprised by two exotic nuclei such as $t$, $^{6,8}$He, $^9$Li, $^{10}$Be, $^{13}$B, 
$^{14}$C, and $^{18,20}$O clusters. Examples are 
$^8$He+$^6$He in $^{14}$Be \cite{KanadaEn'yo:2002ay,Ito:2011zhx} and $^6$He+$t$ states in $^9$Li \cite{KanadaEn'yo:2011nc}.
It is an important task to explore new cluster states near 
cluster-decay thresholds in excited states of various nuclei. 
In  systematics of di-cluster resonances in various unstable nuclei, 
we will obtain an answer to the question, whether Ikeda threshold rule
can be extended to a wide region of the nuclear chart including unstable nuclei.

\section{Monopole and dipole transitions as the probe for clustering}\label{sec:5}
In this section, we focus on the signature of the clustering. Experimentally, several observables
have been utilized as the evidence of clustering. One of the important observable is the $\alpha$
decay width from which the $\alpha$ reduced amplitude or the preformation factor of $\alpha$
particle at the nuclear surface can be extracted. The reduced width amplitudes including those
for the non-alpha clustering can be also measured by the resonant scattering, cluster transfer and
knock-out reactions. We have discussed that the behavior of the proton radii also indicates
the clustering albeit indirectly. However, in general, the measurement of these quantities is not
easy for unstable nuclei, although the rapid development of the experimental technique is enabling
it. Therefore another observable will be very helpful for the discussion of the clustering in
unstable nuclei. 

In this decade, the monopole transition was found to be very sensitive probe for the clustering
\cite{Yamada:2008a} and has been utilized to search for the cluster states in light stable nuclei
such as  $^{11}{\rm B}$ \cite{Kawabata:2005ta}, $^{12}{\rm C}$
\cite{Itoh:2011zz,Freer:2012se,Itoh:2014mwa},
$^{16}{\rm O}$ \cite{Wakasa:2006nt}, $^{24}{\rm Mg}$ \cite{Kawabata:2013xea} and 
$^{32}{\rm S}$ \cite{Itoh:2013lxs}. Recently   
the discussion of the monopole transition and clustering was extended to the neutron-rich Be
isotopes \cite{Yang:2014kxa,Yang:2015lha}. In addition to the monopole transition, dipole
transition is also expected as a promising and unique tool for the study of clustering
\cite{Chiba:2015zxa,Chiba:2015khu,Kanada-En'yo:2015vkg,Kanada-En'yo:2015ttw}. Therefore, we
briefly summarize the recent discussions on the monopole and dipole transitions. We first discuss
the relationship between the clustering and these transitions in $N\simeq Z$ nuclei and introduce
several AMD calculations. Then, we discuss the monopole and dipole responses of neutron-rich
nuclei.

\subsection{Clustering in stable nuclei and isoscalar monopole and dipole transitions}\label{sec:5.1}
It has long been known that many cluster states in light stable nuclei have large monopole
transition strengths from the ground state 
\cite{KanadaEn'yo:1998rf,Uegaki:1977,Kamimura:1981oxj,Descouvemont:1987zzb,Tohsaki:2001an,Funaki:2003af,Chernykh:2010zu,Suzuki:1989zza,Yamada:2011ri,Suzuki:1987jyi}. 
Therefore, isoscalar (IS) monopole transition has been utilized as
a probe to search for the cluster states. Later, Yamada {\it et al.} \cite{Yamada:2008a} gave a
clear explanation of the enhancement mechanism of the monopole transition. Recently, the
enhancement of the IS dipole transition for the cluster states was also pointed out
\cite{Chiba:2015khu}.  Here, we briefly outline these enhancement mechanism and introduce several
AMD results for $N\simeq Z$ nuclei. 

\subsubsection{Estimate of the transition strengths}
\begin{figure}[h]
 \begin{center}
  \resizebox{0.6\textwidth}{!}{
  \includegraphics{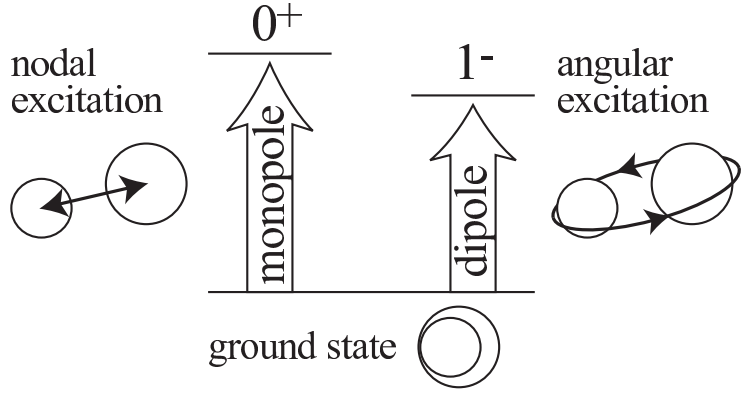}}
  \caption{Schematic figure for the nodal and angular excitations. IS monopole and dipole
  transitions strongly populates these cluster states.}\label{fig:mo1}   
 \end{center}
\end{figure}

It is well known that at the zero limit of the inter-cluster distance, the cluster model wave
function and $SU(3)$ shell model wave function \cite{Elliott:1958zj,Elliott:1958yc} become
mathematically identical \cite{Perring:1956a,Bayman:1958a}. This fact known as the Bayman-Bohr
theorem plays a central role for the discussion of the IS monopole and 
dipole transitions and their relationship to the clustering. 
For example, the ground state of $^{20}{\rm Ne}$ is described well by a $SU(3)$ shell model wave
function. The Bayman-Bohr theorem guarantees that it can be equivalently rewritten in terms of the
cluster model wave function,
\begin{align}
 \Phi(0^+_1) &= \mathcal{A}\set{(0s)^4(0p)^8(1s0d)^4}_{(8,0)}\nonumber\\
 &= n_{N_0}\mathcal{A}
 \Set{R_{N_00}(r)Y_{00}(\hat r)\phi_{\alpha}\phi_{{\rm O}}}, \label{eq:gswf0}
\end{align}
where the first and second lines are the shell model and cluster model representations,
respectively. In the cluster model representation, $\phi_\alpha$ and  $\phi_{\rm O}$ denote the
internal wave functions of clusters, and their relative motion is described by a Harmonic
oscillator wave function $R_{N_00}(r)Y_{00}(\hat r)$ with the principal quantum number $N_0=8$.
This equivalence implies that the degrees-of-freedom of cluster excitation is embedded even in an
ideal shell model state;  {\it i.e.} the excitation of the inter-cluster motion populates the
cluster states from a shell model state. For example, if we increase the nodal quantum number, we
get an excited $0^+$ cluster state,
\begin{align}
 &\Phi(0^+_{ex}) = \sum_{N=N_0+2}^{\infty} f_N n_{N}
 \mathcal A\Set{ R_{N0}(r)Y_{00}(\hat r)\phi_{\alpha}\phi_{{\rm O}}},
 \label{eq:exwf0}
\end{align}
which is a superposition of the wave functions with increased principal quantum number
$N$. Another option is the increase of the orbital angular momentum. With increase
by one, we obtain an excited $1^-$ cluster state,
\begin{align}
 &\Phi(1^-_{ex}) = \sum_{N=N_0+1}^{\infty}  g_N n_{N}
 \mathcal A\Set{ R_{N0}(r)Y_{10}(\hat r)\phi_{\alpha}\phi_{{\rm O}}}.\label{eq:exwf1}
\end{align}
We call these excited cluster states ``nodal excited'' and ``angular excited''
states (Fig. \ref{fig:mo1}). In the case of $^{20}{\rm Ne}$, the $0^+_4$ state at
8.7 MeV is the nodal excited state, while the $1^-_1$ state  at 5.8 MeV is the angular excited
state \cite{Tilley:1998wli}.  As already explained in the section \ref{sec:4.2}, this $1^-_1$
state has importance as the evidence for the parity asymmetric structure of 
$\alpha$+$^{16}{\rm O}$ \cite{Horiuchi:68a}. 

\begin{figure}[h]
 \begin{center}
  \resizebox{0.50\textwidth}{!}{
  \includegraphics{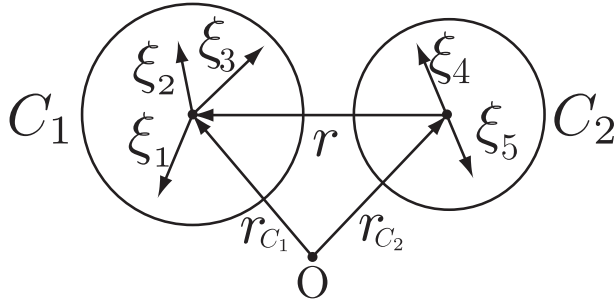}}
  \caption{Definition of the cluster coordinates. $\bm \xi_i$ denote the internal coordinates of
  clusters, while $\bm r$ is the inter-cluster coordinate.}\label{fig:mo2}   
 \end{center}
\end{figure}

By rewriting the IS monopole and dipole operators in terms of the cluster coordinates, it can be
shown that the IS monopole and dipole transitions strongly induce the nodal and angular
excitations. The IS monopole operator is rewritten as \cite{Yamada:2008a}, 
\begin{align}
 \mathcal M^{ IS0}&=\sum_{i=1}^{A}(\bm r_i - \bm r_{\rm cm})^2\nonumber\\
 &= \sum_{i\in C_1}\xi_i^2  +\sum_{i\in C_2}\xi_i^2
 +\frac{C_1C_2}{A}r^2,
 \label{eq:opis0}
\end{align}
where the first line shows the standard definition of the operator in terms of the single-particle 
coordinate $\bm r_i$, while the second line is the representation by the cluster coordinates which
are defined as,
\begin{align}
 &\bm \xi_i = \left\{
 \begin{array}{l}
  \bm r_i - \bm r_{C_1},\quad i\in C_1\\
  \bm r_i - \bm r_{C_2},\quad i\in C_2,\\
 \end{array}
\right.\quad
 \bm r = \bm r_{C_1} - \bm r_{C_2}, \\
 &\bm r_{C_1} = \frac{1}{C_1}\sum_{i\in C_1}\bm r_i,\quad
 \bm r_{C_2} = \frac{1}{C_2}\sum_{i\in C_2}\bm r_i.
\end{align}
Here, we have assumed that the system is composed of the two clusters with masses $C_1$ and $C_2$
as shown in Fig. \ref{fig:mo2}.
Note that this expression makes it clear that $\mathcal M^{ IS0}$ will populate nodal
excited states, because the last term proportional to $r^2$ will induce the nodal excitation of
the inter-cluster motion.

By a similar calculation, one finds that the IS dipole operator is rewritten as follows 
\cite{Chiba:2015khu},
\begin{align}
&\mathcal M^{IS1}_\mu =\sum_{i=1}^{A}(\bm r_i - \bm r_{\rm c.m.})^2
 \mathcal Y_{1\mu}(\bm r_i-\bm r_{\rm c.m.}) \nonumber\\
 &=  \sum_{i\in C_1}\xi_i^2\mathcal Y_{1\mu}(\bm \xi_i)
 +\sum_{i\in C_2}\xi_i^2\mathcal Y_{1\mu}(\bm \xi_i)\nonumber\\
 &-\frac{\sqrt{32\pi}}{3}
 \biggl[\left(\frac{C_2}{A}\sum_{i\in C_1}\mathcal Y_{2}(\bm \xi_i)
 -\frac{C_1}{A}\sum_{i\in C_2}\mathcal Y_{2}(\bm \xi_i)\right)
  \mathcal Y_{1}(\bm r)\biggr]_{1\mu}
 \nonumber\\
 &+\frac{5}{3}\left( \frac{C_2}{A}\sum_{i\in C_1}\xi_i^2 
 -\frac{C_1}{A}\sum_{i\in C_1}\xi_i^2 \right)\mathcal Y_{1\mu}(\bm r) \nonumber\\
 &+\frac{C_1C_2(C_1-C_2)}{A^2}r^2\mathcal Y_{1\mu}(\bm r). \label{eq:opis1}
\end{align}
From this expression we see that  the terms depending on $\mathcal Y_{1\mu}(\bm r)$ and
$r^2\mathcal Y_{1\mu}(\bm r)$ will induce the angular excitation to populate angular excited
cluster states. 

It is possible to derive analytic formulae for IS monopole and dipole transition matrices by using
the wave functions Eqs. (\ref{eq:gswf0})-(\ref{eq:opis0}) and (\ref{eq:opis1}) 
\cite{Yamada:2008a,Chiba:2015khu}. In the case of $^{20}{\rm Ne}$,  they read
\begin{align}
 M^{IS0} =& \braket{\Phi(0^+_{ex})|\mathcal M^{IS0}|\Phi(0^+_{gs})}\nonumber\\
 =&f_{N_0+2}\sqrt{\frac{\mu_{N_0}}{\mu_{N_0+2}}}
 \braket{R_{N_00}|r^2|R_{N_0+20}}, \label{eq:mat0}\\
 M^{IS1} =&\sqrt{3}\braket{\Phi(1^-_{ex})|\mathcal M^{IS1}|\Phi(0^+_{gs})}\nonumber\\
 =&\sqrt{\frac{3}{4\pi}}\Biggl[g_{N_0+1}\sqrt{\frac{\mu_{N_0}}{\mu_{N_0+1}}}
 \biggl\{\frac{48}{25}\braket{R_{N_00}|r^3|R_{N_0+11}}\nonumber\\
 &+\frac{16}{3}\left(\braket{r^2}_{\alpha} - \braket{r^2}_{\rm O}\right)
\braket{R_{N_00}|r|R_{N_0+11}}\biggr\} \nonumber\\
 &+\frac{3}{5}
 g_{N_0+3}\sqrt{\frac{\mu_{N_0}}{\mu_{N_0+3}}}\braket{R_{N_00}|r^3|R_{N_0+31}}\Biggr],
 \label{eq:mat1}
\end{align}
where $\braket{r^2}_{\alpha}$ and $\braket{r^2}_{\rm O}$ are the mean-square radii of the
clusters. $\mu_N$ is defined as,
\begin{align}
 \mu_N = \braket{R_{Nl}(r)Y_{lm}(\hat r)\phi_{\alpha}\phi_{\rm O}|
 \mathcal A\set{R_{Nl}(r)Y_{lm}(\hat r)\phi_{\alpha}\phi_{\rm O}}}.
\end{align}
By using the amplitudes $f_N$ and $g_N$ calculated by AMD, the formulae can be easily
estimated as  
\begin{align}
 &M^{IS0} = 7.67f_{N_0+2} = 5.48\ \rm fm^2, \label{eq:es0}\\
 &M^{IS1} = 3.08 g_{N_0+1} -7.36 g_{N_0+3} = 5.82\ \rm fm^3, \label{eq:es1}
\end{align}
which are as large as the Weisskopf estimates,
\begin{align}
 &M_{\rm WU}^{IS0} = \frac{3}{5}(1.2A^{1/3})^2 \simeq 0.864 A^{2/3}
 \simeq 6.37\ \rm fm^2,\\
 &M_{\rm WU}^{IS1} = \sqrt{\frac{3}{4\pi}}\frac{3}{6}(1.2A^{1/3})^3 \simeq 0.422 A
 \simeq 8.44\ \rm fm^3.
\end{align}
Thus, the IS monopole and dipole transition strengths from the ground state to the excited cluster
states are as strong as the Weisskopf estimates, even though the ground state is an ideal shell model
state.  It is well known that the giant resonances have very strong transition strengths and
exhaust most of the sum rule. But, they cannot appear at small excitation energy because they
involve the change of the matter density as illustrated in Fig. \ref{fig:int1} (a). On the other
hand, the cluster excitations illustrated in Fig. \ref{fig:int1} (c) occur at relatively small
energies, because  their excitation energies are governed by Ikeda threshold rule
(Fig. \ref{fig:int2}). This naturally explains why there exist many narrow resonances in the
response functions of light nuclei in the IS monopole and dipole channels 
\cite{Itoh:2011zz,Kawabata:2013xea,Itoh:2013lxs,Youngblood:1997zz,Youngblood:1998zza,Lui:2001xh,Chen:2009zzp,Gupta:2016yon,Peach:2016yop}.
Hence, the low-lying IS monopole and dipole strengths can be regarded as
good signature of clustering. 

\subsubsection{AMD results for light stable nuclei}
\begin{figure}[h]
 \begin{center}
  \resizebox{0.60\textwidth}{!}{
  \includegraphics{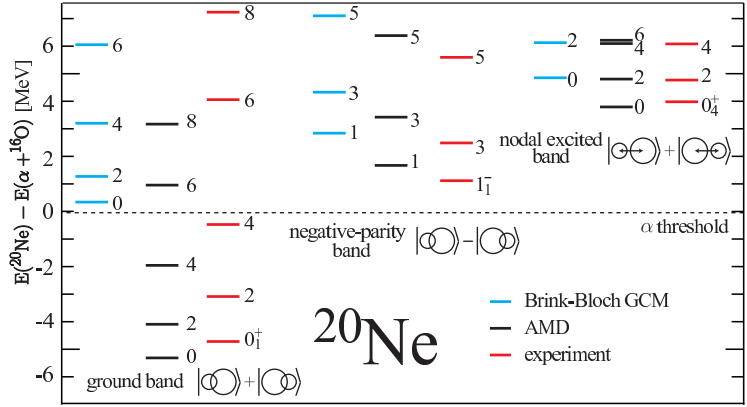}}
  \caption{Observed and calculated $\alpha+{}^{16}{\rm O}$ cluster bands in $^{20}{\rm Ne}$. The
  Brink-Bloch GCM calculation assumes the $\alpha+^{16}{\rm O}$ cluster structure.
  This figure is reconstructed from the data presented in Ref. \cite{Chiba:2015khu}.
  }\label{fig:mo3}   
 \end{center}
\end{figure}

The discussion above neglects the cluster correlation in the ground state and the distortion of
the clusters in the excited states, which will affect the transition strengths. To obtain accurate
results by taking these effects into account, AMD calculations were performed for several nuclei
\cite{Kimura:2003uf,Chiba:2015zxa,Chiba:2015khu}.
As an example, Fig. \ref{fig:mo3} shows the observed and calculated $\alpha+^{16}{\rm O}$ cluster
bands in 
$^{20}{\rm Ne}$.  
Compared to the Brink-Bloch GCM which assumes the $\alpha+^{16}{\rm O}$ cluster structure, 
AMD reasonably describes the ground band, which means that the distortion of clusters is
important for the ground band. On the other hand, both models reasonably describe the excited
cluster bands including the nodal and angular excited states. This indicates that these excited
cluster bands have almost ideal $\alpha+{}^{16}{\rm O}$ cluster structure. It is also confirmed
from the large distance between $\alpha$ and $^{16}{\rm O}$ clusters listed in Tab. \ref{tab:mo1}
which are estimated in the same way as explained in the section \ref{sec:3.2}. The calculated
transition 
matrices are also given in Tab. \ref{tab:mo1}. One sees that both of the IS monopole and dipole
transitions are very strong and much more enhanced than the Weisskopf estimates and the cluster
estimates given in Eqs. (\ref{eq:es0}) and (\ref{eq:es1}). This enhancement owes to the
$\alpha+{}^{16}{\rm O}$ clustering in the ground state as we see from the large inter-cluster
distance ($R(0^+_1)=4.0$ fm) in the ground state. As a result, the transition matrices are further
amplified.  

\begin{table}[h]
\caption{The estimated distances $d$ between $\alpha$ and $^{16}{\rm O}$ clusters in the ground
 state, nodal and angular excited states of $^{20}{\rm Ne}$ in unit of fm. The IS monopole and
 dipole transition  matrices from the ground state to the nodal and angular excited states are given
 in units of   $\rm fm^2$ and $\rm fm^3$, respectively. Numbers in parenthesis are ratio to the
 Weisskopf  estimates. The data is taken from Ref. \cite{Chiba:2015khu}.}  \label{tab:mo1}       % Give a unique label
% For LaTeX tables use
\begin{tabular}{cccccc}
 \hline\noalign{\smallskip}
 $^{20}{\rm Ne}$& $d(0^+_1)$ & $d(0^+_4)$ & $d(1^-_1)$ & $M^{IS0}$ & $M^{IS1}$  \\
 \noalign{\smallskip}\hline\noalign{\smallskip}
 BB GCM & 5.0        & 6.5        & 5.5        & 46 (7.3)     & 90 (11)      \\
 AMD    & 4.0        & 6.0        & 5.0        & 16 (2.5)     & 38 (4.5)       \\
\noalign{\smallskip}\hline
\end{tabular}
\end{table}

\begin{table}[h]
\caption{The estimated distances between $\alpha$ and $^{18}{\rm O}$ clusters in the ground state, 
 molecular-orbit and atomic-orbit states of $^{22}{\rm Ne}$ in unit of fm. The IS monopole and
 dipole transition  matrices from these cluster states are given in units of  $\rm fm^2$ and $\rm
 fm^3$,  respectively. These results are obtained by AMD. Numbers in parenthesis are ratio to the 
 Weisskopf  estimates.}  \label{tab:mo2}  % Give a unique label 
% For LaTeX tables use
\begin{tabular}{ccccccc}
 \hline\noalign{\smallskip}
 $^{22}{\rm Ne}$& $0^+_1$ & $0^+_2$ & $1^-_1$ & $1^-_2$ & $0^+_3$ & $1^-_3$  \\
 \noalign{\smallskip}\hline\noalign{\smallskip}
 $d$    & 3.25    & 3.75 & 5.0 & 5.5 & 6.5  & 6.75     \\
 $M^{IS0/IS1}$ &  & 1.7  & 5.8 & 1.9  & 24 & 22    \\
               &  &(0.26)& (0.62) &(0.2) & (3.5)  &(2.4)  \\
\noalign{\smallskip}\hline
\end{tabular}
\end{table}
We also comment on the transitions in $^{22}{\rm Ne}$. As already discussed in the section
\ref{sec:4.2}, the valence neutrons yield two different kinds of cluster states; the
molecular-orbit states ($0^+_2$, $1^-_1$ and $1^-_2$) and the atomic-orbit states ($0^+_3$ and
$1^-_3$). Table \ref{tab:mo2} 
summarizes the inter-cluster distances and transition strengths of these cluster states. The
atomic-orbit states have large transition strengths from the ground state, because they are
regarded as an ordinary di-cluster states composed of $\alpha$ and $^{18}{\rm O}$ clusters, and
hence, the discussion made in the previous section can be applied. On the other hand, the
molecular-orbits states have hindered transition strengths in spite of their pronounced
clustering. This difference is explained as follows. The transitions from the ground state to the
molecular-orbit states involve the rearrangement of two valence neutron configuration in addition
to the cluster excitation of the core. However, the IS monopole and dipole operators cannot excite
the cluster core and valence neutrons simultaneously. As a result, the transitions to the
$\sigma$-bond cluster states (the cluster states with the rearrangement of valence neutrons) are
hindered. In other words, the IS monopole 
and dipole transitions are sensitive to the excitation of the inter-cluster motion, but insensitive
to the valence neutron excitation. These property has been utilized for the search of the
atomic-orbit states in Be isotopes \cite{Yang:2014kxa,Yang:2015lha}, and also discussed in the
next section.

\subsubsection{sAMD results for light stable nuclei}

In IS monopole strengths in $^{16}$O, significant percentages
of the energy-weighted sum rule (EWSR) has been found in a low-energy region. 
Yamada {\it et al.} pointed out that 
two different types of IS monopole excitations exist in $^{16}$O
\cite{Yamada:2011ri}:
the low-energy IS monopole strengths of excitations into cluster states in $E\lesssim 16$ MeV are separated from
high-energy IS monopole strengths for the IS giant monopole resonance (ISGMR) in $E> 16$ MeV, which corresponds to the collective vibration mode described by coherent one-particle and one-hole (1p-1h) excitations in a mean field.
The separation of the low-energy IS monopole strengths from the ISGMR was also found in $^{12}$C \cite{Youngblood:1998zz,John:2003ke}. 
In order to theoretically describe these two kinds of monopole modes, 
the cluster excitation and collective breathing mode
(coherent 1p-1h excitations), an extended version 
of antisymmetrized molecular dynamics
called ``shifted basis AMD (sAMD)'' combined with the cluster GCM 
has been constructed and applied to IS monopole and dipole
excitations in light nuclei 
\cite{Kanada-En'yo:2013dma,Kanada-En'yo:2015vkg}.

\begin{figure}[htb]	
\begin{center}
\resizebox{0.5\textwidth}{!}{%
\includegraphics{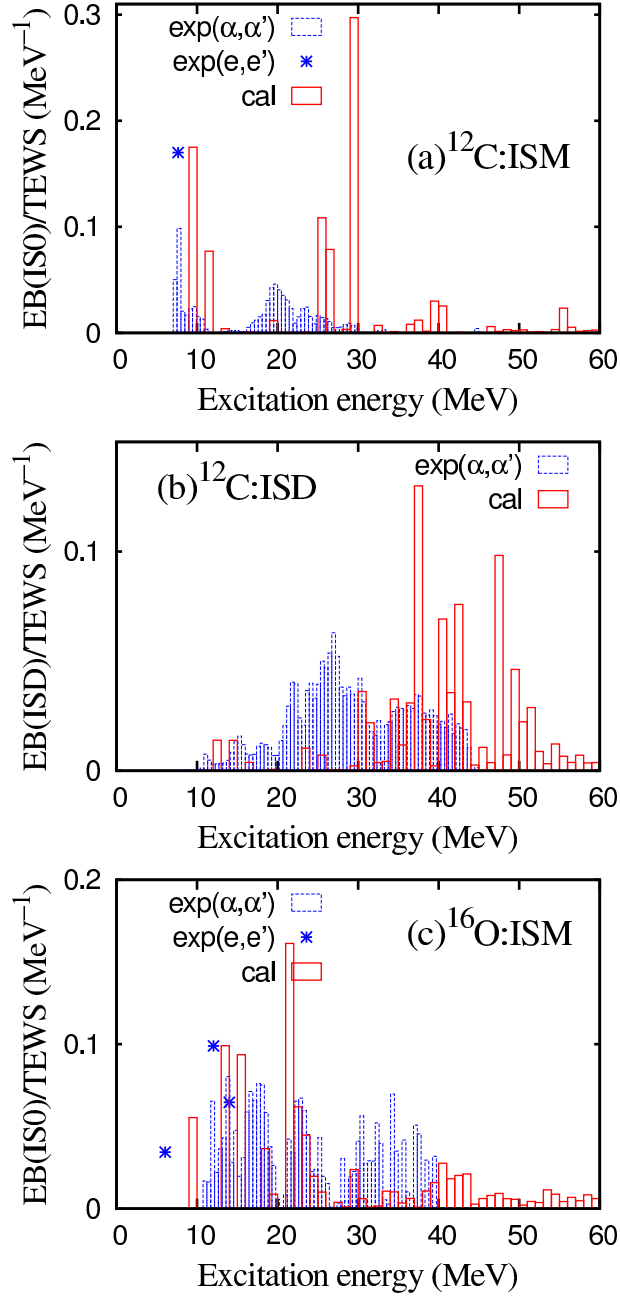} }	
%\includegraphics[width=9.0cm]{beiso-e1-fig-kimura.eps} 	
%\vspace{0.5cm}
  \caption{
Energy-weighted strengths of IS monopole and dipole transitions of
$^{12}$C and $^{16}$O calculated using the sAMD combined with the 
cluster GCM, 
and those measured by
$(\alpha,\alpha')$\cite{John:2003ke,Lui:2001xh} and $(e,e')$\cite{Chernykh:2010zu} scatterings.
Figures are taken from Refs.~\cite{Kanada-En'yo:2013dma,Kanada-En'yo:2015vkg}.
\label{fig:c12-o16-ism}}
\end{center}
\end{figure}

Figure \ref{fig:c12-o16-ism} shows
IS monopole and dipole strengths for $^{12}$C
calculated using the sAMD with the $3\alpha$ GCM, and
IS monopole strengths for $^{16}$O calculated using 
the sAMD with the $^{12}$C+$\alpha$ GCM. The 
MV1  ($m=0.62$, $b=h=0$) central and G3RS ($u_1=-u_2=-3000$ MeV) 
spin-orbit interactions are used.
Experimental data measured by 
$(\alpha,\alpha')$\cite{John:2003ke,Lui:2001xh} and $(e,e')$\cite{Chernykh:2010zu} scatterings are also shown in the figure.

In the calculated IS monopole strengths of $^{12}$C, we found 
two kinds of excitations: low-energy strengths around 
$E_x=10$ MeV for cluster excitations and high-energy strengths of the 
ISGMR for the collective breathing mode.
The experimentally observed IS monopole strengths of $^{12}$C 
show that the low-energy IS monopole strengths in $E\lesssim 12$ MeV
exhaust significant percentages of the EWSR comparable to the high-energy strengths in $E> 12$ MeV of the ISGMR. 
The calculation describes this separation of the low-energy 
cluster modes and the high-energy ISGMR.  
Although the quantitative reproduction of the 
ISGMR energy and width is not satisfactory, the 
ratio of the IS monopole strengths of the low-energy to high-energy parts 
is reproduced well  in the present calculation. 
Also in the IS dipole strengths of $^{12}$C, we obtain 
low-energy strengths in $E_x=10-15$ MeV 
which are contributed by $1^-$ states of 
cluster excitations. These strengths may correspond to 
the significant IS dipole strengths experimentally measured 
in $E_x=10-20$ MeV region. 

In the calculated IS monopole strengths of $^{16}$O, 
three peaks for cluster states are obtained in the 
low-energy part ($E_x < 16$ MeV).
This is consistent with the 
$4\alpha$ orthogonality condition model calculation by Yamada {\it et al.} \cite{Yamada:2011ri}.
In the high-energy part, significant strengths exist in $E_x=20-25$ MeV
region corresponding to the ISGMR for the collective breathing mode.

%%%%%%%%%%%%%%%%%%%%%%%%%%%%%%
%%%%%%%%%%%%%%%%%%%%%%%%%%%%%

\subsection{Monopole and dipole transitions in neutron-rich nuclei}\label{sec:5.2}

\subsubsection{Isovector dipole excitations in Be isotopes}
Low-energy isovector (IV) dipole excitation is one of the current issues 
concerning exotic excitation modes in neutron-rich nuclei. 
In neutron-rich nuclei, low-energy IV dipole resonances (LEIVDRs) are expected 
to arise from excess neutron motion against a core nucleus. 
On the other hand, IV giant dipole resonances (IVGDRs), which have been systematically 
observed in stable nuclei,  
are understood as a collective mode of the opposite
motion between protons and neutrons.
In prolately 
deformed stable nuclei,  a two peak structure of the IVGDRs has been observed 
because of longitudinal and transverse oscillations.
In deformed neutron-rich nuclei such as Be isotopes, 
a variety of dipole excitation modes may exist because of excess neutron motions 
on the top of the deformed ground states.
Questions to be answered are whether LEIVDRs appear and
how the IVGDR is affected in the presence of excess neutrons. 

\begin{figure}[htb]
\begin{center}
\resizebox{0.7\textwidth}{!}{%
\includegraphics{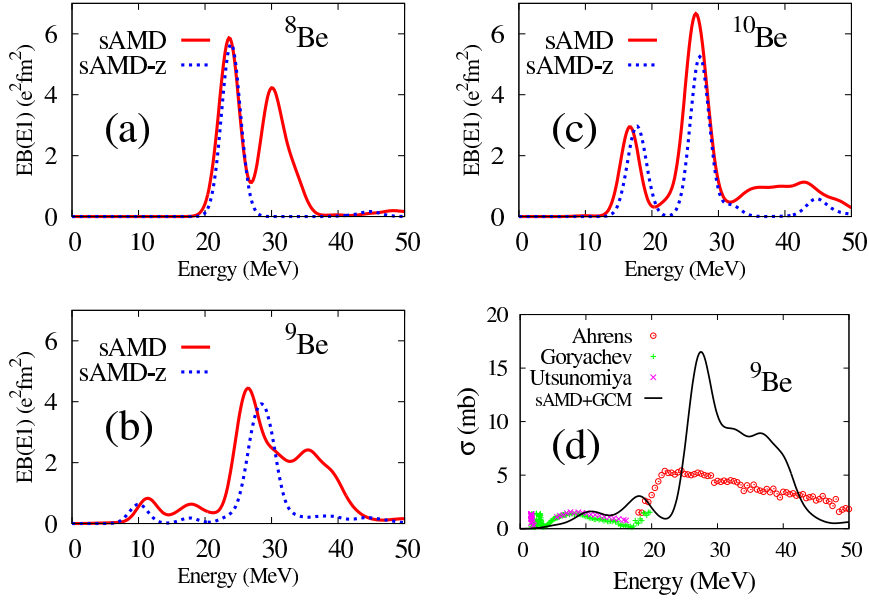} }	
%\includegraphics[width=9.0cm]{beiso-e1-fig-kimura.eps} 	
%\vspace{0.5cm}
  \caption{
Energy-weighted E1 strengths of (a)$^8$Be, (b)$^9$Be, and (c)$^{10}$Be obtained by the sAMD calculations
using the MV1  ($m=0.62$, $b=h=0$) central and G3RS ($u_1=-u_2=-3000$ MeV) 
spin-orbit interactions. 
Results of the truncated calculations for the longitudinal mode (sAMD-z) are also shown. (d)
The $E1$ cross sections of $^9$Be calculated by 
sAMD+$\alpha$GCM+cfg \cite{Kanada-En'yo:2015ttw} compared with 
the experimental photonuclear cross sections.
The experimental data are those by Ahrens {\it et al.} \cite{Ahrens:1975rq},
Goryachev {\it et al.} \cite{Goryachev92}, and 
Utsunomiya {\it et al.} \cite{Utsunomiya:2015bus}.
The theoretical strengths are smeared by Gaussian with a width 
$\gamma=2$ MeV. 
Figures are taken from Ref.~\cite{Kanada-En'yo:2015ttw}.
\label{fig:beiso-e1}}
\end{center}
\end{figure}

The IV and IS dipole excitations in $^{9}$Be and $^{10}$Be 
have been investigated using the sAMD+GCM \cite{Kanada-En'yo:2015ttw}.
IV dipole excitations for 
$^8\textrm{Be}(0^+_1)\to ^8\textrm{Be}(1^-)$, 
$^9{\rm Be}(3/2^-_1)\to ^9{\rm Be}(1/2^+, 3/2^+, 5/2^+)$, and 
$^{10}\textrm{Be}(0^+_1)\to ^{10}\textrm{Be}(1^-)$ calculated by the sAMD 
using the MV1  ($m=0.62$, $b=h=0$) central and G3RS ($u_1=-u_2=-3000$ MeV) 
spin-orbit interactions 
are shown in Fig.~\ref{fig:beiso-e1}(a)-(c). In Fig.~\ref{fig:beiso-e1} (d), 
the calculated 
$E1$ cross sections 
%$\sigma(E)= \frac{16\pi^3}{9}\frac{e^2}{\hbar c}E\frac{dB(E1)}{dE}$
of $^9$Be are compared with the experimental photonuclear cross sections.

The IVGDR in $^8$Be splits into two peaks as expected from the $2\alpha$ cluster structure 
with a prolate deformation. The lower peak of the IVGDR
at $E = 20-25$ MeV is contributed by the longitudinal mode, whereas the higher 
peak of the IVGDR around $E=30$ MeV comes from the transverse mode of the $2\alpha$
cluster. In $^9$Be and $^{10}$Be, the $E1$ strengths for the IVGDR are obtained 
in the $E>20$ MeV region. The IVGDRs in $^9$Be and $^{10}$Be are 
contributed by the IV dipole strengths in the $2\alpha$ core, and they show 
the two peak structure because of the prolately deformed core. However, 
the higher peak for the transverse mode is significantly affected by excess neutrons.
The higher peak somewhat broadens in $^9$Be and it is highly fragmented 
in $^{10}$Be.
In contrast, the shape of the lower peak is almost same as 
that of $^8$Be indicating that excess neutrons do not affect so much the longitudinal mode.
In the ground states of $^9$Be and $^{10}$Be, 
excess neutrons dominantly occupy the $\pi$ orbits as discussed 
previously.  The $\pi$-orbit neutrons are distributed in the transverse region, and therefore they affect only the transverse mode but not 
the longitudinal mode in the IVGDR strengths.

In the $E<20$ MeV region below the IVGDR energy, 
the low-energy IV dipole strengths appear in $^9$Be and $^{10}$Be 
because of the valence neutron motion against the $2\alpha$ core. 
In $^9$Be, the low-energy IV dipole strengths are well separated from the IVGDR strengths
and exhaust about 10\% of the energy weighted sum of the 
calculated $E1$ strengths (20\% of the Thomas-Reiche-Kuhn sum rule). This result is consistent with the
experimental strength distributions,
though the calculation underestimates the width of the IVGDR (see Fig.~\ref{fig:beiso-e1} (d)).
In $^{10}$Be, the strong IV dipole excitation is found at $E\sim 15$ MeV.
This IV dipole resonance 
is caused by the negative-parity excitation of the inter-cluster motion between $^6$He and 
$\alpha$ clusters and regarded as the parity partner of the 
$0^+$ state of the $^6$He+$\alpha$ cluster resonance
predicted in $10<E<15$ MeV. In single-particle description, 
the $E1$ strength from the ground state to the dipole resonance is 
enhanced remarkably by the coherent contribution of two valence neutrons.

It should be noted that there are low-lying states having relatively weak
$E1$ strengths,
$^9$Be($1/2^+_1$), $^9$Be($3/2^+_1$), and $^9$Be($5/2^+_1$) in $E<5$ MeV 
and $^{10}$Be($1^-_1$) in $E<10$ MeV.
These low-lying states are dominated by single-particle
excitations from the $\pi_{3/2}$ orbit to the $\sigma_{1/2}$ orbit
in the molecular orbit configurations. Since the $\pi_{3/2}$ and $\sigma_{1/2}$ orbits
have different node numbers for the transverse ($n_{\perp}$) and longitudinal ($n_z$) directions:
the $\pi_{3/2}$  orbit has node numbers $(n_{\perp},n_z)=(1,0)$, whereas the 
$\sigma_{1/2}$ orbit has $(n_{\perp},n_z)=(0,2)$. From the difference in the 
node numbers, it is clear that 
$E1$ transition is forbidden between these orbits. 
%As shown later,
%the $^6$He+$\alpha$ cluster components are somewhat mixed
%with the dominant MO configurations in the low-lying states of  $^{10}$Be, 
%and contribute to the $E1$ strength in $E<10$ MeV.

%%%%%%%%%%%%%%%%%%%%%%%%%%%%%%
%%%%%%%%%%%%%%%%%%%%%%%%%%%%%

\subsubsection{Isoscalar monopole and dipole excitations in $^{10}$Be}
IS monopole excitations are more direct 
probes for cluster states because the IS monopole operator strongly 
excites the inter-cluster mode as discussed previously.
The IS dipole excitations can be also 
useful probes to search for cluster states, 
because the IS dipole operator excites 
compressive modes similar to the IS monopole operator and 
it is expected to cause strong transitions into 
opposite parity cluster states \cite{Kanada-En'yo:2015vkg,Chiba:2015khu}. 

\begin{figure}[htb]
\begin{center}
\resizebox{0.55\textwidth}{!}{%
\includegraphics{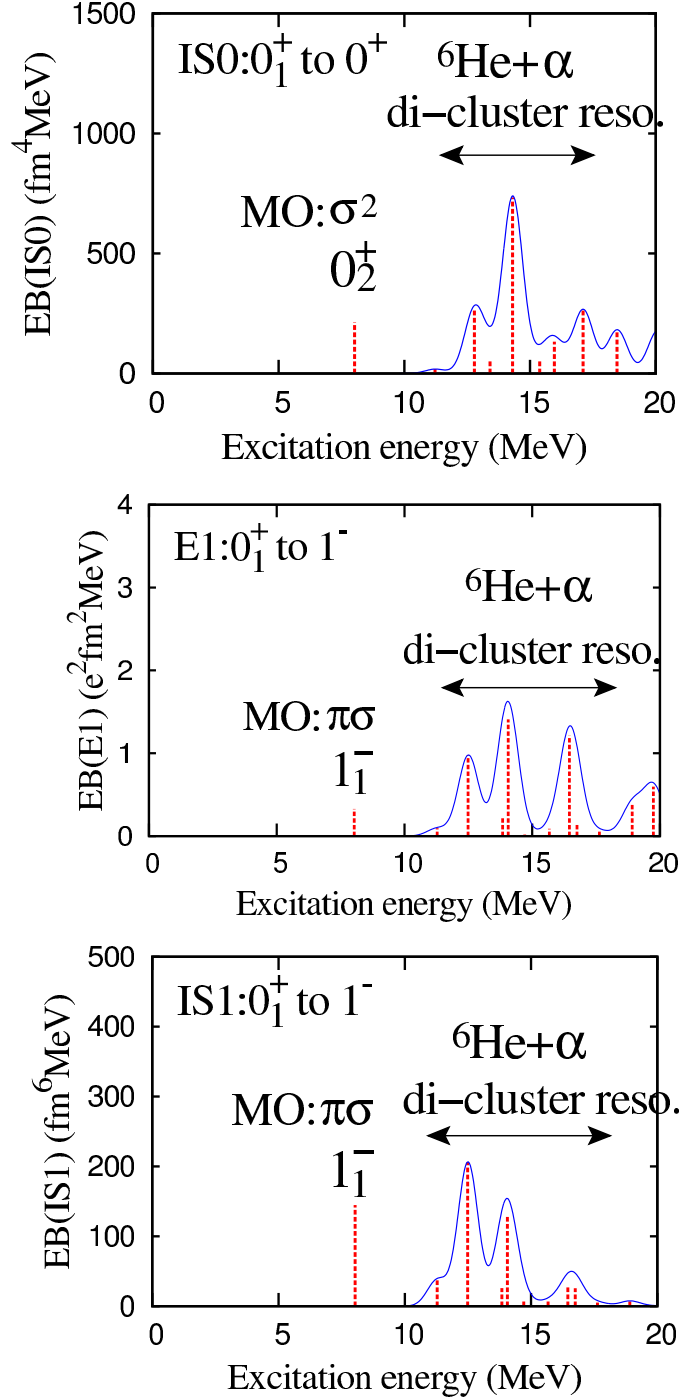} }	
%\includegraphics[width=12.0cm]{be10-is0-e1.eps} 	
%\vspace{0.5cm}
  \caption{IS monopole, $E1$, and IS dipole strengths of $^{10}$Be 
obtained by the $^6$He+$\alpha$ cluster model 
calculation using 
the Volkov No.2  ($m=0.6$, $b=h=0.125$) central and G3RS  ($u_1=-u_2=-1600$ MeV)  spin-orbit interactions \cite{KanadaEn'yo:2011nc,Kanada-En'yo:2016usa}.
Energy weighted strengths are shown by dashed lines.
For states above the $\alpha$-decay energy (10.1 MeV), 
energy weighted strength distributions, 
$EB(IS0,E1,IS1)$, smeared by 
Gaussian with $\gamma=1/\sqrt{\pi}$ MeV are shown by solid lines. The figures are taken from Ref.~\cite{Kanada-En'yo:2016usa}.
\label{fig:beiso-is0}}
\end{center}
\end{figure}

The IS monopole, $E1$, and IS dipole excitations from the $^{10}$Be ground state 
are investigated with the $^6$He+$\alpha$ GCM model using 
the Volkov No.2  ($m=0.6$, $b=h=0.125$) central and G3RS  ($u_1=-u_2=-1600$ MeV)  spin-orbit 
interactions.
In the calculation, the inter-cluster 
distance $R$ between $^6$He and $\alpha$ is treated as the generator coordinate, and 
$p$-shell configurations of $^6$He are taken into account.
At a moderate inter-cluster distance, the $^6$He+$\alpha$ GCM model 
describes approximately the molecular orbit configuration because of the 
antisymmetrization between single-particle wave functions. It also contains
$^6$He+$\alpha$ cluster resonances and discretized continuum 
in a finite box boundary, $R\le 15$ fm. The details of the 
$^6$He+$\alpha$ GCM calculation are described in Refs.~\cite{KanadaEn'yo:2011nc,Kanada-En'yo:2016usa}.
The calculated IS monopole, $E1$, and IS dipole strengths are shown in Fig.~\ref{fig:beiso-is0}. 
$^6$He+$\alpha$ cluster resonances 
above the $\alpha$-decay threshold (10.1 MeV) are 
fragmented into several states because of  coupling with discretized 
continuum states as well as channel coupling. 
Nevertheless, the IS monopole strengths are concentrated around $E\sim
15$ MeV. The peak in the IS monopole strengths is dominantly contributed 
by the $^6$He($0^+$)+$\alpha$ resonance, whereas the broad distribution in 
$E=12\sim 20$ MeV contains contributions from both 
$^6$He($0^+$)+$\alpha$ and $^6$He($2^+$)+$\alpha$ cluster resonances. 
It should be commented that
the $^6$He($0^+$)+$\alpha$ and $^6$He($2^+$)+$\alpha$ channels can be excited by the
IS monopole operator
because both components are already contained in the  $^{10}$Be ground state.
In contrast to the significant IS monopole strengths for the cluster resonances, 
the IS monopole strength for the $0^+_2$ state with the dominant $\sigma^2_{1/2}$ configuration  
is relatively weak because the IS monopole operator
does not excite the $\sigma^2_{1/2}$ configuration from the ground state
$\pi^2_{3/2}$ configuration.
The small but finite IS monopole strength for the $0^+_2$ state originates in 
a minor component of $^6$He($0^+$)+$\alpha$ cluster 
mixed in the dominant $\sigma^2_{1/2}$ configuration. 

In the $E1$ strength distribution, remarkable strengths in $E=10-20$ MeV 
correspond to the LEIVDR of the 
$^6$He+$\alpha$ cluster mode, which is obtained at $E\sim 15$ MeV
in the sAMD calculation shown in Fig.~\ref{fig:beiso-e1} (b).
In the $^6$He+$\alpha$ GCM calculation, this IV dipole resonance spreads
over the $E=10-20$ MeV region.
From the point of view of neutron 
configurations, the enhanced $E1$ strength for the $^6$He+$\alpha$ cluster mode
can be understood by coherent contributions of two valence 
neutrons around the $2\alpha$.
For the lowest $1^-$ state at $E=8$ MeV,
which is assigned to the experimental $1^-$ state at 5.96 MeV, the $E1$ strength 
is relatively weak consistently with the sAMD result 
because the state is dominated by the $\pi_{3/2}\sigma_{1/2}$ configuration,
for which the $E1$ transition is suppressed, as mentioned previously. 

A more direct probe for negative-parity cluster resonances is the IS dipole
excitation. Similarly to the IS monopole operator, 
the IS dipole operator contains the $r^3$ term as the leading order contribution,
and hence, the IS dipole strength is sensitively enhanced by the excitation of the inter-cluster motion.
As expected, the IS dipole strengths are remarkable in $E=10-15$ MeV
for the $^6$He+$\alpha$ cluster resonances. Also for the $1^-_1$ state at $E=8$ MeV, 
the IS dipole strength is significant 
because of some mixing of the $^6$He+$\alpha$ cluster component in the 
dominant $\pi_{3/2}\sigma_{1/2}$ configuration in the $1^-_1$ state. 
Consequently, the low-energy IS dipole strengths for 
the $^6$He+$\alpha$ cluster mode are distributed in
several states in the $E=5-15$ MeV region including the $1^-_1$ state
through the state mixing.

The present results indicate that 
the di-cluster resonances in $^{10}$Be shows the enhanced 
IS monopole, $E1$, and IS dipole strengths.
Similar discussions of the enhanced IS monopole strengths for di-cluster resonances 
have been made for the 
$^8$He+$\alpha$ cluster resonance in $^{12}$Be($0^+_3$) by Ito {\it et al.} \cite{Ito:2011zza}.
In contrast to the di-cluster resonances in highly excited states, 
the strengths to low-lying states are relatively suppressed because 
they are dominated by the particle-hole excitations 
in the molecular orbit configurations and have 
neutron configurations different from the initial ground states.
In general, even though an excited state in neutron-rich nuclei 
has a developed cluster structure, the IS monopole and dipole strengths can be relatively 
weak if the excited state has an excited neutron configuration.
It means that the IS monopole and dipole strengths are useful observables to 
identify excitation modes of cluster states; the single-particle excitation on the top of the cluster state
or the excitation of the inter-cluster motion.
We can conclude that the IS monopole and dipole strengths can be good probes for mode analysis of 
cluster states and also useful to discover new cluster states in neutron-rich nuclei as well as stable nuclei.

%%%%%%%%%%%%%%%%%%%%%%%%%%%%%%
%%%%%%%%%%%%%%%%%%%%%%%%%%%%%
\section{Summary}\label{sec:6}
In summary, we have reviewed the clustering phenomena in unstable nuclei from a point-of-view
provided by the studies based on AMD. In particular, we put focus on the molecule-like
states generated by the excess neutrons. 

Whether the excess neutrons (protons) will reduce or enhance the clustering of unstable nuclei is
non-trivial and important question, because the ordinary Ikeda threshold rule cannot be applied
to unstable nuclei. 

The AMD studies showed that all Be isotopes have $2\alpha$
cluster core surrounded by the excess neutrons, and the degree of the $2\alpha$ clustering  
(inter-cluster distance between $2\alpha$ clusters) are changed depending on the neutron
number. The clustering is first reduced toward $N=6$ system, but then, it is enhanced from $N=7$
system toward neutron drip line. It is also shown that the clustering plays an important role for
the breaking of the neutron magic number $N=8$ in $^{11}{\rm Be}$ and $^{12}{\rm Be}$. 
The B isotopes manifest more drastic change. The stable nucleus $^{11}{\rm B}$ has a compact shell
model like ground state. But the addition of excess neutrons induces the clustering toward the
neutron drip line. Thus, the excess neutrons reduce the clustering near the $N=Z$ nuclei, but
enhance it toward neutron drip line. It is pointed out that these reduction and enhancement of
the clustering are well correlated to the behavior of the proton radii. The proton radii is also
reduced near the $N=Z$ nuclei, then it is increased toward the drip line. The experiments are
now confirming this characteristic behavior of proton radii in Be and B isotopes. 

The discussion is also extended to the Ne and Mg isotopes in which the breaking of the $N=20$ magic
number is well known. By combining the AMD and the density-folding model, it is shown that the
reaction cross section, the matter radii and  quadrupole deformation are correlated to each
other. They are reduced toward the $N=14$ and 16 isotones, then they are enhanced toward the
island of inversion and the trend continues until the neutron drip line. The discontinuous
behavior of the reaction cross section at $N=19$ isotones shows that the west border of the island
of inversion is located at $N=19$. On the other hand, no discontinuity was found in the
neutron-rich side indicating that the island of inversion ($N\simeq 20$) and the other region of
deformation ($N\simeq 28$) are merged. It is also noted that an extended version of AMD called
AMD+RGM reasonably explains the observed reaction cross section of $^{31}{\rm Ne}$ and the
formation of the neutron halo. Similar to the Be isotopes, the deformation change in Ne isotopes
as function of neutron number is attributed to the clustering. By the estimation of the
overlap between proton distributions, it is shown that the inter-cluster distance between $\alpha$
and O clusters are reduced toward $N=16$, but enlarged in the island of inversion. Again the AMD
predicts that the proton radii is well correlated with the clustering. 

Behind the change of the clustering in the ground states of unstable nuclei, the excess neutrons
play the central role. The AMD studies showed the formation of the molecular-orbits around
$2\alpha$ cluster core in Be isotopes without any {\it a priori} assumption. The molecular-orbits
are classified to so-called $\pi$- and $\sigma$-orbits. The former reduces the clustering, while
the latter enhances it. The combinations of molecular-orbits naturally explain the reduction and
enhancement of the clustering in Be isotopes mentioned above. In addition to the clustering of
the ground states, the AMD study  predicted many excited rotational bands with different
configurations of excess neutrons, and showed that the concept of the molecular-orbits is
applicable and useful for the understanding of the excited states.  By further increase of the
excitation energy, the existence of the di-cluster bands is also predicted in which the excess
neutrons are confined either of the clusters analogous to the ionic bonding of molecules.

The concept is also successful for the O, Ne and F isotopes. In the $^{22}{\rm Ne}$, it is found
that the excess neutrons in $\sigma$-orbit induce the clustering of the excited states. Analogous
molecular-orbits are also predicted in O isotopes. Evolution of the molecular orbits in the island
of inversion is examined in the F isotopes. It is pointed out that the cooperative effects of the
clustering and the quenching of $N=20$ shell gap greatly reduce the energies of the
molecular-orbit states near the neutron drip line of F isotopes. The extension of the
molecular-orbit to the three center system is also discussed for the case of C isotopes. The
linear-chain formation in $^{14}{\rm C}$ having the $^{10}{\rm Be}+\alpha$ intrinsic structure
is predicted, which looks consistent with the recent observations. It is also found that the
orthogonality effect is important for the stabilization of the linear chain. 

We also discussed the monopole and dipole transitions as  probe for nuclear clustering in the
excited states. It was shown that the monopole and dipole transitions strongly and selectively
populate the excited cluster states in the stable nuclei. To describe the monopole and dipole
responses, an extended version of AMD named shifted-basis AMD is introduced. The application of
the shifted-basis AMD to neutron-rich Be isotopes revealed that the coupling of the inter-cluster
motion and valence neutrons yields novel types of the excitation modes.

\begin{acknowledgements}
We would like to thank A. Dot\'e, Y. Taniguchi, A. Ono and H. Horiuchi
for fruitful discussions and collaborations. Part of the numerical calculations were
performed by using the supercomputers at RCNP in Osaka University, in the High Energy Accelerator
Research Organization (KEK), and at Yukawa Institute for Theoretical Physics (YITP) in Kyoto
University. 
The supports by the Grants-in-Aid for Scientific Research on Innovative Areas from MEXT Grant
No. 2404:24105008 and JSPS KAKENHI Grant Nos. 16K05339, 26400270 and 15K17662 are acknowledged.
\end{acknowledgements}

\bibliography{refk,refy,refs}

\end{document}